\documentclass[twocolumn]{aastex631}

\usepackage{graphicx}	
\usepackage{amsmath}	
\usepackage{nccmath}
\usepackage{soul} 
\usepackage{afterpage} 
\usepackage{xcolor} %
\usepackage{booktabs}  
\usepackage{lastpage} 
\usepackage{bm} 
\usepackage{float}
\usepackage{wasysym}

\def\eqp#1{(\ref{eq:#1})}
\def\eql#1{\label{eq:#1}}
\newcommand{\be}{\begin{equation}}
\newcommand{\ee}{\end{equation}}
\newcommand{\ba}{\begin{eqnarray}}
\newcommand{\ea}{\end{eqnarray}}
\newcommand{\bas}{\begin{eqnarray*}}
\newcommand{\eas}{\end{eqnarray*}}
\definecolor{purple}{rgb}{0.7,0.0,0.7}

\newcommand\epsfil{\epsilon_\mathrm{filter}}
\newcommand\bfu{\bm{u}}
\newcommand\bfuh{\widehat{\bfu}}
\newcommand\rmi{\mathrm{i}}
\newcommand\rmK{\mathrm{K}}
\newcommand\rms{\mathrm{s}}
\newcommand\rmt{\mathrm{t}}
\newcommand\rmRe{\mathrm{Re}}

\newcommand\nut{\nu_\rmt}
\newcommand\cs{c_\rms}
\newcommand\rmin{r_\mathrm{min}}
\newcommand\rmax{r_\mathrm{max}}
\newcommand\zmin{z_\mathrm{min}}
\newcommand\zmax{z_\mathrm{max}}
\newcommand\p{\partial}
\newcommand\pp{{\prime\prime}}
\newcommand\urpp{u_r^{\pp}}
\newcommand\uzpp{u_z^{\pp}}

\newcommand\bfx{\mathbf{x}}
\newcommand\bfk{\mathbf{k}}

\newcommand\bfF{\mathbf{F}}

\newcommand\ft{\widetilde{f}}
\newcommand\ftilde{\widetilde{f}}
\newcommand\fpp{f^{\prime\prime}}
\newcommand\rhob{\overline{\rho}}
\newcommand\pb{\overline{p}}
\newcommand\ut{\widetilde{u}}
\newcommand\rc{r}
\newcommand\bfut{\widetilde{\bfu}}
\newcommand\urt{\ut_r}
\newcommand\upt{\ut_\phi}
\newcommand\uzt{\ut_z}
\newcommand\uppp{u_\phi^{\prime\prime}}
\newcommand\bfupp{\bfu^{\prime\prime}}
\newcommand\uapp{u_a^\pp}
\newcommand\ubpp{u_b^\pp}
\newcommand\mdot{\dot{m}}
\newcommand\ppr{\frac{\p}{\p r}}
\newcommand\ppz{\frac{\p}{\p z}}
\newcommand\Tpr{T_{\phi r}}
\newcommand\Tpz{T_{\phi z}}
\newcommand\ci{c_\mathrm{i}}
\newcommand\bfe{\mathbf{e}}
\newcommand\bfer{\bfe_r}
\newcommand\bfez{\bfe_z}
\newcommand\bfephi{\bfe_\phi}
\newcommand\ppt{\frac{\p}{\p t}}
\newcommand\ubar{\overline{u}}
\newcommand\rhobar{\overline{\rho}}
\newcommand\vb{\overline{v}}
\newcommand\up{u^\prime}
\newcommand\vp{v^\prime}
\newcommand\vt{\widetilde{v}}
\newcommand\upp{u^{\prime\prime}}
\newcommand\vpp{v^{\prime\prime}}
\newcommand\bfomega{\bm{\omega}}
\newcommand\bfzeta{\bm{\zeta}}
\newcommand\bfn{\bm{n}}
\newcommand\X{\times}
\newcommand\tauc{\Omega_\rmK t_\mathrm{rad}}
\newcommand\jzt{\breve{j}_z}

\newcommand\calL{\mathcal{L}}

\newcommand\lammax{\lambda_{r,\mathrm{max}}}
\DeclareMathOperator\sgn{sgn}
\DeclareMathOperator\real{Re}

\submitjournal{ApJ}

\shorttitle{Vertical shear instability}
\shortauthors{Shariff \& Umurhan}

\begin{document}

\title{A high resolution simulation of protoplanetary disk turbulence driven by the vertical shear instability}

\author[0000-0002-7256-2497]{Karim Shariff}
\affiliation{NASA Ames Research Center \\
Moffett Field, CA 94035, USA}

\author{Orkan M. Umurhan}
\affiliation{SETI Institute\\
Mountain View, CA 94043, USA}



\begin{abstract}

A high resolution fourth-order Pad\'e scheme is used to simulate locally isothermal 3D disk turbulence driven by the vertical shear instability (VSI) using 268.4 M points. In the early non-linear period of axisymmetric VSI, angular momentum transport by vertical jets creates correlated N-shaped radial profiles of perturbation vertical and azimuthal velocity.  This implies dominance of positive perturbation vertical vorticity layers and a recently discovered angular momentum staircase with respect to radius ($r$). These features are present in 3D in a weaker form.
The 3D flow consists of vertically and azimuthally coherent turbulent shear layers containing small vortices with all three vorticity components active.
Previously observed large persistent vortices in the interior of the domain driven by the Rossby wave instability are absent. We speculate that this is due to a weaker angular momentum staircase in 3D in the present simulations compared to a previous simulation. The turbulent viscosity parameter $\alpha(r)$ increases linearly with $r$.  At intermediate resolution, the value of $\alpha(r)$ at midradius is close to that of a previous simulation. The specific kinetic energy spectrum with respect to radial wavenumber has a power law region with exponent $-1.84$, close to the value $-2$ expected for shear layers. The spectrum with respect to azimuthal wavenumber has a $-5/3$ region and lacks a $-5$ region reported in an earlier study. Finally, it is found that axisymmetric VSI has artifacts at late times, including a very strong angular momentum staircase, which in 3D is present weakly in the disk's upper layers.

\end{abstract}



\section{Introduction} \label{sec:intro}
\subsection{General introduction and motivation}

A question posed by astronomical observations of protoplanetary disks (PPDs), the history of our own solar system, and the results of theoretical and numerical modeling is: what types of flow, temperatures, and magnetic field configurations occur at different locations and during different periods of disk evolution, and how do they impact the dynamics and chemistry of solid particles which eventually form planets? The traditional motivation for studying flow turbulence in PPDs was to explain the so-called ``anomalous viscosity'' that gives rise to an accretion flow into the central star.  More recently, however, the study of PPD turbulence is also very much motivated by the desire to explore its role in the pathway from micron sized grains to asteroid-sized bodies, called planetesimals, that are believed to be the progenitors of rocky planets, the putative solid cores of gas giant planets, and whose remnants are today's asteroids, comets, and Kuiper belt objects. This motivation is reflected in a wide ranging 19-author review \citep{Lesur_etal_2023} examining various aspects of disk turbulence both in isolation and in interaction with particles and grains embedded in the gaseous nebula. 
 
With respect to turbulence in isolation, it has come to be realized that that the ionized fraction needed to sustain magnetorotational instability (MRI) occurs only at small radii near the star, large radii where the surface density is sufficiently low, and the upper layers of the disk \citep{Lesur_etal_2023}.  At the same time, several hydrodynamic modes of instability have been identified.  These are the sub-critical baroclinic instability (SBI), convective over-stability (COS), zombie vortex instability (ZVI), and, vertical shear instability (VSI), which is the subject of the present work.  Each of these is active for a certain range of radiative relaxation time, which depends on the number density and size distribution of dust grains.

VSI has received the lion's share of interest over the last few years since it is thought to be the most active turbulence generating mechanism within the solar nebula's midplane regions
during the first million years 
\citep[e.g.,][]{Fukuhara_etal_2021},
a time when the first planetesimals were made.  Of the many open questions in this respect, perhaps the most pressing is the role that turbulence plays in bringing about or thwarting the formation of planetesimals \citep[e.g., see recent discussions in][]{Schaefer_etal_2020,Estrada_and_Umurhan_2023}.  In order to answer this, we must develop an understanding of the strength and structure of turbulence at different length scales in PPD settings.  For example, how much of the turbulent energy generated by VSI at large scales propagates down to the scales where particles coalesce to form planetesimals?  How does it subsequently influence particle dynamics at those scales?
We take a much-needed step in this direction by examining VSI (without particles) at sufficiently high resolution to capture a larger range of its small scale structure.

\subsection{Physics of VSI and previous theoretical work}\label{sec:VSI_physics}

The present work employs cylindrical coordinates $(z, r, \phi)$ which form a right-handed system with unit vectors satisfying $\bfephi = \bfez \times \bfer$, and $\bfephi$ pointing \textit{into} the page.  The spherical radius is denoted $R = \left(z^2 + r^2\right)^{1/2}$.

Research on VSI in protoplanetary disks has its roots in work involving the radiative zones of differentially rotating stars \citep{Goldreich_and_Schubert_1967, Fricke_1968}.  In that context, thermal diffusivity is much larger than kinematic viscosity, allowing displaced fluid elements to maintain their angular momentum but not their entropy.  As explained by \citet{Nelson_etal_2013}, the latter removes the stabilizing effect of vertical buoyancy, while the presence of a vertical shear allows the stabilizing effect of a positive radial gradient of angular momentum to be overcome for modes that have a sufficiently short radial wavelength.

\citet{Urpin_and_Brandenburg_1998} and \cite{Urpin_2003} were the first to suggest that VSI could be a source of turbulence in PPDs and present a local stability analysis, modeling radiative transfer using Fick's law for thermal conduction.  The first simulations of VSI were performed by \cite{Arlt_and_Urpin_2004} using {\sc Zeus-3D}; they also present a simple local stability analysis for isothermal flow.  Their simulations capture the correct growth-rate of the instability and extend into the nonlinear regime, providing the first estimates of the turbulent $\alpha$ parameter arising from VSI. However, their nonlinearly saturated states seem inconclusive, particularly when comparing turbulent transport characteristics between their small and large radial domain simulations.

As such, the aforementioned work failed to generate interest until \citet{Nelson_etal_2013} accidentally encountered the instability while trying to test the basic state for an MRI simulation.  They presented the first plots of the flow structure in axisymmetric simulations, as well as in a 3D simulation with an azimuthal domain size of $\Delta\phi = \pi/4$. These results demonstrate that well-developed VSI consists of radially narrow, vertically oriented structures of vertical velocity with alternating sign.
They modelled radiative effects as a relaxation to the basic state temperature and found that the flow is unstable only when the relaxation time, $t_\mathrm{rad}$, is comparable to, or shorter than the orbital time.  They also extended earlier local stability analyses into the vertically global regime.

\cite{Nelson_etal_2013} provided the impetus for many subsequent analytical and numerical investigations.  \citet{Lin_and_Youdin_2015} performed a detailed linearized analysis of the effect of radiative relaxation time and obtain the following local criterion for VSI to be active:
\be
   t_\mathrm{rad} \lesssim \frac{|r (\p\Omega/\p z)|}{N_z^2}.
\ee
Since the (stabilizing) Brunt-V\"ais\"al\"a frequency $N_z$ increases more rapidly with $|z|$ than the vertical shear $\p\Omega/\p z$, surface modes in the upper regions of the disk are the first to be damped as $t_\mathrm{rad}$ increases.  They obtain a key criterion for global instability:
\be
   \Omega_\rmK t_\mathrm{rad} < \frac{h |q|}{\gamma - 1}, \eql{Lin}
\ee
where $h \equiv H/r$ is the ratio of disk scale height to radius, $q$ is the radial temperature exponent, and $\gamma$ is the ratio of specific heats.  Equation \eqp{Lin} says that VSI in thicker disks and those with a steeper radial temperature is less constrained by the relaxation time.  
Based on the assumption of
a minimum mass nebula model,
\citet{Lin_and_Youdin_2015} express the relaxation time in terms of an effective thermal opacity
and find that VSI should be active in the intermediate region of the disk from $r = 5$ to 50 au.  However, follow up studies show that
models with more massive disks or reduced depletion in the small grain population can sensitively alter the medium's opacity, leading to a larger range of the VSI active region
\citep[e.g.,][\S 5.3 of \cite{Lesur_etal_2023}]{Fukuhara_etal_2021}.

In the limit where 
$\Omega_\rmK t_\mathrm{rad} \ll h|q|/(\gamma-1)$, linear disturbances are isothermal.  Several authors
\citep{Nelson_etal_2013,Barker_and_Latter_2015,Umurhan_etal_2016} have shown that
the resulting system of linear perturbation equations may then be analytically examined in the
radially-local but vertically-global limit,
which extends previous fully local analyses
\citep{Goldreich_and_Schubert_1967,Fricke_1968,Arlt_and_Urpin_2004} and Boussinesq
equation treatments \citep[e.g.,][]{Urpin_2003}. The main new feature contained in these semi-global treatments is that unstable modes are exponentially growing oscillations.
\cite{Umurhan_etal_2016} show that the most unstable disturbances have a radial wavelength $\lambda_r$ given by
\begin{equation}
\lammax= \pi |q| h(r) H(r),
\label{lambda_r_def}
\end{equation}
where $H(r)$ is the local disk scale height.
Moreover, they show that modes become purely oscillatory (neither growing nor decaying) for disturbance wavelengths $\lambda_r>\lambda_{r,{\rm marginal}} = 2\lambda_{r,\mathrm{max}}$.
 
Latter and Papaloizou's (\citeyear[][henceforth L\&P]{Latter_and_Papaloizou_2018}) theoretical study begins with a simple but informative set-up, namely, incompressible uniform density flow in a local box that is uniformly sheared radially as well as vertically.  This set-up permits Fourier analysis with the wave vector dependent on time ($\mathbf{k} = \mathbf{k}(t)$), a trick due to Lord Kelvin \citep{Craik_etal_1986} that eliminates terms that are linear in the coordinates.  As is known (\textit{op.~cit.}), for such a set-up the non-linear term for each individual Fourier mode is zero, and therefore each Fourier mode is also a solution to the non-linear equations.  Thus, L\&P note that a single mode can grow to arbitrary amplitude.  In reality, a perturbation will consist of many modes and therefore non-linear interaction among them can lead to a saturated turbulent state.  An elegant result (near their Equation 15) is presented for the angle with respect to the vertical of the axisymmetric VSI jets.  These jets are nearly vertical and their small tilt depends on the vertical shear.  Another simple result is that the growth rate, $\sigma$, of the most unstable mode is proportional to the vertical shear:
\be
   \sigma \approx r |\p\Omega/\p z|,
\ee
for small vertical shear.  Next, L\&P go beyond the uniform density case to consider uniform stratification with the Boussinesq approximation and include temperature and momentum diffusivity.  They find that radially short waves exploit thermal diffusion to circumvent buoyant stabilization.  Finally, L\&P consider secondary (or parasitic) instabilities generated by a primary one with velocity amplitude $V > 0$.  The axisymmetric parasites take the form of Kelvin-Helmholtz (KH) modes generated on vertical jet shear layers of the primary VSI mode.  A necessary condition for instability for moderate $V$ is that the Rossby number
\be
   \mathrm{Ro}_V \equiv \frac{k V}{\Omega} \gtrsim 0.9262,
\ee
where $k$ is the wavenumber of the primary VSI mode.  The presence of $\Omega$ in the denominator expresses the fact that rotation is stabilizing.  If $\mathrm{Ro}_V \gg 1$ such that rotation does not play a role, then the characteristics of the secondary instability are like those of classical KH instability.  For the non-axisymmetric parasites, one must be mindful of the fact that the primary VSI mode also has \textit{azimuthal} jets.  In fact, if we make the approximation that $k_z \approx 0$, i.e., that the VSI mode has a negligible tilt relative to the vertical, then, Equations (11) and (15) in L\&P imply that the amplitude of the azimuthal jets equals that of the vertical jets.  The azimuthal jets are also susceptible to KH instability.  However, the presence of background shear, rotation, and presence of both vertical and azimuthal vorticity modify its characteristics from classical KH instability; this will be seen in \S\ref{sec:at_two_H}.

\citet{Cui_and_Latter_2022} continue the L\&P analysis of parasitic modes and begin by noting that the primary VSI mode observed in global simulations is a standing wave in the vertical direction and radially propagating.  They then look for a pair of inertial modes that grow by resonant interaction with the primary and find that there is an infinity of such pairs that can have much smaller radial wavelength than the primary mode.  Numerical simulations will require many grid cells per primary wavelength $\lambda_{r,\mathrm{max}}$ to resolve them.  

\begin{table*}\centering
\begin{tabular}{l c c c c c c c c c}
\toprule
Reference& $p$ &$q$ & $h$ & $\tauc$ & $n_R\X n_\theta\X n_\phi$&$n_R\X n_\theta\X n_\phi$ per $H_0$&$\Delta R$&$R_0\Delta\theta$&$\Delta\phi$\\
       &        &       &        &              &                  or                       &                 or                                      & or             &     or      &                  \\
       &          &       &       &              & $n_r \X n_z \X n_\phi$         & $n_r \X n_z \X n_\phi$ per $H_0$         & $\Delta r$&$\Delta z$           &\\
\midrule
\cite{Richard_etal_2016}  & $-1.5$ &$-2$ & $0.20$ &0.05-0.5& $500\X  200\X 300$  & $250\X20\X38.2$      &  $2H_0$& $10H_0$&$2\pi/4$\\
\cite{Stoll_and_Kley_2016}&$-1.5$&$-1$&0.05       &0               & $1024\X 256 \X64$   & $38\X 25.6\X 4.1$   & $27H_0$&$10H_0$&$2\pi/8$\\
\cite{Stoll_etal_2017}&$-1.5$&$-1$&0.05       &0                & $600\X 128 \X1024$   & $21\X 12.8\X 8.1$   & $29H_0$&$10H_0$&$2\pi$\\
\cite{Manger_and_Klahr_2018}&$-2/3$&$-1$&$0.10$&$0$&$256\X 128\X 768$ &$17.1^\dagger\X 14.6\X 9.8$ & $15H_0$ & $8.75H_0$ & $2\pi$\\
\cite{Manger_etal_2020} & $-1.5$ &$-1$& $0.03$   & 0  &$256\X 128\X 3402$ & $19.2\X 18.3\X 16.2$ &$13.3H_0$&$7H_0$&$2\pi$\\
\cite{Manger_etal_2020} & $-1.5$ &$-1$& $0.10$   & 0  &$256\X 128\X 1024$ & $17.1\X 18.3\X 13.0$ &$15 H_0$&$7H_0$&$2\pi$\\
 Present & $-1.5$ & $-1$ & $0.10$ &0& $512\X 512\X 1024$ & $74\X 74\X 16$ & $7 H_0$ & $7H_0$ & $2\pi$ \\
\bottomrule
\end{tabular}
\caption{Resolution comparison between representative previous 3D simulations (without radiative transport) and the present one.  $H_0$: scale height at the midradius of the computational domain.  $p$: radial density exponent.  $q$: radial temperature exponent.  $h \equiv H_0/R_0$, the disk aspect ratio.  $R$ and $r$ denote the radius in spherical and cylindrical coordinates, respectively.  $\Delta R$ and $\Delta r$ denote the width of the radial domain and include sponge regions.  The resolution per scale height ($H_0$) is assessed at the midradius of the computational domain. \textdagger: The radial mesh for this simulation is logarithmic in $R$ and the value $17.1$ represents an average resolution per $H_0$.}
\label{tab:previous}
\end{table*}

\subsection{Previous numerical work}
Table~\ref{tab:previous} lists parameters in some previous 3D simulations without radiative transport and compares them with the present one.  The metric for numerical resolution given in the table, introduced by \cite{Manger_and_Klahr_2018}, is the number of grid points per scale height $H_0$ at the midradius of the domain.

The computational study of \cite{Richard_etal_2016} was geared to the study of large vortices with vertical vorticity.  They chose parameters to excite both VSI and sub-critical baroclinic instability which is known to generate such vortices.  Indeed, their simulation produced elliptical vortices via the Rossby wave instability (RWI) which was studied by \citet{Lovelace_etal_1999} for razor thin adiabatic disks.  A necessary condition for the instability is that there be a local maximum of the quantity
\be
   \mathcal{L} \equiv \frac{\Sigma}{2 \omega_z} \left(\frac{P}{\Sigma^\gamma}\right)^{2/\gamma},
\ee
where $\Sigma$ is the surface density, $\omega_z$ the basic state vertical vorticity, $P$ the vertically integrated pressure, and $\gamma$ the adiabatic exponent.  The resulting vortex aspect ratio was found to depend on $t_\mathrm{rad}$.  Table~\ref{tab:previous} shows that their radial resolution is superlative.  Even though their radial domain width is only 2 scale heights, a sufficient number of structures is accommodated.

There is a series of 3D {\sc Pluto} simulations with radiative transport and stellar irradiation \citep{Stoll_and_Kley_2014, Stoll_and_Kley_2016}.  The primary results are: (i) Persistence of VSI with radiative transport.  (ii) A peak in the $T_{r\phi}$ Reynolds stress away from the midplane.
\citet{Stoll_etal_2017} performed vertically isothermal simulations and studied the existence of a meridional mean flow due to Reynolds stress gradients which we confirm in \S\ref{sec:mrmf}. Table~\ref{tab:previous} shows that these simulations are marginally resolved in the azimuthal direction, having $\approx 5$ grid points per scale height along the circumference at midradius.  
\cite{Flock_etal_2020} also performed {\sc Pluto} simulations with radiative transport and stellar irradiation.  Parameters were chosen to represent a typical T Tauri system.  They employed a large number of grid points, $1024 \times 512 \times 2044$ ($n_R \times n_\theta \times n_\phi$), which gives 70 grid cells per scale height.  Their main findings were that: (i) VSI is able to vertically loft 1 mm grains such that their scale height ratio $H_\mathrm{grains}(r)/r = 0.037$; this is much larger than the value $H/r = 0.007$ for the HL Tau system, which is better fit by nonideal MHD simulations \citep{Flock_etal_2017}. (ii) A large persistent vortex is formed by the RWI at a radial location near the inner radial boundary, where the thermal relaxation times becomes low enough that VSI becomes viable.  This causes a peak in the turbulence $\alpha(r)$ parameter, which in turn leads to a dip in surface density and a maximum in the RWI indicator function $\calL(r)$.

\cite{Manger_and_Klahr_2018} performed 3D simulations using {\sc Pluto's} piecewise parabolic scheme.  They observed multiple elliptic eddies in the midplane with anticyclonic vertical vorticity, lifetimes of hundreds of orbits, and aspect ratios $> 8$.  Such vortices are formed only for sufficiently large azimuthal domains with $\Delta\phi=\pi$ and $\Delta\phi = 2\pi$.  Their specific kinetic energy spectrum with respect to the azimuthal wavenumber $m$ has a steeper than $-5/3$ power law region followed by a $-5$ power law region.  Our simulations obtain only the $-5/3$ range
(\S\ref{sec:spectra}).

\citet[][henceforth M20]{Manger_etal_2020} constitutes the latest 3D simulations prior to the present work.  One of their cases has the same disk aspect ratio, and exponents of density and temperature as the present case, namely, $h = 0.1, p = -3/2$, and $q = -1$, respectively.  Compared to their resolution per scale height in the radial and vertical directions, ours is about four times improved, while being comparable in the azimuthal direction.
In addition, the Pad\'e scheme we employ provides better resolution for the same number of grid points \citep{Lele_1992}.
Another metric particularly suited to assess resolution in the meridional plane is the number of grid points per radial wavelength of the fastest growing mode $\lambda_{r,\mathrm{max}}$ introduced in Equation (\ref{lambda_r_def}).  
Most of the simulations listed in Table~\ref{tab:previous} have a temperature exponent of $q = -1$, for which case $h(r) = H_0/r_0$, and Equation (\ref{lambda_r_def}) becomes
\be
   \lambda_{r,\mathrm{max}}(r) = \pi |q| \frac{H_0}{r_0} H(r).
   \eql{lambda_r}
\ee
Here $q$ is the radial temperature exponent such that $T(r) \propto (r/r_0)^q$, $H_0$ is the scale height at midradius ($r_0$), and $H(r)$ is the local scale height (see \S\ref{sec:basic_state}).  To obtain the above stated metric (at midradius), multiply the tabulated number of grid points per scale height in the Table \ref{tab:previous} by $0.31$ for the $h \equiv H_0/r_0 = 0.10$ cases, and by $\approx 0.10$ for the $h = 0.03$ case.  Doing this implies that for $h = 0.10$, we have $22.9$ points in the meridional plane per $\lambda_{r,\mathrm{max}}(r_0)$, while M20 have $5.7$ points.  For their $h = 0.03$ case, M20 have $1.9$ points.  

M20 varied the disk aspect ratio $h\equiv H_0/R_0$ and the density exponent $p$ (see Equation \ref{midplane_density}) and found that the eddy viscosity parameter $\alpha \propto h^{2.6}$.  
They also bolstered their 2018 findings concerning the specific energy spectrum and the formation of large persistent vortices in the midplane.

Most recently, \cite[][henceforth MF24a and MF24b, respectively]{Melon_Fuksman_etal_2024_II, Melon_Fuksman_etal_2024_I} performed high resolution axisymmetric simulations with two-moment radiative transport and frequency dependent opacities.  They find that as dust is depleted, VSI is excited only in the upper layers of the disk.  They find that the specific angular momentum $j_\phi(r) = u_\phi r$ develops a staircase profile with respect to $r$, i.e., regions of flattened $j_\phi(r)$ separated by jumps.  We confirm that this occurs in the axisymmetric case and weakly in the 3D case.

\section{Set-up, basic state, and simulation parameters}

\subsection{Basic state}\label{sec:basic_state}

The initial condition for the present simulations consists of a basic state and perturbation. The basic state follows \citet{Nelson_etal_2013}.  The temperature, or equivalently the isothermal sound speed squared, varies as a power law with cylindrical radius,
\be
\ci^2(r) = c_0^2 \left(r/r_0\right)^q, \eql{cs}
\ee
where $r_0$ is a reference radius.  The temperature gradient is due to stellar heating.
The midplane density is also assumed to follow a power law:
\be
   \rho_\mathrm{mid}(r) = \rho_0 \left(r/r_0\right)^p.
   \label{midplane_density}
\ee
The vertical scale height of the disk is
\be
   H(r) = \ci(r) / \Omega_\rmK(r), \eql{H}
\ee
where the Keplerian angular velocity is
\be
   \Omega_\rmK(r) = \Omega_0 (r / r_0)^{-3/2}, \hskip 0.5truecm \Omega_0 \equiv \left(GM / r_0^3\right)^{1/2}. \eql{Omega}
\ee
Substituting \eqp{cs} and \eqp{Omega} into \eqp{H} we obtain
\be
   H(r) = H_0 (r / r_0)^{(3+q)/2}, \hskip 0.5truecm H_0 \equiv c_0 / \Omega_0.
\ee 
We choose $q = -1$; for this case $H(r)$ grows linearly with $r$.  Therefore, $h(r) \equiv H(r)/r = H_0/r_0$.

Defining the spherical radius $R \equiv (r^2 + z^2)^{1/2}$, the density and azimuthal velocity profiles that satisfy vertical hydrostatic balance and radial centrifugal balance are given by \cite{Nelson_etal_2013} as
\begin{align}
& \rho(r, z)    = \rho_0 \left(\frac{r}{r_0}\right)^p \exp\left[\frac{GM}{\ci^2}\left(\frac{1}{R} - \frac{1}{r}\right)\right], \\
& u_\phi(r, z) = u_\rmK(r) \left[\left(p + q\right)\left(\frac{H}{r}\right)^2 + \left(1 + q\right) - \frac{qr}{R}\right]^{1/2},
\end{align}
where $u_\rmK(r) = \left(GM/r\right)^{1/2}$ is the local Keplerian velocity.

\begin{figure}
\centering
\includegraphics[width=3.5truein, trim=2 2 2 2,clip]{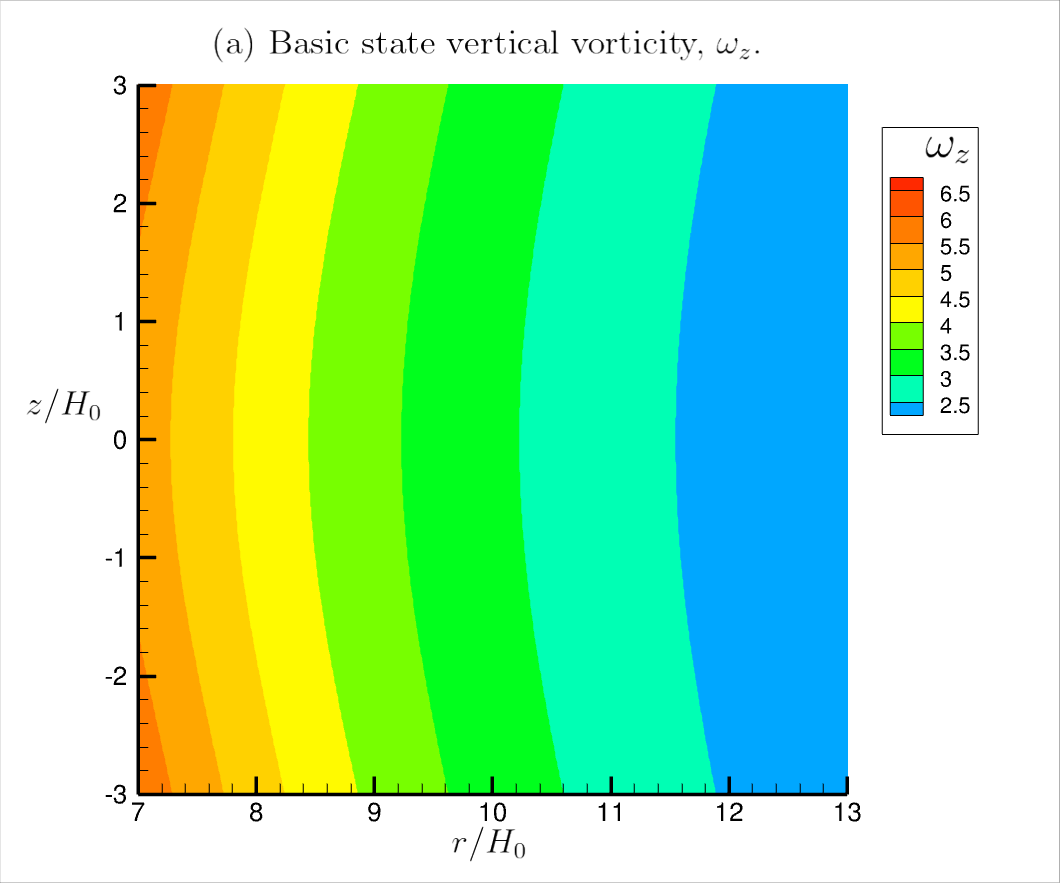}
\includegraphics[width=3.5truein, trim=2 2 2 2,clip]{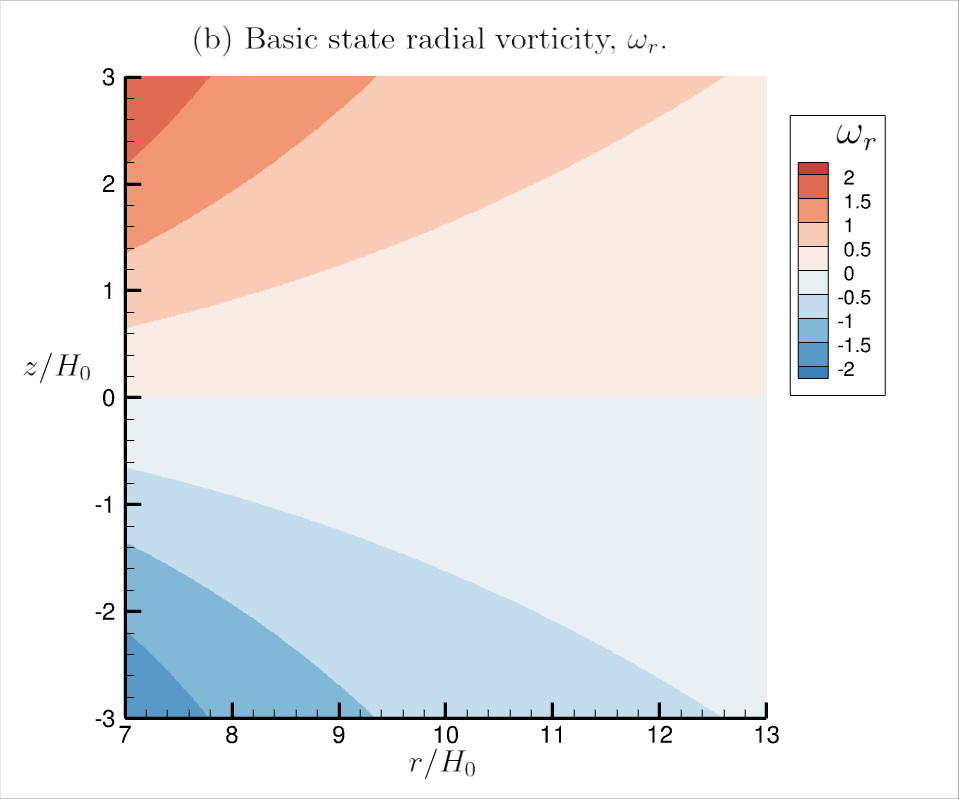}
\caption{Basic state vorticity.}
\label{fig:vort_basic}
\end{figure}
In order to later compare values of the fluctuation vorticity relative to those in the basic state, Figure~\ref{fig:vort_basic} plots the two non-zero components of the basic state vorticity.

\subsection{A remark on the basic state for VSI}

The VSI basic state has an angular velocity that varies not only with cylindrical radius $r$ but also with the vertical direction, $z$: $u_\phi = u_\phi(r, z)$.  As the name implies, VSI is driven by the vertical gradient $\p_z u_\phi$ of angular velocity.  This gradient corresponds to the existence of a \textit{radial} vorticity, $\omega_r$, 
\be
   \omega_r = \frac{1}{r} \frac{\p u_z}{\p\phi} - \frac{\p u_\phi}{\p z} = - \frac{\p u_\phi}{\p z}, \eql{omr}
\ee 
since $u_z = 0$ in the basic state.  The presence of $\omega_r$ is attributed to a baroclinic torque that arises from the radial temperature gradient.  However, since baroclinic torque produces \textit{azimuthal} rather than radial vorticity, the effect is indirect as we now explain.  For the present case of infinitely rapid cooling, the equation of state is vertically isothermal
\be
   p = \rho \ci^2(r). \eql{p}
\ee
The rate of baroclinic vorticity production then becomes
\be
   \frac{1}{\rho^2} \left(\nabla\rho \times \nabla p\right) = \frac{1}{\rho} \frac{\p\ci^2}{\p r} \frac{\p\rho}{\p z}\bfephi.
   \eql{baro}
\ee
Equation \eqp{baro} says that the baroclinic term applied to the basic state produces {\it azimuthal} rather than radial vorticity.  Considering the signs of the rhs, we conclude that $\omega_\phi > 0$ is produced in the upper disk,  while $\omega_\phi < 0$ is produced in the lower disk.  This vorticity induces a radial rather than azimuthal velocity.

To investigate the reason for the apparent conflict, consider the $\phi$ vorticity equation assuming axisymmetry; this is given as Equation \eqp{vort_phi_eq}:
\be
\frac{D}{Dt}\left(\frac{\omega_\phi}{\rho r}\right) = \frac{1}{\rho r^2}\frac{\p u_\phi^2}{\p z} + \frac{1}{\rho^2 r}\frac{\p\ci^2}{\p r}\frac{\p\rho}{\p z},
\ee
where the second term is the baroclinic torque.  If we assume both hydrostatic and centrifugal balance ($u_z = u_r = 0$) then $\omega_\phi = 0$ and we obtain:
\be
   \frac{\p u_\phi^2}{\p z} = -\frac{r}{\rho}\frac{\p \ci^2}{\p r} \frac{\p\rho}{\p z}. \eql{vert_shear}
\ee
Equation \eqp{vert_shear} shows that vertical shear ($=$ minus the radial vorticity) is proportional to the rate of baroclinic generation of azimuthal vorticity.  However, it should be clear from the derivation, that \eqp{vert_shear} arises from the \textit{constraint} of vertical and centrifugal balance.  In an actual disk, this constraint will never be satisfied identically.  Therefore, in the future it may be worthwhile to perform simulations starting with temperature and Keplerian gradients alone, and allow all other vorticity components to be freely generated.  Presumably, the final stationary turbulent state will have the same statistics as with the conventional approach.

For the basic state given by \eqp{p}, the vertical Br\"unt-V\"ais\"al\"a frequency squared $N_z^2 > 0$.  This implies that buoyancy has a stabilizing effect when the flow is adiabatic.  Buoyancy can be suppressed when the temperature of displaced parcels relaxes sufficiently rapidly to the basic state.  In the present work we assume that the radiative relaxation time is zero, i.e., that the vertically isothermal equation of state $p = \rho \ci^2(r)$ remains valid in the perturbed state.

\subsection{Simulation parameters} \label{sec:simpar}

\begin{table}
\centering
\begin{tabular}{l c c c c}
\toprule
 Parameter & Value\\
\midrule
No. of grid points, $n_r \times n_z \times n_\phi$ & $512 \times 512 \times 1024$\\
Density exponent, $p$ &$-3/2$\\
Temperature exponent, $q$               &$-1$\\
Disk aspect ratio, $H_0/r_0$ & $0.10$\\
Orbital period, $T_0$, at $r_0$ & $1$\\
Density, $\rho_0$ at $r_0$ & $1$\\
Sound speed $c_0$ at $r_0$ & $2\pi$\\
Radial domain, $[\rmin/H_0, \rmax/H_0]$ & $[6.5, 13.5]$\\
Vertical domain, $[\zmin/H_0, \zmax/H_0]$ & $[-3.5, 3.5]$\\
Azimuthal domain & $\phi\in[0,2\pi]$\\
Sponge width, $\delta_\mathrm{sponge}/H_0$ & 0.5\\
Decay period, $t_\mathrm{sponge}$, for sponge, & 20 time steps\\
Number of processors & 2048\\
cpu time per step & 1.785 s\\
Intel Processor & Haswell\\
Coeff. of artificial bulk viscosity, $C_\beta$ & 1.3\\
Strength of Pad\'e filter, $\epsfil$:&\\
$t/T_0 \in [0, 300.27)$ & 0.125\\
$t/T_0 \in [300.27, 365.84)$ & 0.06\\
$t/T_0 \in [365.84, 372.33)$ & 0.03\\
$t/T_0 \in [372.33, 479.4]$ & 0.015\\
No. of time steps & 862,584 \\
\bottomrule
\end{tabular}
\caption{Parameters for the main 3D run.}
\label{tab:params}
\end{table}
The simulation code, {\sc Pad\'e} \citep{Shariff_2024}, uses fourth order Pad\'e differentiation \citep{Lele_1992} which maintains accuracy for higher wavenumbers than conventional central finite-difference schemes of the same order.  
The simulations solve the hydrodynamic equations for compressible flow in cylindrical coordinates $(r,z,\phi)$ with a point mass source of gravity at $r = z = 0$.
The equation of state is locally isothermal:
\be
   p = \rho \ci^2(r),
\ee
where the isothermal sound speed $\ci(r)$ does not vary with time and has radial dependence given by Equation \eqp{cs}.  This assumes rapid radiative relaxation of thermal fluctuations to the basic state.

Table~\ref{tab:params} lists simulation parameters.  The number of grid points is $512 \times 512 \times 1024$ ($n_r \times n_z \times n_\phi$).  
We choose $r_0$ to be the midradius of the computational domain.  The density and temperature exponents are set to $p = -3/2$ and $q = -1$, respectively.  The disk aspect ratio is chosen to be $h \equiv H_0 / r_0 = 0.1$.  These three parameters were chosen to be the same as for case {\tt p1.5h0.1} in M20.

We are free to set three quantities to unity in the code: $H_0 = \rho_0 = T_0 = 1$, where $T_0$ is the Keplerian period at $r_0$.  Therefore, the unit of velocity is $H_0/T_0$, i.e., scale heights per orbital period at $r_0$, and in code units the sound speed
$c_0 = H_0 \Omega_0 = 2\pi$.

The radial and vertical domains are both $7 H_0$ long with a sponge strip of width $\delta_\mathrm{sponge}/H_0 = 0.5$ adjacent to each radial and vertical boundary.  The flow field is relaxed to the basic state in the sponge regions with a characteristic relaxation time $t_\mathrm{sponge} = 20$ time steps.  Zero normal velocity boundary conditions are applied at the radial and vertical boundaries.  The azimuthal domain is a full circle with periodic boundary conditions.  The number of grid points per $H_0$ is therefore $74 \times 74 \times 16$ in $r, z,$ and $\phi$ (at $r_0)$, respectively.  The smaller resolution in $\phi$ is justified by the fact that Keplerian shear elongates vortical structures in this direction.  However, a future check on this assumption should be made.  To capture possible weak shocks in upper layers of the disk, artificial bulk viscosity, which acts on the dilatation $\nabla\cdot\bfu$, is activated with coefficient $C_\beta = 1.3$.  However, we now believe this is unnecessary with a sponge layer at the upper and lower boundaries, and deactivated it in test simulations not reported here.

The initial condition is seeded with velocity perturbations of the form
\be
   \delta u = \epsilon_\mathrm{pert} M(z) \sum_\bfk \real\{A(\bfk) e^{\rmi \bfk \cdot \bfx}\}
\ee
for each component, where $A(\bfk)$ is a complex number with unit amplitude and random phase, $\bfk \equiv (k_r, k_z, k_\phi)$, $\bfx \equiv (r, z, \phi)$, and $M(z)$ is a modulation.
The radial wavenumbers, $k_r$, consist of a set of eleven: a fundamental, seven harmonics, and three subharmonics.  The fundamental has a wavelength of the most amplified mode:
\be
   \lambda_{r,\mathrm{fund}} = \pi |q| \frac{H_0}{r_0} H_0.
\ee
The wavenumbers $k_z$ and $k_\phi$ both consist of a fundamental and eleven harmonics.  The fundamental wavelengths in the $z$ and $\phi$ directions equal the domain size in these directions, i.e., $L_z$ and $2\pi$, respectively.  The amplitude was set to $\epsilon_\mathrm{pert} = 0.001 c_0$.  Finally, a half cosine modulation,
\be
   M(z) = \cos(\pi z / L_z), \hskip 0.5truecm 
   z \in [-L_z/2, L_z/2],
\ee
is applied to make perturbations vanish at the top and bottom boundaries:

\subsection{Varying the strength of Pad\'e filter}

\begin{figure}
\centerline{\includegraphics[width=3.5truein]{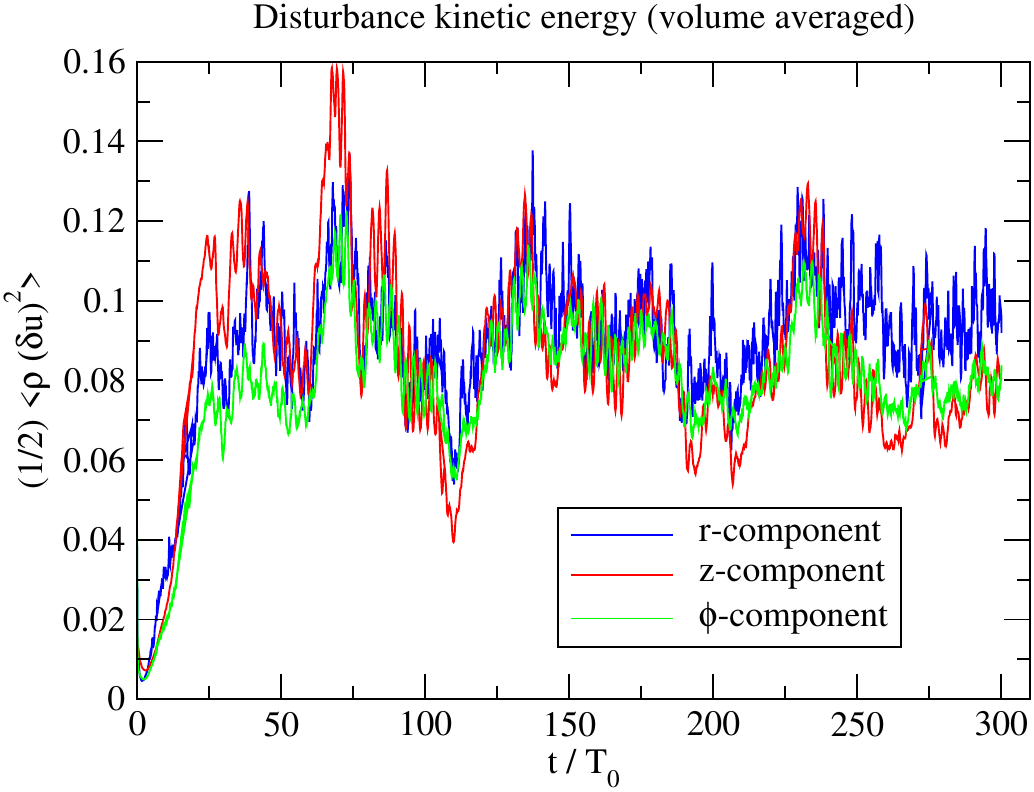}}
\caption{Time history of the components of the volume-averaged disturbance kinetic energy (relative to the basic-state) for the 3D run in the period $t/T_0 \in [0, 300.27]$ during which the strength of the Pad\'e filter was set to $\epsilon_\mathrm{filter} = 0.125$.}
\label{fig:ke}
\end{figure}
The kinematic molecular viscosity in disks is so low that one cannot hope to perform a direct numerical simulation (DNS) wherein all spatial scales down to molecular dissipation are resolved.   Therefore, some treatment of unresolved scales is needed.  For this, all simulations to date rely on the dissipation inherent in shock-capturing schemes to damp small scale fluctuations.  This is referred to as an ``implicit'' sub-grid treatment \citep{Boris_etal_1992, Ritos_etal_2018} to distinguish it from the use of an explicit sub-grid model such as that of \cite{Smagorinsky_1963}.  Here, we adopt an 
implicit sub-grid treatment using a Pad\'e filter \citep{Lele_1992} whose details and application procedure are summarized in Appendix \ref{PadeFilterDetails}.

The Pad\'e filter requires the setting of a parameter ($\epsfil$) which determines the strength of the filter.  The main part of the run was for $t/T_0 \in [0, 300.27]$ during which the strength of the Pad\'e filter was set to $\epsilon_\mathrm{filter} = 0.125$.  This value was indicated as being conservative from test axisymmetric VSI runs presented in \cite{Shariff_2024}.  Figure \ref{fig:ke} shows the components of the disturbance kinetic energy for the 3D run during this period.  Time and $\phi$ averaged statistics of Reynolds stresses were taken for $t/T_0 \in [54.53, 300.10]$ during which the flow is statistically stationary.  These stresses, together with the \cite{Shakura_and_Sunyaev_1973} $\alpha(r)$ they imply, will be shown in \S\ref{sec:stresses}.

After the main part of the run was complete, we experimented with reducing $\epsfil$ to see if we could capture more fine scale features and extend the power-law range in spectra without producing spurious $2\Delta$ oscillations; energy spectra (\S\ref{sec:spectra}) and inspections of the flow field confirmed that this was possible.  The variation of $\epsfil$ with time is documented at the bottom of Table~\ref{tab:params}.  The structure of the 3D flow field and associated energy spectra will be presented in the period when $\epsfil = 0.015$.

\section{Axisymmetric simulation: early non-linear stage}

For all color contour plots in the paper, pure white pixels corresponds to values that exceed the range of the color map.

The dynamics of axisymmetric VSI at early times are also present in the 3D case, although with less intensity and coherence.   At later times, axisymmetric VSI contains strong artifacts (\S\ref{sec:axi_artifacts}) that are not present in 3D.  To enable a one-to-one comparison with the 3D case, the axisymmetric run was chosen to have the same parameters as the 3D run (with $\epsfil = 0.125$), except for a slight change in the number of grid points in order to evenly fit the 24 processors available on one node.  The grid is $576^2$ ($n_r \times n_z$).

\begin{figure}
\centering
\includegraphics[width=3.2truein]{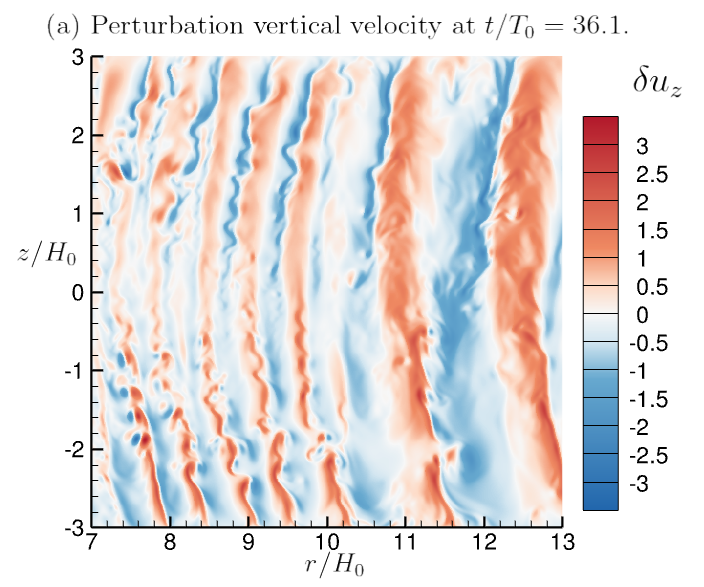}
\includegraphics[width=3.2truein]{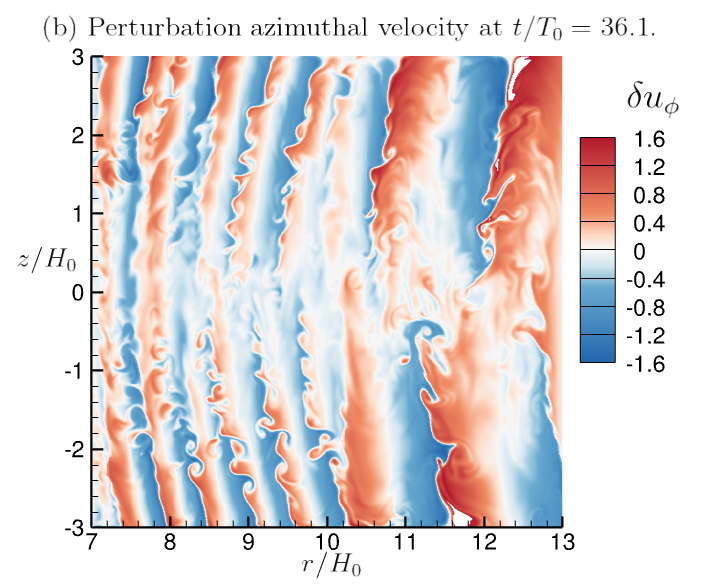}
\includegraphics[width=3.2truein]{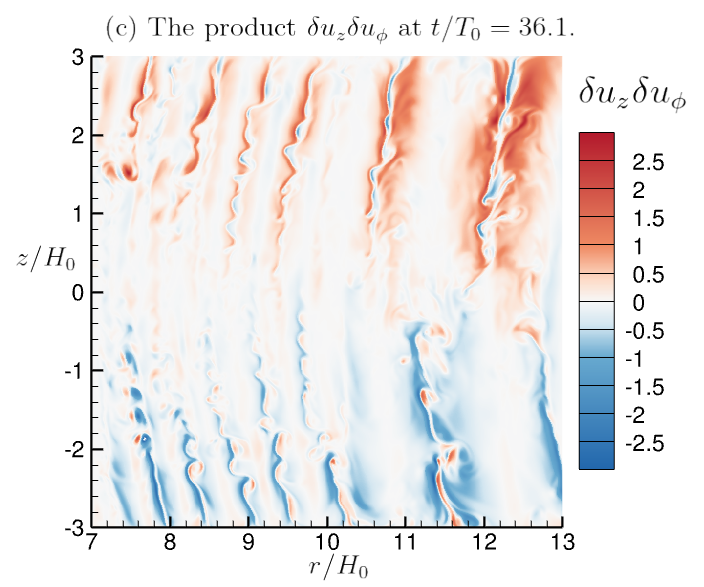}
\caption{Perturbation velocity for the axisymmetric simulation on a $572^2$ mesh at $t/T_0 = 36.1$.  (a) Vertical velocity, $u_z$.  (b) Perturbation azimuthal velocity, $\delta u_\phi$. (c) The product $\delta u_z \delta u_\phi$.  Pure white regions represent values that are outside of the range of the color bar/legend.}
\label{fig:vel_axi}
\end{figure}

\subsection{Perturbation velocity}\label{sec:axi_vel}
The characteristic feature of axisymmetric VSI is the formation of counter flowing vertical jets that remain coherent over the entire vertical extent of the disk; see Figure~\ref{fig:vel_axi}(a).  It should be obvious that buoyancy suppression is needed for the maintenance of such jets.  Upward jets are usually narrow and fast in the lower half of the disk and then widen and slow down as they travel into in the upper half of the disk. Downward jets are narrow and fast for $z > 0$, then widen and slow down as they travel towards the lower half of the disk.  This phenomenon may simply be due to the larger driving vertical shear for larger $|z|$.  As a result, a wider jet is located next to a narrow counter flowing jet.

Figure~\ref{fig:vel_axi}(b) shows the azimuthal velocity perturbation, $\delta u_\phi$.  As mentioned earlier, the local linear stability analysis of L\&P shows that $\delta u_\phi$ has the same magnitude as $\delta u_z$ when tilt is neglected.  The same is approximately true here.

Figure~\ref{fig:vel_axi}(c) shows that the product $\delta u_z \delta u_\phi$ is mostly $\gtrless 0$ for $z \gtrless 0$.  This important fact is explained as follows.  The specific angular momentum $j_z \equiv u_\phi r$ follows fluid elements for axisymmetric flow\footnote{For $p = \rho \ci^2(r)$, the baroclinic term does not enter the $\omega_z$ transport equation \eqp{vort_eq_z} for axisymmetric flow, and therefore the circulation $\Gamma_z = 2 \pi j_z$ is unaffected by baroclinic torque.}:
\be
\frac{D j_z}{Dt} = 0.
\ee
Since the basic state $j_z$ decreases away from the midplane, upward jets in $z > 0$ bring in higher $j_z$ leading to a positive product.  Downward jets in $z > 0$ bring in lower $j_z$, also leading to a positive product.  Similarly, upward jets in $z < 0$ bring in low $j_z$, leading to a negative product.
Finally, downward jets in $z < 0$ bring in high $j_\phi$, again leading to a negative product.  Note that this explanation is linear because it involves the advection of the mean $j_z$ by a perturbation $u_z$.  This pattern of the $\delta u_z \delta u_\phi$ product will be reflected in the $T_{z\phi}$ Reynolds stress in 3D (\S\ref{sec:stresses}).

\begin{figure}
\centering
\includegraphics[width=3.4truein]{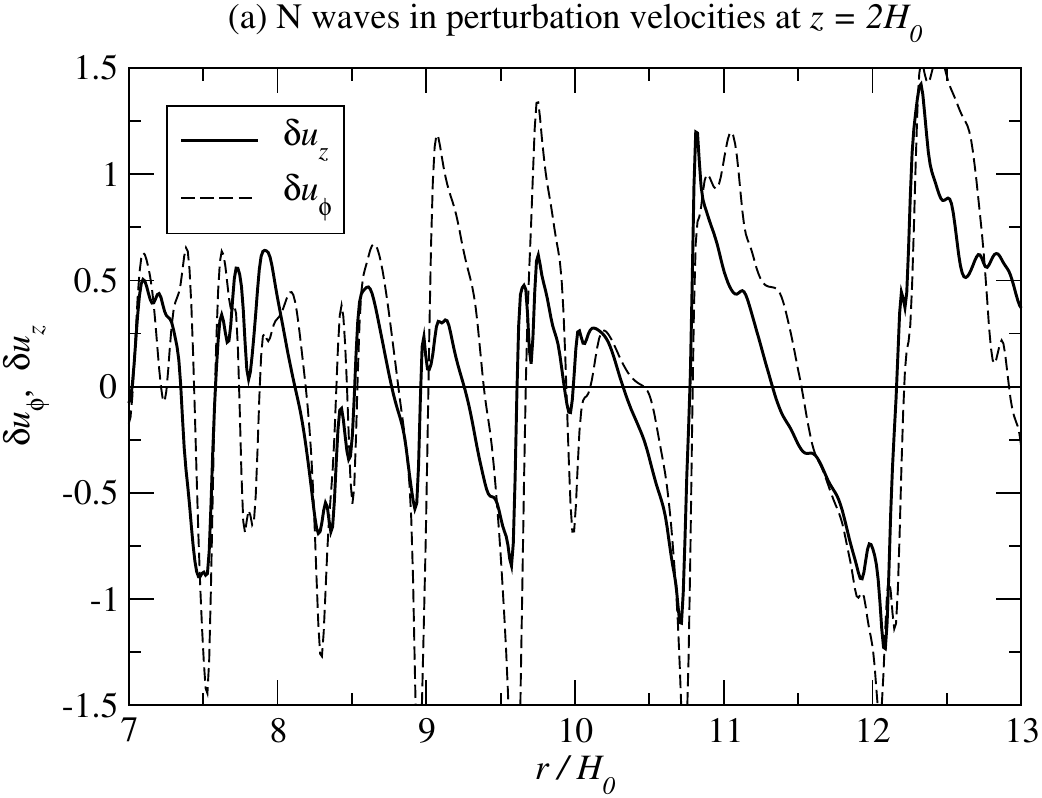}
\includegraphics[width=3.4truein]{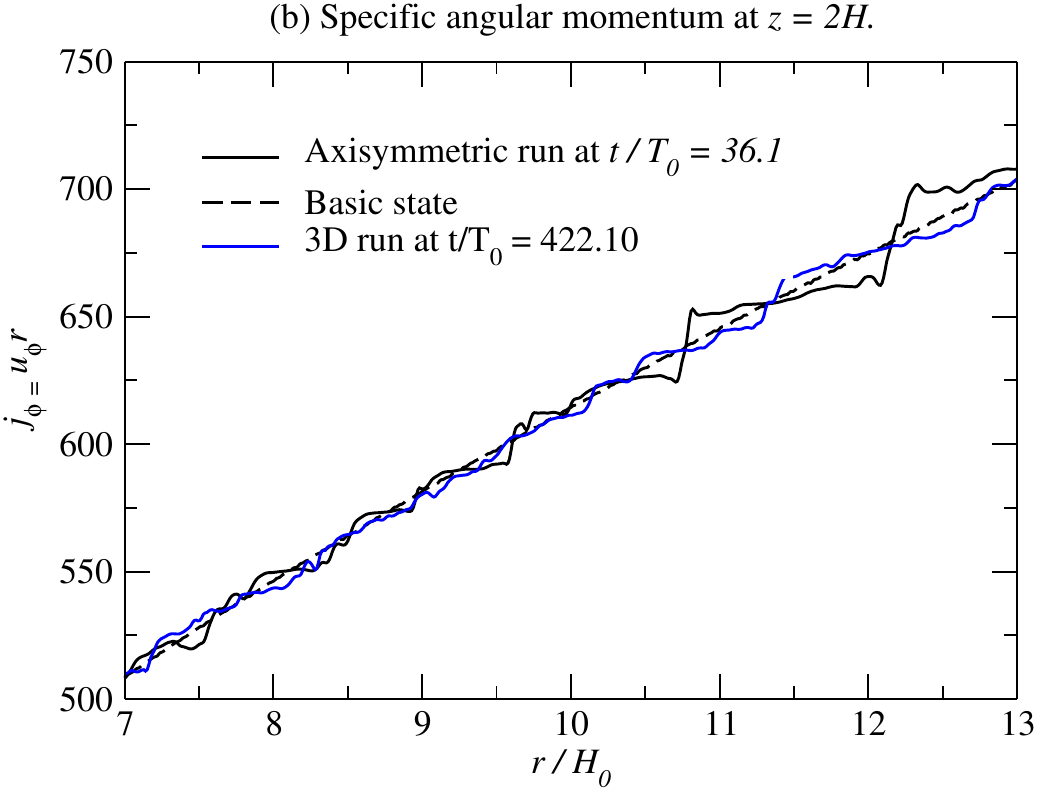}
\includegraphics[width=3.4truein]{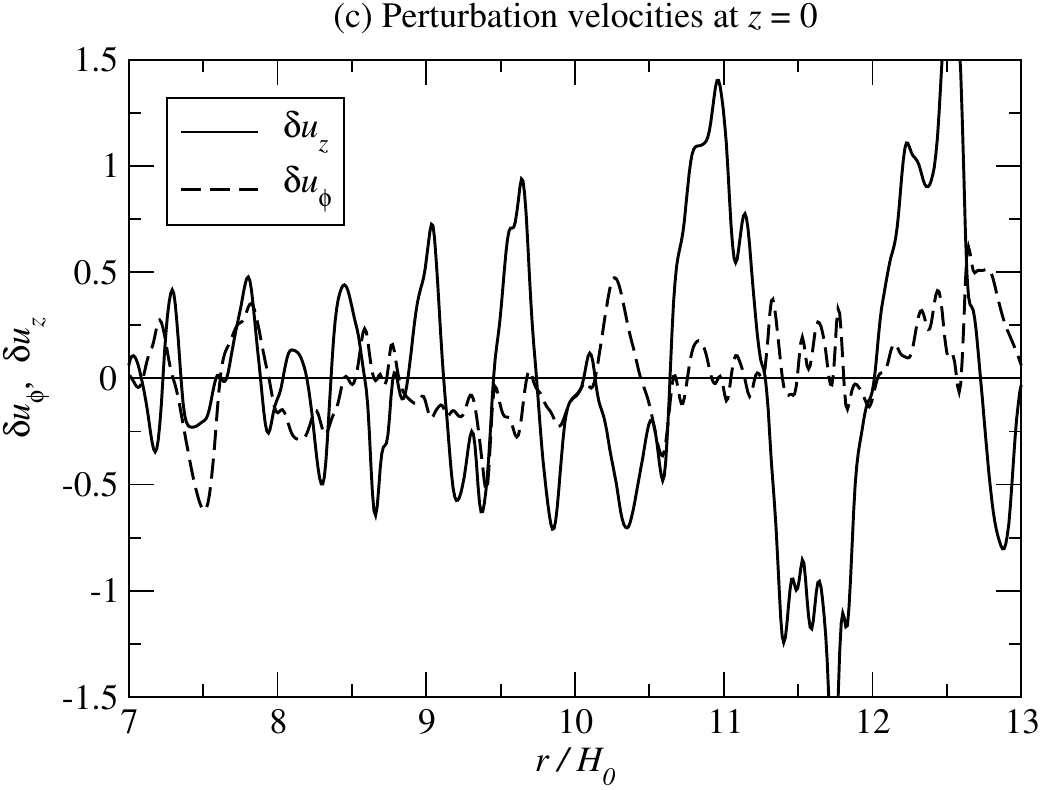}
\caption{(a) Perturbation vertical and azimuthal velocities at $t/T_0 = 36.1$ showing their correlation and their N-wave form. (b) Specific angular momentum $j_z \equiv u_\phi r$ at $z = 2H_0$.  Solid black line: At $t/T_0 = 36.1$ for the axisymmetric run.  Solid blue line: At $t/T_0 = 422.10$ for the 3D run (azimuthal location $\phi = 0$). (c) Similar to panel (a) except evaluated at $z=0$.}
\label{fig:jz_axi_early}
\end{figure}
Figure~\ref{fig:jz_axi_early}(a) was constructed to provide an alternate way of illustrating the fact, which a contour plot will soon make clear, that the perturbation ($\delta\omega_z$) of vertical vorticity,
\be
   \omega_z \equiv \frac{1}{r}\frac{\p}{\p r} (u_\phi r) = \frac{1}{r}\frac{\p j_z}{\p r},
   \eql{omz_def_axi}
\ee
is dominated by thin positive layers. (Equation \eqp{omz_def_axi} is valid for axisymmetric flow only.)
Figure ~\ref{fig:jz_axi_early}(a) shows that the correlation between $\delta u_\phi$ (solid) and $\delta u_z$ (dashed) is more remarkable than revealed in Figure~\ref{fig:vel_axi}.  It also shows that both velocity components have N shaped profiles such that regions of positive slope are much steeper than regions of negative slope.  We note that in the lower half of the disk (not plotted), $\delta u_z$ has mirrored N waves while $\delta u_\phi$ has normal N waves.  
The azimuthal vorticity is defined as
\be
   \omega_\phi \equiv \frac{\p u_r}{\p z} - \frac{\p u_z}{\p r},
\ee
where the second term dominates in VSI.
Therefore, for $z > 0$, an N profile for $\delta u_z$ implies that $\delta\omega_\phi < 0$ shear layers
dominate over $\delta\omega_\phi > 0$ shear layers.  The opposite in true for $z < 0$.

N-shaped profiles for $\delta u_\phi$ (in both halves of the disk) implies that layers of positive $\delta\omega_z$ dominate.  This feature is present in 3D but with lower intensity and regularity.
The preference for $\delta\omega_z > 0$ is explained as follows.  The basic state $j_z(r)$ profile increases monotonically with $r$; see the dashed line in Figure \ref{fig:jz_axi_early}b.  Consider the upper disk ($z > 0$) and an interface that separates a downward jet on the left and an upward jet on the right ($\downarrow\uparrow$).  
%
Recalling that the basic state $j_z$ is a maximum at the midplane, the downward jet on the left will lower $j_z$ while the upward jet on the right will increase $j_z$.  This will eventually lead to a positive jump in $j_z$ with respect to $r$ and an increase in positive $\omega_z$ since
\be
\omega_z = \frac{1}{r}\frac{\p}{\p r}
\left(u_\phi r\right) = \frac{1}{r}
\frac{\p j_z}{\p r}.
\ee
Similarly, consider an interface that separates an upward jet on the left and a downward jet to its right ($\uparrow\downarrow$).  This will reduce the slope of $j_z(r)$, leading to the flatter part of the $j_z$ staircase and a lowering of $\delta\omega_z$.
Each type of interface behaves in the opposite manner for $z < 0$.  The continuation of a $\downarrow\uparrow u_z$ interface into the lower disk will now cause $j_z(r)$ to flatten and $\omega_z$ to decrease.  Similarly, the continuation of a $\uparrow\downarrow u_z$ interface into the lower disk will case $j_z(r)$ to steepen and $\omega_z$ to increase.  This means that regions of $\omega_z > 0$ intensification in the upper disk will appear staggered relative to those in the lower half.

The staircase structure of $j_z(r)$ is shown as the solid line in Figure \ref{fig:jz_axi_early}(b).
In the 3D case (solid blue line), this pattern is less intense and coherent. 
\citet{Klahr_etal_2023} and MF24a,b were the first to point out the formation of a $j_z(r)$ staircase for axisymmetric flow.  The N waves in $u_\phi(r)$ simply reflect this staircase structure.

Figure \ref{fig:jz_axi_early}(c) repeats Figure \ref{fig:jz_axi_early}(a) at the midplane.  While $\delta u_z$ is comparable to its value at $z = 2H_0$, $\delta u_\phi$ is much weaker, and the correlation between $\delta u_z$ and $\delta u_\phi$ observed at $z = 2H_0$ is absent.  This is because the vertical gradient of $j_z$ which drives the creation of $\delta u_\phi$ vanishes at $z = 0$.

\subsection{Perturbation vorticity}
\label{sec:axi_vort}
\begin{figure}
\centering
\includegraphics[width=3.5truein,  trim=2 2 2 2,clip]{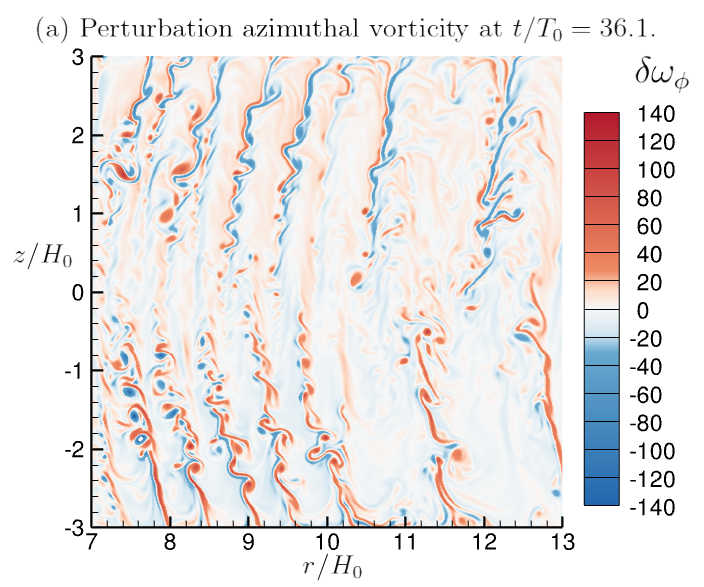}
\includegraphics[width=3.5truein,  trim=2 2 2 2,clip]{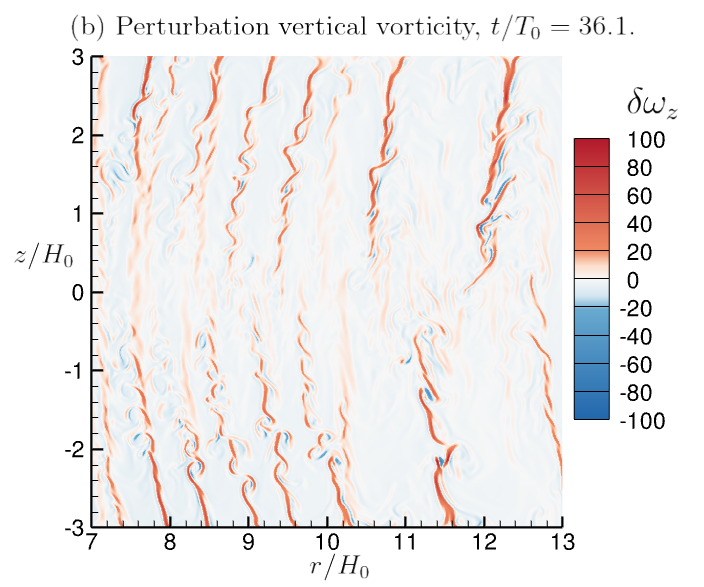} 
\includegraphics[width=3.5truein,  trim=2 2 2 2,clip]{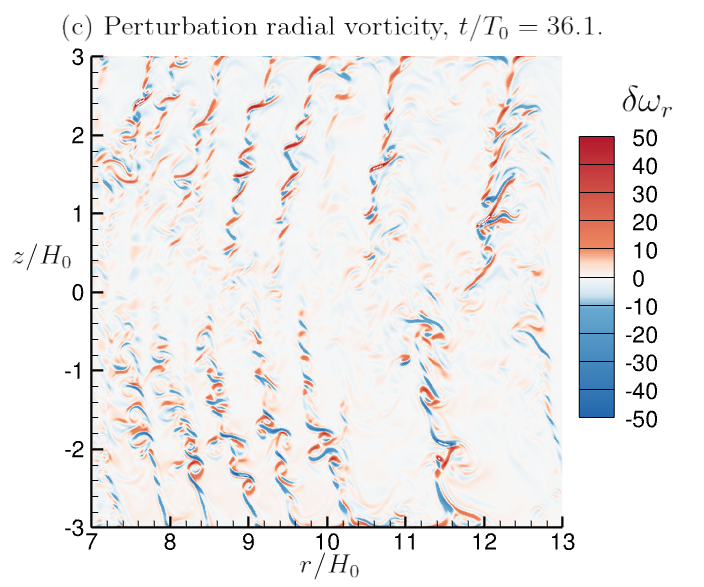}
\caption{Perturbation vorticity for the axisymmetric run.}
\label{fig:vor_axi}
\end{figure}
Figure~\ref{fig:vor_axi} shows the perturbation vorticity field in the axisymmetric simulation.  It attains values many times that of the basic state vorticity with larger values as $|z|$ increases.  From the range of the data, we conclude that the dominant component is azimuthal (Figure~\ref{fig:vor_axi}(a)).  This will be reinforced below where we present the PDF (probability density function) of each perturbation vorticity component.
The azimuthal component is organized into shear layers that form the boundaries of the vertical jets.

In the upper disk, $\omega_\phi < 0$ (blue) shear layers are dominant.  Such layers are adjacent to a thin sliver of opposite sign (red) vorticity, followed by a diffuse red region in between the shear layers.  The opposite is true in the lower disk.  These patterns can also be seen in Figure 9 (lower right panel) of MF24a. 
The dominant sign pattern of $\omega_\phi$ correlates with the N waves in $\delta u_z(r)$ discussed in \S\ref{sec:axi_vel}.  
The shear layers display vortices created by the KH instability.

Figure~\ref{fig:vor_axi}(b) shows the perturbation, $\delta\omega_z$, of the vertical vorticity relative to the basic state value.  It is dominated by layers of $\delta\omega_z > 0$ consistent with the N waves in $\delta u_\phi$ and positive jumps in $j_\phi(r)$ as discussed previously (\S\ref{sec:axi_vel}).  These layers spatially coincide with the shear layers of azimuthal vorticity.

The radial perturbation vorticity
(Figure~\ref{fig:vor_axi}(d)) is the weakest of the three components.  Positive values are dominant in the upper disk, the opposite being true for $z < 0$.

\begin{figure}
\centering
\includegraphics[width=3.2truein,  trim=2 2 2 2,clip]{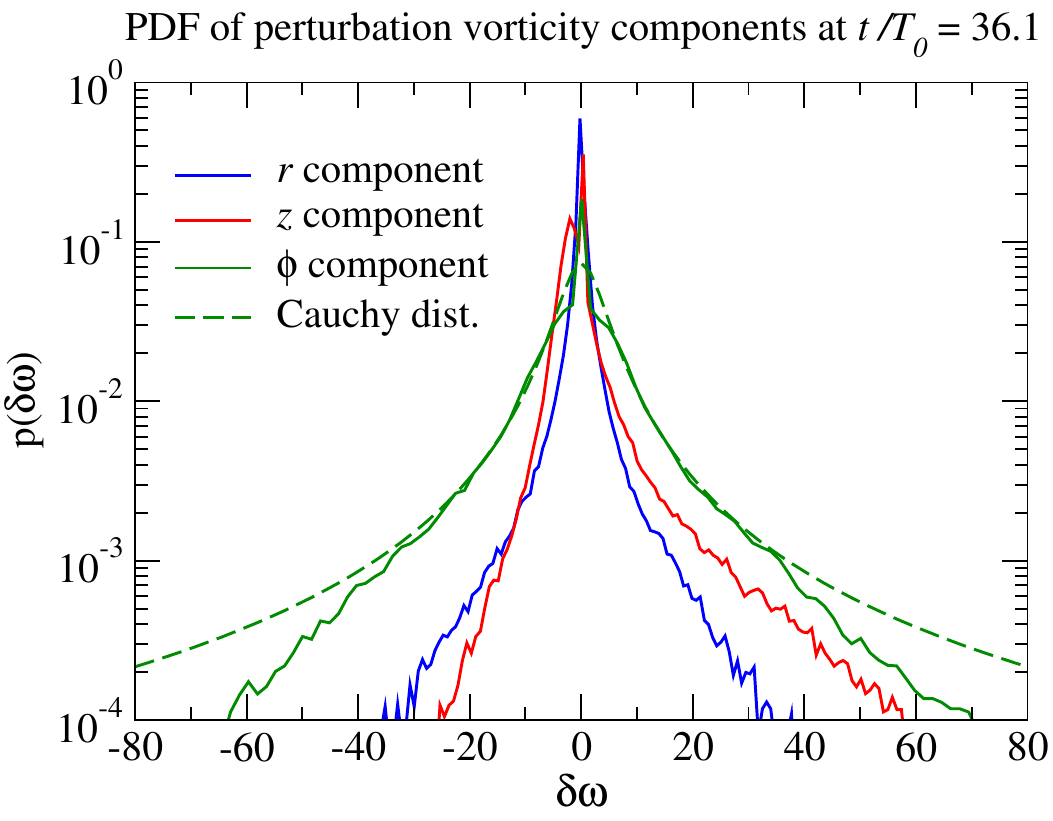}
\caption{Probability density function of perturbation vorticity components for the axisymmetric run at $t/T_0 = 36.1$.
The parameter $c$ in the Cauchy distribution \eqp{Cauchy} was set to $c = 4.35$ to obtain the green dashed curve.}
\label{fig:pdf}
\end{figure}

Some of the above assertions are confirmed by the PDF of each vorticity component; see Figure~\ref{fig:pdf}, which uses semi-log axes.  One observes that $\delta\omega_\phi$ (solid green curve) is indeed dominant.  The inner part of its PDF is fit reasonably well by a Cauchy distribution (green dashed curve)
\be
   p(s) = \frac{c}{\pi \left(c^2 + s^2\right)}, \eql{Cauchy}
\ee
for $c = 4.35$.  The choice of this distribution was inspired by \citet{Jimenez_1996} who obtains a similar result for the inner part of the velocity gradient PDF in 2D isotropic turbulence and for a collection of 2D Gaussian core vortices.  For large vorticity values in 2D isotropic turbulence, the analysis of \citet{Falkovich_and_Lebedev_2011} predicts exponential tails, which would show up as linear behavior on a semi-log plot.  It is unclear whether this is true in the present case for $\delta\omega_\phi$.  The red curve in Figure~\ref{fig:pdf} confirms that $\delta\omega_z$ is skewed toward positive values consistent with Figure \ref{fig:vor_axi}(b).
To confirm the sign pattern for $\omega_\phi$, we plotted $p(\sgn(z)\delta\omega_\phi)$ (not shown to avoid clutter) and found it to be skewed toward negative values consistent with
the color contour plot (Figure~\ref{fig:vor_axi}(a)).  

\subsection{Origin of the sign pattern of azimuthal vorticity}\label{sec:vort_phi}

To understand the origin of the sign pattern of $\omega_\phi$, consider the two source terms, denoted $T_1$ and $T_2$, on the rhs of the $\omega_\phi$ transport equation \eqp{vort_phi_eq}
multiplied through by $\rho r$ for convenience:
\be
   T_1 = \frac{1}{r} \frac{\p u_\phi^2}{\p z} \hskip 0.25truecm
   \mathrm{and} \hskip 0.25truecm T_2 = \frac{\p \ci^2}{\p r}\frac{\p\log\rho}{\p z},
\ee
where $T_1$ is due to vertical shear and $T_2$ is the baroclinic term.  The two terms balance in the basic state, i.e., $\breve{T}_1 + \breve{T}_2 = 0$, where a breve accent denotes a basic state quantity.
Therefore, it is more helpful to consider the deviations, $\delta T_1$ and $\delta T_2$, from the basic state.
\begin{figure}
\centering
\includegraphics[width=3.5truein, trim=2 2 2 2,clip]{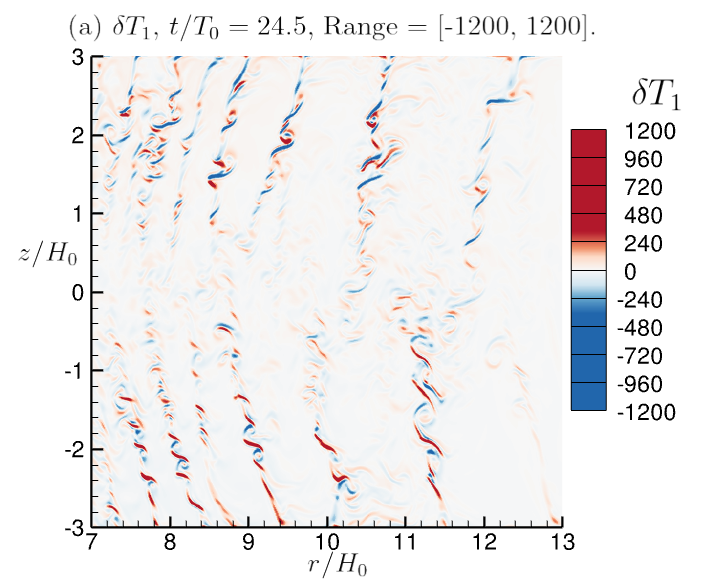} 
\includegraphics[width=3.5truein,  trim=2 2 2 2,clip]{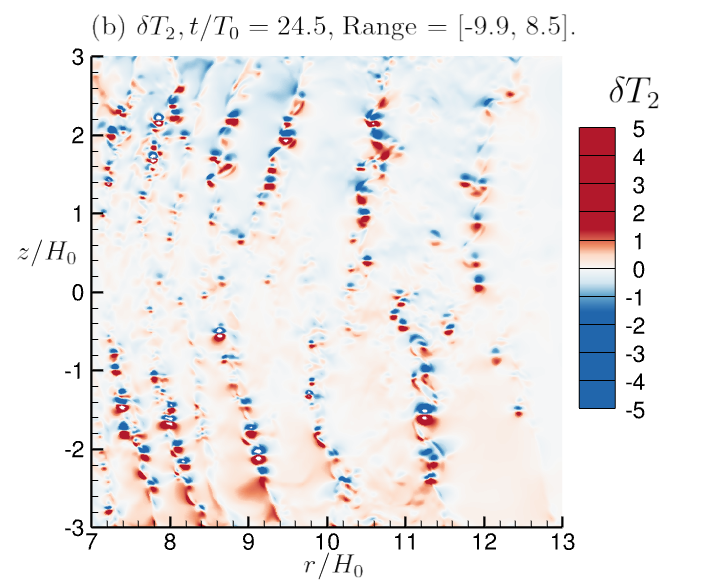}
\caption{(a) Deviations from the basic state of the vertical shear term $T_1$ on the rhs of the transport equation \eqp{vort_phi_eq} for azimuthal vorticity. (b) Similarly, the deviation $T_2$ (baroclinic term). In panel (b), values exceeding the range of the color legend are rendered in white.  The actual range of values is indicated in the headings.}
\label{fig:T1_T2}
\end{figure}
Figure \ref{fig:T1_T2}(a) shows that the vertical shear term $\delta T_1$ has the same overall sign pattern in shear layers as $\omega_\phi$.  Its values lie in the range $\delta T_1 \in [-1200, 1200]$, which is much larger than the range $T_1 \in [-50, 50]$ in the basic state.  On the other hand, Figure \ref{fig:T1_T2}(b) shows that the baroclinic term $\delta T_2$ has a range, $[-9.9, 8.5]$ that is much weaker than the basic state and does not show the same pattern.  We therefore attribute the behavior of $\omega_\phi \approx \p_r u_z$ to $\delta T_1$.  
Let us further ask: what determines $\delta T_1$? We have
\be
\delta T_1 = \frac{1}{r}\frac{\p}{\p z}\left(\delta u_\phi^2\right) \approx \frac{2}{r} \breve{u}_\phi
\frac{\p}{\p z}(\delta u_\phi) = \frac{2}{r^2} \breve{u}_\phi
\frac{\p}{\p z}(\delta j_z).
\ee
Hence, the critical quantity is $\p_z(\delta j_z)$, the perturbation in the vertical gradient of angular momentum, since the factor that multiplies it is $>0$ and known from the basic state.  The quantity $\p_z(\delta j_z)$ is obtained from the transport equation for $j_z$ 
\be
   \frac{\p j_z}{\p t} + u_z \frac{\p \jzt}{\p z} \approx 0, \eql{jz_lin}
\ee
which is linearized about the basic state and assumes that vertical transport dominates radial transport.  Differentiating \eqp{jz_lin} with respect to $z$ gives
\be
\frac{1}{\delta t}\frac{\p}{\p z}(\delta j_z) = -\frac{\p u_z}{\p z}\frac{\p\jzt}{\p z} - u_z\frac{\p^2\jzt}{\p z^2}.
\eql{ddz_delta_jz}
\ee
We denote the two terms on the rhs of \eqp{ddz_delta_jz} by $T_3$ and $T_4$.
\begin{figure}
\centering
\includegraphics[width=3.5truein, trim=2 2 2 2,clip]{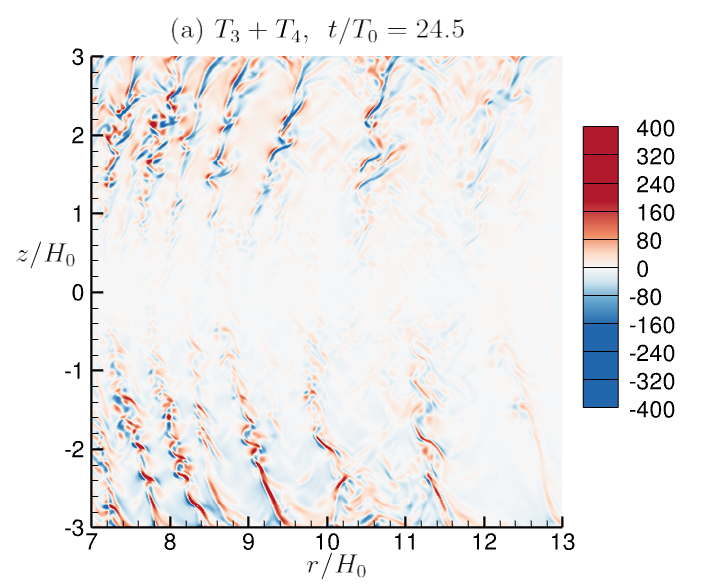} 
\includegraphics[width=3.5truein,  trim=2 2 2 2,clip]{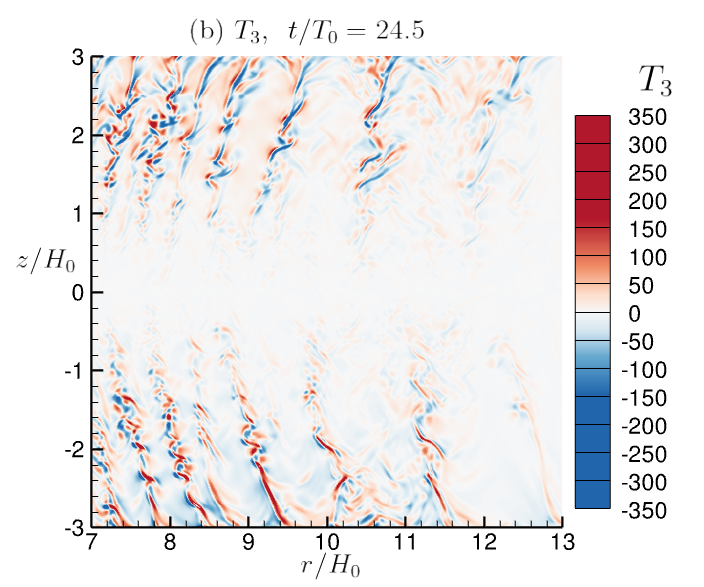}
\includegraphics[width=3.5truein,  trim=2 2 2 2,clip]{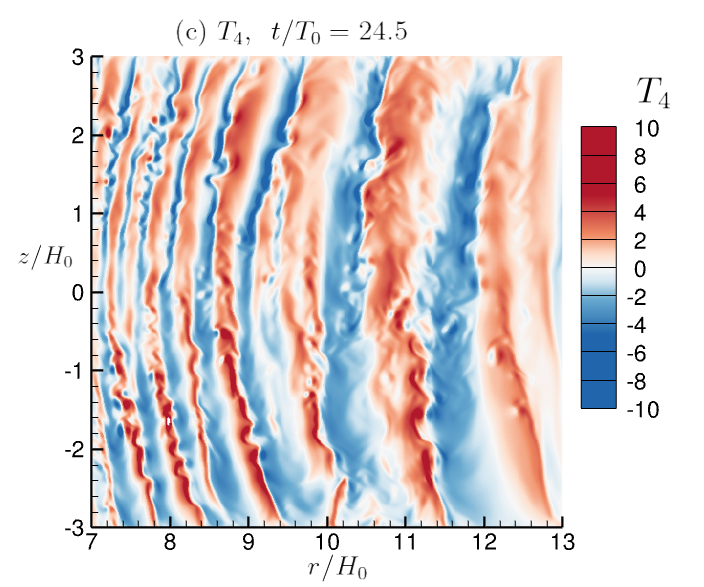}  
\caption{Terms in the transport equation \eqp{ddz_delta_jz} for $\p_z(\delta j_z)$.}
\label{fig:ddz_delta_jz}
\end{figure}
Figure \ref{fig:ddz_delta_jz} shows that $T_3 + T_4$ does indeed have the sign pattern of $\delta T_1$.  By far, the greater contribution is from $T_3$.  
Finally, note that since $\omega_r = - \p u_\phi/\p z$, we can also write
\be
   \delta T_1 \approx - \frac{2}{r}\breve{u}_\phi\delta\omega_r.
\ee
If we refer back to Figure~\ref{fig:vor_axi}(c), we see that the sign pattern of $\delta\omega_r$ is indeed the opposite of $\delta T_1$.

\subsection{Increase of wavelength with time}

For the present case of $q = -1$ and $H_0/R_0 = 0.1$, Equation \eqp{lambda_r} for the most amplified linear instability wavelength becomes
\be
   \lammax = \frac{\pi}{100} r, \eql{lambda_r_2}
\ee
which equals $0.24 H_0$ at $r = 7.5 H_0$, the midpoint of the interval to be plotted below.  On the other hand, if we inspect Figure~\ref{fig:vor_axi}(b) we find that the radial wavelength is more than twice larger, being about $0.59 H_0$ at the same radial location.  To investigate this further, we ran the same case with a white noise initial perturbation instead of a superposition of waves with random phases as described in \S\ref{sec:simpar}.  This change was made to eliminate any artifact that might be present due to the presence of subharmonics of $\lambda_{r,\mathrm{max}}$ in the wavy initial condition.
\begin{figure}
\centering
\includegraphics[width=3.4truein]{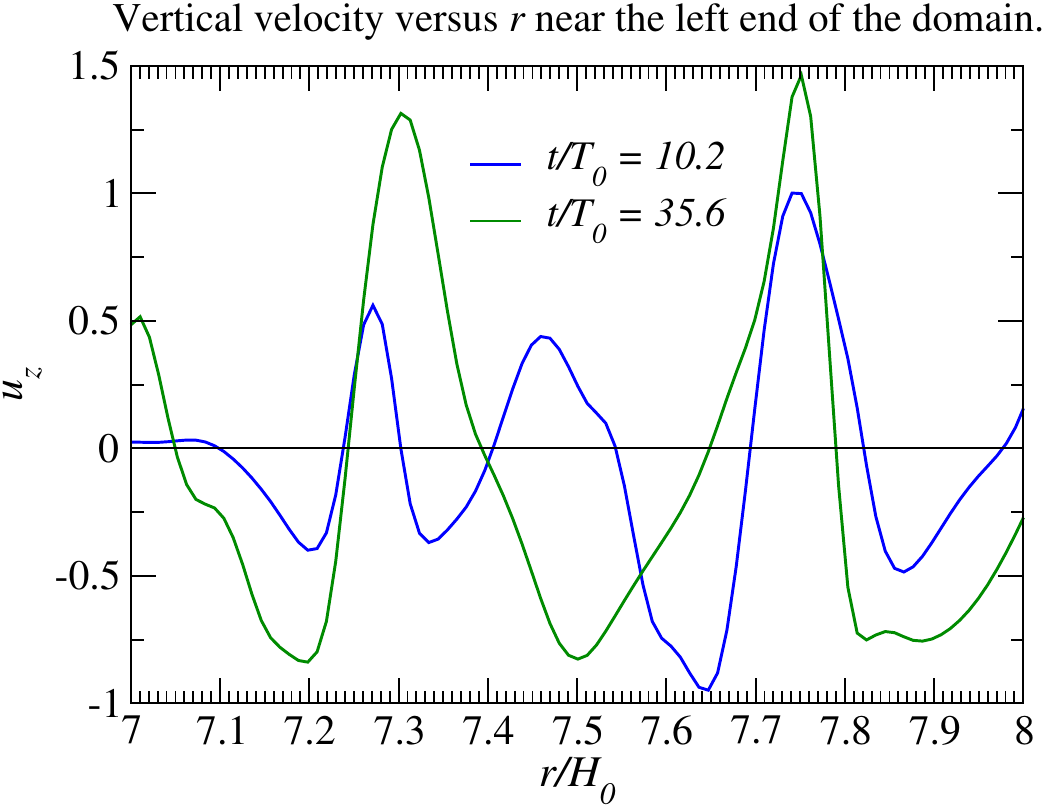} 
\caption{Profiles of $u_z(r)$ near the $r_\mathrm{min}$ boundary at $z/H_0 = -2$ for the axisymmetric run with a white noise initial perturbation.}
\label{fig:uz_profiles}
\end{figure}

We find that wavelength increase is also observed with a white noise perturbation.
Figure~\ref{fig:uz_profiles} depicts profiles of $u_z(r)$ at two instants (for $z/H_0 = -2$) near the left end of the domain where the growth rate is the largest.  An increase of wavelength with time is visually apparent.  A value for the wavelength was obtained as the distance between downward zero crossings.  At $t/T_0 = 10.2$ (blue curve), the average of the three wavelengths in the interval plotted is $\lambda_r = 0.24$ which equals the most linearly amplified wavelength at the midpoint of the interval.  On the other hand, at $t/T_0 = 35.6$, the average of the two wavelengths is larger, namely, $\lambda_r = 0.37$.  A temporal increase of wavelength can occur either due to nonlinearity or non-constant coefficients in the linear phase, i.e., spatial variation of the basic state gradient.  Our guess is that the wavelength increase is due to nonlinearity and is related to a putative inverse cascade in axisymmetric VSI discussed further in \S\ref{sec:axi_artifacts}.  Radial wavelength increase is also observed in the radiative hydrodynamic axisymmetric simulations of MF24b (their Figure 11).

\citet{Pfeil_and_Klahr_2021} performed an axisymmetric radiative hydrodynamics simulation using flux limited diffusion and observed a  sinusoidal vertical velocity at the midplane with amplitude increasing with radius; see their Figures 19 and 20.  Their Figure 20 shows that the wavenumber $k_r H(r)$ decreases from 24 and 14 with increasing radius.  On the other hand, the wavenumber from linear stability is
\be
   k_r H(r) = \frac{2}{|q|} \frac{R_0}{H_0} \approx 37,
\ee
since their $q = -1$ and $H_0/R_0 = 0.054$.  Hence their wavelengths are between a factor of 1.5 to 2.6 larger than for linear instability (for the locally isothermal equation of state).  Curiously, their Figure 19 does not show any tendency for the wavelength to increase with time.  This is at odds with the axisymmetric radiative hydrodynamic simulations of M24b and should be investigated.
 
\section{3D simulation} \label{sec:3d_sim}
\subsection{Perturbation vorticity in a meridional plane}\label{sec:meridional_vorticity}

\begin{figure}
 \centering
\includegraphics[width=3.4truein,  trim=2 2 2 2,clip]{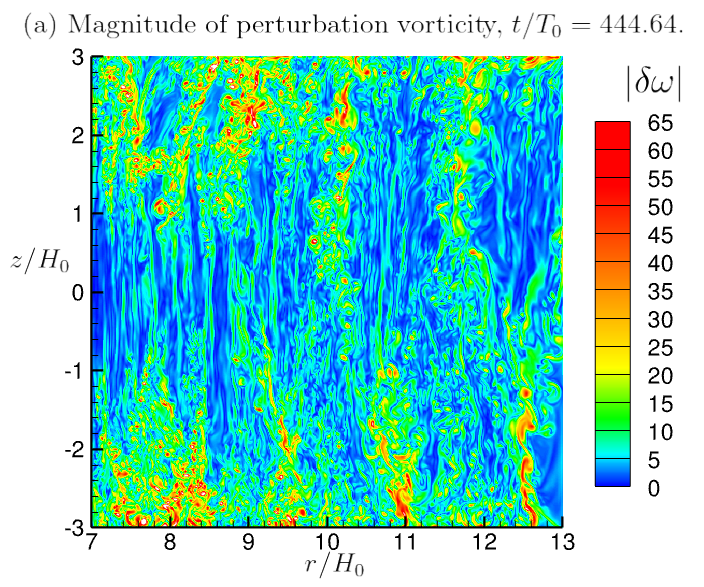} 
\includegraphics[width=3.4truein,  trim=2 2 2 2,clip]{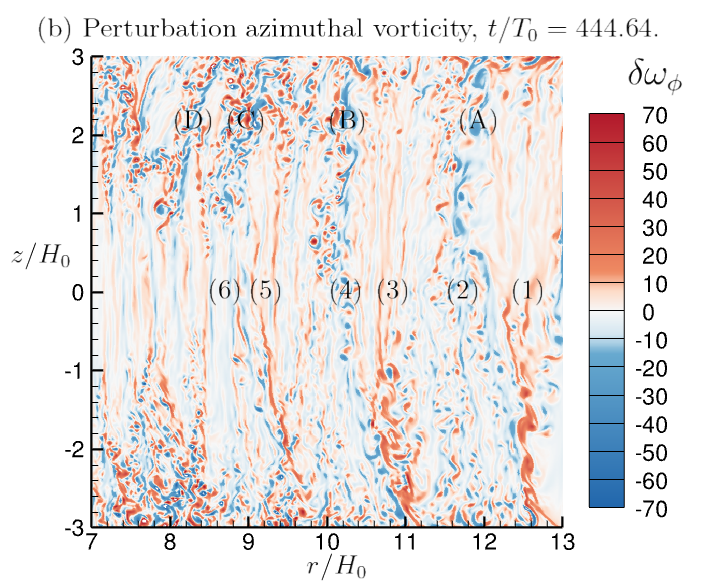}
\includegraphics[width=3.4truein,  trim=2 2 2 2,clip]{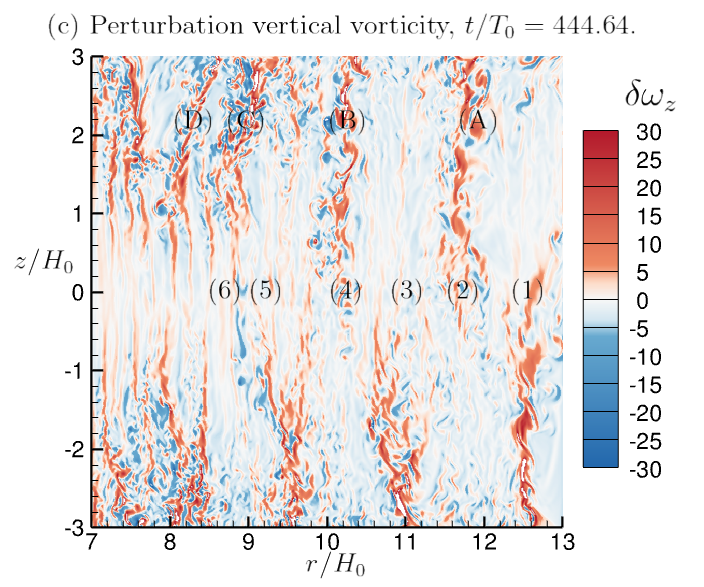} 
\caption{3D run.  Magnitude and components of the perturbation vorticity (relative to the basic state) in the $\phi = 0$ meridional plane at $t/T_0 = 444.64$. Turbulent shear layers are numbered (1)--(6) at the midplane and (A)--(B) at $z/H_0 = 2$.}
\label{fig:vort_merid}
\end{figure}
We now turn to the 3D simulation.  Figure~\ref{fig:vort_merid} shows components of the perturbation vorticity ($\delta\vec{\omega}$) in a meridional ($rz$) plane at $t/T_0 = 444.64$ during the period when the strength of the Pad\'e filter has been reduced to $\epsfil = 0.015$.
As in the axisymmetric case, the strongest component is $\delta\omega_\phi$ (panel (b)) and consists of at least 6 turbulent shear layers (TSLs) that extend across the vertical extent of the disk; they are numbered from right to left and alternate in the dominant sign of $\delta\omega_\phi$.  
The predilection for negative layers for $z > 0$ and positive layers for $z < 0$ observed in the axisymmetric case (Figure~\ref{fig:vor_axi}(a)) is still noticeable.
A more accurate description is that half of each TSL with predominantly $\delta\omega_\phi\lessgtr 0 $ is more organized for $z \gtrless 0$, undergoes KH instability, and becomes disorganized or filamentary for $z \lessgtr0$.  This was also true in the axisymmetric case.

The vertical component ($\delta\omega_z$, Figure~\ref{fig:vort_merid}(c)) is the next strongest; however, its values are about $1/3$ as large as in the axisymmetric case (compare with Figure \ref{fig:vor_axi}(b)).
Inspection of the vertical velocity (not shown) indicates that each TSL forms the boundary of a vertical jet.
The numerical labels in panel (c) are placed at the same locations as in panel (b).  
One observes that $\delta\omega_z$ is dominated by positive values and exists on either the upper or lower half of each TSL.  This makes $\delta\omega_z$  regions in the upper and lower halves staggered relative to each other.
The explanation for this was given in \S\ref{sec:axi_vel}.
To the extent that the $\delta\omega_z$ TSLs remain coherent in the azimuthal direction (which is the case in the upper layers of the disk as will be seen later), they correspond to a change in angular momentum as shown earlier (Figure~\ref{fig:jz_axi_early}, blue line).  This is much weaker than in the axisymmetric case.   We will see later that some of these layers undergo Kelvin-Helmholtz type instability as predicted in the L\&P analysis.

One can define the Rossby number ($\mathrm{Ro}$) as the ratio of $|\delta\vec{\omega}|$ to the local Keplerian vorticity; in code units we have
\be
   \mathrm{Ro} \equiv \frac{|\delta\vec{\omega}|}{\pi (r/10)^{-3/2}}.
\ee
We find that $\mathrm{Ro} \approx 10$ in filaments that cross the midplane and reaches values of 20 in KH eddies near the midplane.  In the upper layers of the disk, $\mathrm{Ro}$ attains values as high as 125. A high value of $\mathrm{Ro}$ in a vortex or vortex layer indicates that it is minimally unaffected by Keplerian rotation and mean shear.  

\begin{figure}
\centering
\includegraphics[width=3.2truein]{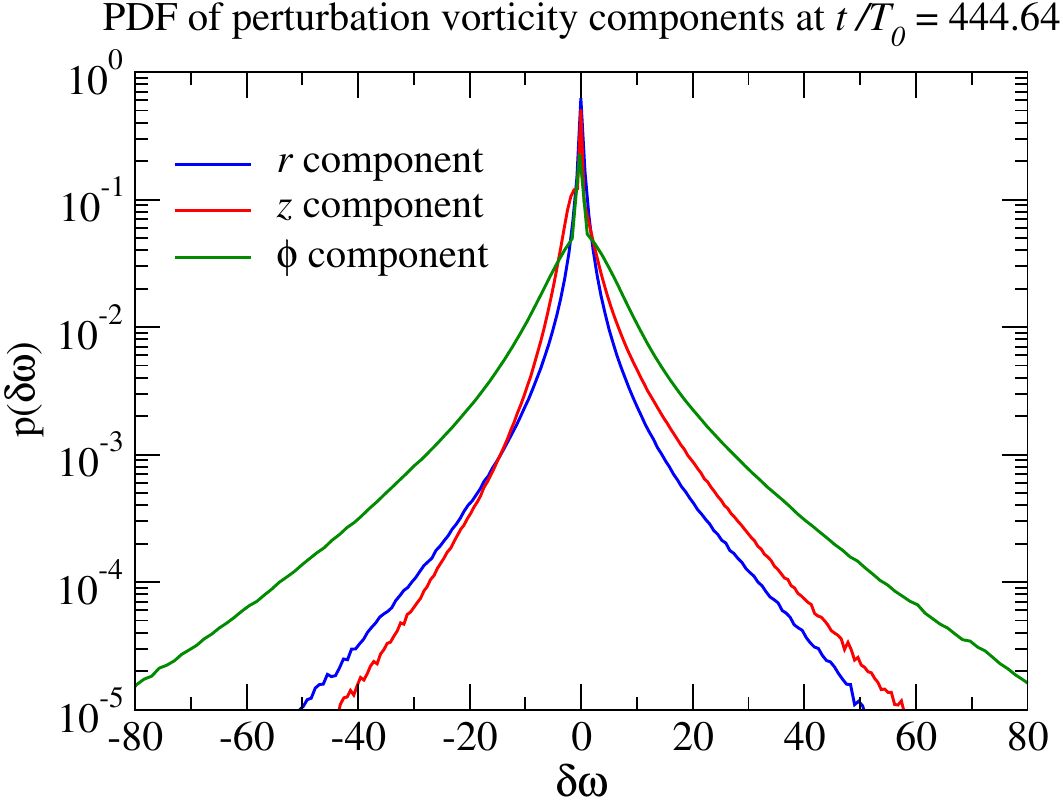}
\caption{Perturbation vorticity PDF for the 3D run at $t/T_0 = 444.64$ sampled at 128 meridional planes.}
\label{fig:pdf3d}
\end{figure}
Figure \ref{fig:pdf3d} plots the PDF of perturbation vorticity components at $t/T_0 = 444.64$ sampled on 128 $\phi = \mathrm{constant}$ planes.  As in the axisymmetric case, $\delta\omega_\phi$ (green curve) is dominant.  The preference for $\delta\omega_z > 0$ (red curve) is still present but to a smaller extent than in the axisymmetric case. 

\subsection{Midplane vorticity}\label{sec:midplane_vort}

\begin{figure}
 \centering
\includegraphics[width=3.5truein,trim=2 2 2 2,clip]{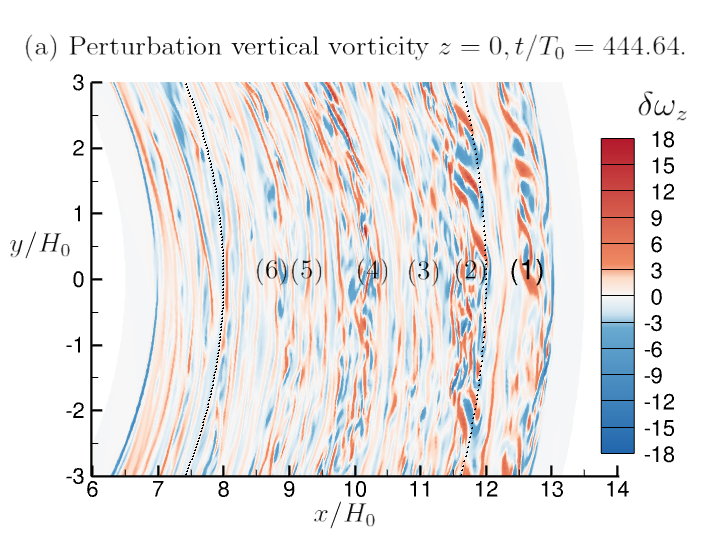}
\includegraphics[width=3.5truein,trim=2 2 2 2,clip]{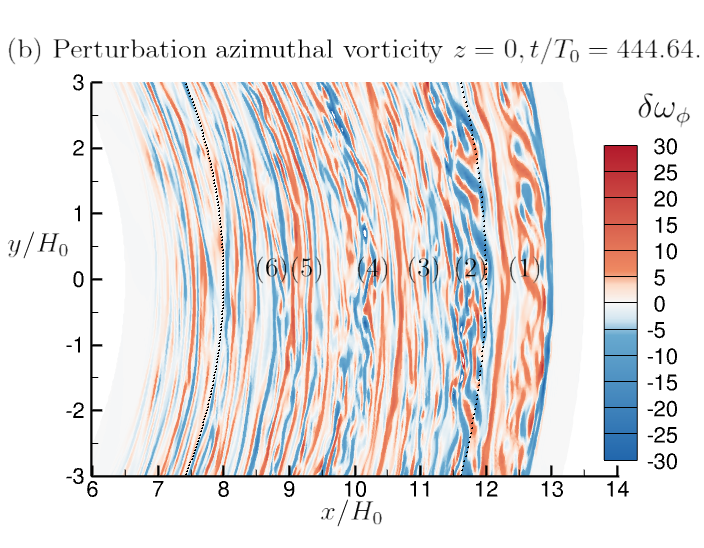}
\includegraphics[width=3.5truein,trim=2 2 2 2,clip]{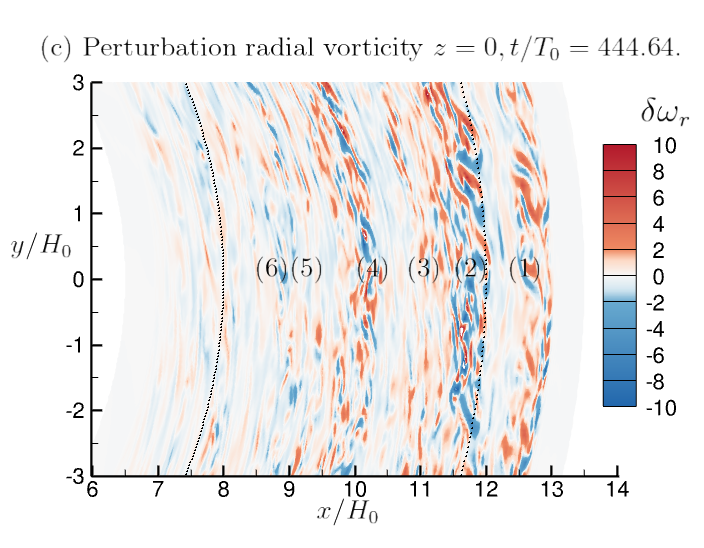}
\caption{Components of the perturbation vorticity (relative to the basic state) in the midplane at $t/T_0 = 444.64$.  The dotted lines indicate regions of width $\Delta r/H_0$ adjacent to each radial boundary that are affected by boundary conditions as indicated later in the plot of the turbulent $\alpha(r)$ parameter.}
\label{fig:vort_horiz}
\end{figure}
Since particles settle to the midplane, the turbulence there is of special interest.  Figure~\ref{fig:vort_horiz} shows components of the perturbation vorticity relative to the basic state.  
The plot window is centered at $\phi = 0$ to allow comparison with the $\phi = 0$ meridional plane shown in Figure~\ref{fig:vort_merid}.  The numerical labels are at the same positions as in Figure~\ref{fig:vort_merid}.
The dotted lines indicate regions of width $\Delta r/H_0=1$  near the radial boundaries where the flow is affected  by boundary conditions as indicated later (\S\ref{sec:alpha}) by anomalous local peaks in the turbulent $\alpha(r)$.
The flow consists of bands where the flow is most non-axisymmetric, i.e., not sheared along the azimuth.  
The bands correspond roughly to where the 6 TSLs observed in the meridional plane cross the midplane.
Each band consists of lower aspect ratio $\delta\omega_z$ vortices of length $\lesssim 0.4 H_0$.
At outer radii, $\delta\omega_z$ vortices of both signs having similar aspect ratio are observed.  However, at inner radii, $\delta\omega_z > 0$ is more filamentary because cyclonic $\delta\omega_z$ is easily sheared \citep{Kida_1981} by the Keplerian mean (which is stronger at inner radii).
Adjacent to each band is a region where perturbation vorticity is more sheared.  One also observes that $\delta\vec\omega$ weakens with decreasing $r$; it is also more sheared, likely because the Keplerian shear increases as $r$ decreases.  It is emphasized that the $\delta\omega_z$ structures that have low aspect ratio (i.e. are not elongated) are often associated with large values of $\delta\omega_\phi$ and $\delta\omega_r$.  

The perturbation vorticity is dominated by the azimuthal component (Figure \ref{fig:vort_horiz}(b)), consistent with what is observed in a meridional plane and in PDFs (Figure~\ref{fig:pdf3d}).  This component is associated with the jets of vertical velocity and is generally more sheared than $\delta\omega_z$.  However, there are intermittent locations where $\omega_\phi$ has smaller aspect ratio structures where, as mentioned above, other vorticity components are also active. 

Figure \ref{fig:vort_horiz}(c) shows the perturbation radial vorticity: it has the lowest values among the three and is occasionally associated with other vorticity components.  It most likely arises from the tilting and stretching of $\omega_z$ by $\omega_\phi$ KH vortices.

An important difference in our simulations compared to M20 is the lack of large persistent vortices.  This difference will be discussed in \S\ref{sec:long_lived}.
\subsection{Vorticity in a horizontal plane at two scale heights}\label{sec:at_two_H}

Perturbation vorticity components increase in magnitude away from the midplane and are therefore able to withstand the stabilizing effects of Keplerian rotation and shear.
\begin{figure}
\centering
\includegraphics[width=3.5truein,trim=2 2 2 2,clip]{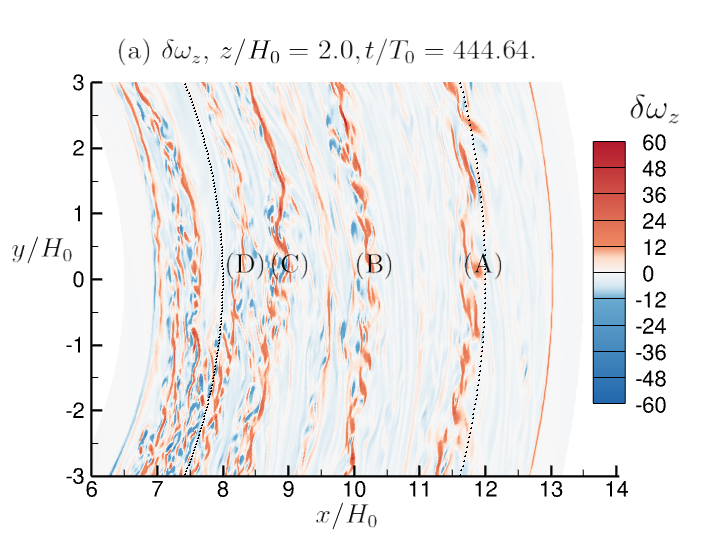}
\includegraphics[width=3.5truein,trim=2 2 2 2,clip]{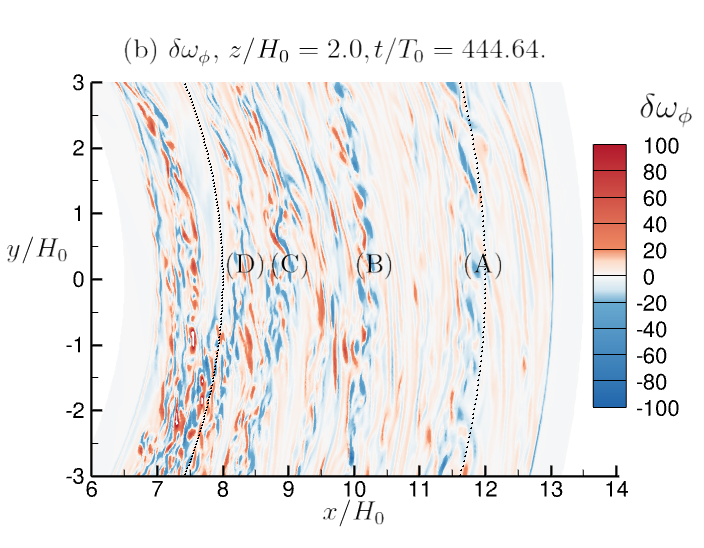} 
\includegraphics[width=3.5truein,trim=2 2 2 2,clip]{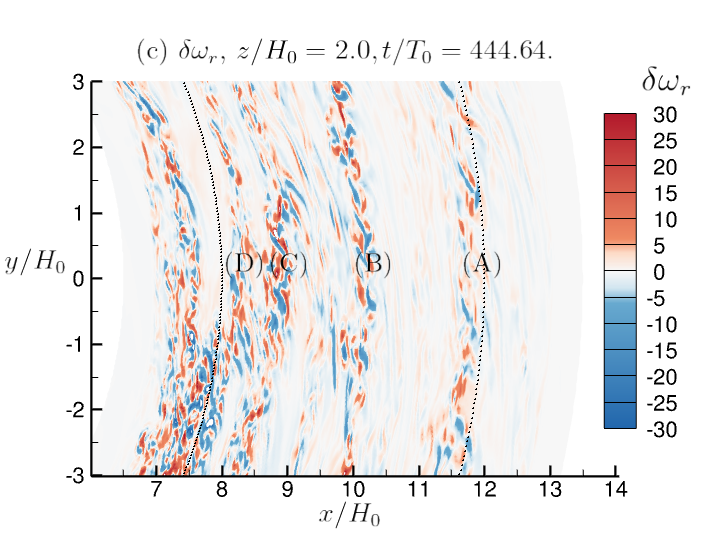}
\caption{Components of the perturbation vorticity (relative to the basic state) at $z = 2 H_0$ and $t/T_0 = 444.64$.}
\label{fig:vort_2H}
\end{figure}
Figure~\ref{fig:vort_2H} shows components of $\delta\vec\omega$ at $z = 2H_0$.  
The letters (A)--(D) are placed at the same locations as in the meridional plane plots (Figure~\ref{fig:vort_merid}).
Panel (a) depicts the vertical component; it is more than three times stronger than at the midplane.  It is dominated by $\delta\omega_z > 0$ layers and displays KH-like instability for larger $r$, presumably because rotation and shear weaken as $r$ increases.  A few small elliptical $\delta\omega_z < 0$ structures are also present. Figure~\ref{fig:vort_2H}(b) shows that the $\delta\omega_z > 0$ layers are associated with negative $\omega_\phi$, just as in the axisymmetric case (\S\ref{sec:axi_vort}).  Recall from the axisymmetric case that the lower half of the disk has $\omega_\phi >0$ in these layers.  The values of $\omega_\phi$ are about three times larger than at the midplane.
Figure~\ref{fig:vort_2H}(c) shows that the weakest component, namely, $\delta\omega_r$ is also about thrice as large as it is at the midplane.

\subsection{Radial wavelength in 3D} \label{sec:3D_wavelength}
\begin{figure}
 \centering
\includegraphics[width=3.5truein,trim=2 2 2 2,clip]{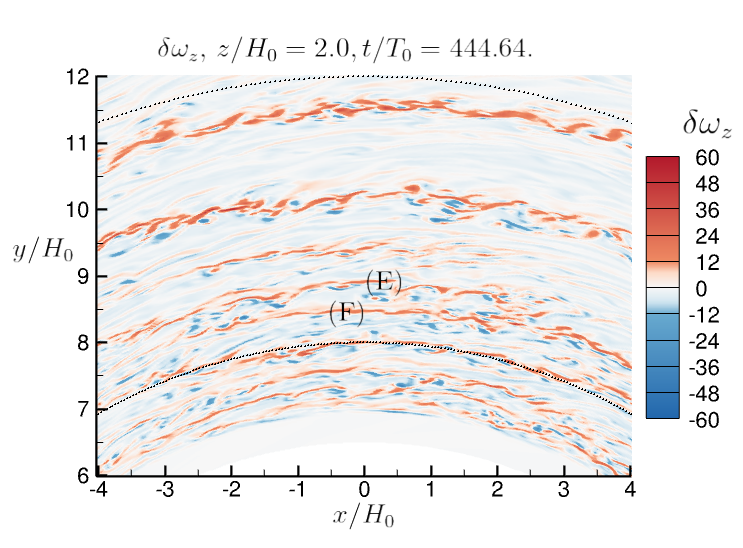}
\caption{$\delta\omega_z$ at $z = 2H_0$ centered at $\phi = 90^\circ$ ($t/T_0 = 444.64$).}
\label{fig:vort_z_2H_phi_90}
\end{figure}

The $\delta\omega_z$ layers in Figure~\ref{fig:vort_2H}(a) (labeled (A)--(D)) can be used to infer a wavelength.  These layers are coherent over a large azimuthal extent.  
We find that the ratio of the interlayer spacing $\ell_r$ to the local instability wavelength is 4.4 for layers (A) and (B), 4.5 for (B) and (C), and 2.4 for (C) and (D).
Figure \ref{fig:vort_z_2H_phi_90} shows $\delta\omega_z$ centered at $\phi = 90^\circ$: the spacing ratio for layers (E) and (F) is 1.5.  Therefore the spacing ratio increases with radius.  This means that the spacing $\ell_r(r)$ increases with $r$ faster than $H(r)$ does.
The inexorable radial wavelength growth with time present in axisymmetric VSI is halted in 3D and a time invariant local radial length scale is established.  An interesting question is whether $\ell_r(r)$ is an appropriate length scale for determining the turbulent viscosity $\nu_\rmt$.

\subsection{Specific kinetic energy spectra in the midplane}\label{sec:spectra}
Spectra of velocity squared (which we refer to as the specific kinetic energy) with respect to the radial and azimuthal directions are defined as
\begin{align}
  S_{uu}(k_r) &= \bfuh(k_r) \cdot \bfuh^*(k_r),\\
  S_{uu}(m) &= \bfuh(m) \cdot \bfuh^*(m),
\end{align}
where the hat denotes a real to complex Fourier transform, a star denotes the complex conjugate, and $m$ is the azimuthal wavenumber.  
To obtain $S_{uu}(k_r)$, a Hanning window was applied in $r$ since it is not a periodic direction, after which an average was taken in $\phi$.
For $S_{uu}(m)$, an average was taken in $r$.
Note that spatial inhomogeneity of the flow in $r$ precludes a rigorous mathematical interpretation of these spectra.

\textit{Remark.} Dr. J. David Melon Fuksman (private communication) has suggested that since the flow has nonuniform density, it would be more appropriate to consider the quantity
\be
   E(\bfk) \equiv \rmRe\left\{\widehat{\rho\bfu}(\bfk) \cdot\bfuh^*(\bfk)\right\}. \eql{Ek}
\ee
Note that since $\rmRe(fg^*) = \rmRe(f^* g)$ it does not matter whether we conjugate $\widehat{\rho\bfu}$ or $\bfuh$.  \citet[][pg. 110]{Dutton_1963} has shown that the integral of \eqp{Ek} with respect to $\bfk$ is proportional to the kinetic energy when the support of the velocity field $\bfu(\bfx)$ is compact.

\begin{figure}
 \centering
\includegraphics[width=3.3truein]{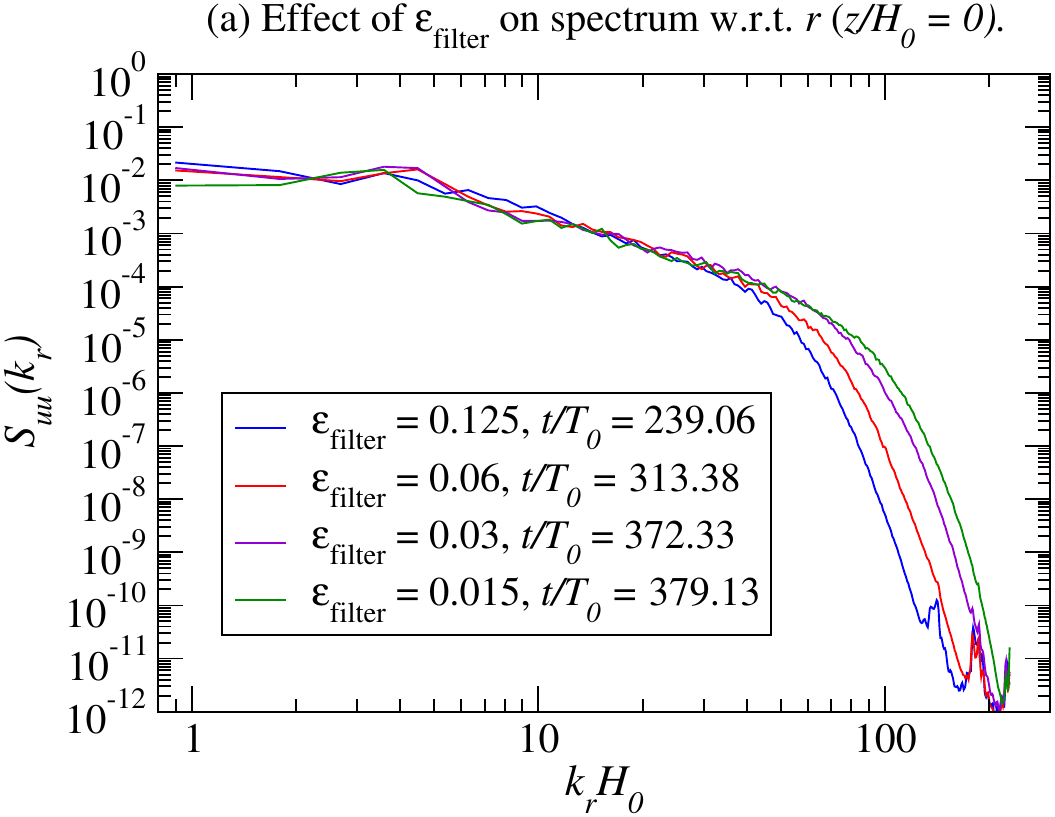}
\vskip 0.2 truecm
\includegraphics[width=3.3truein]{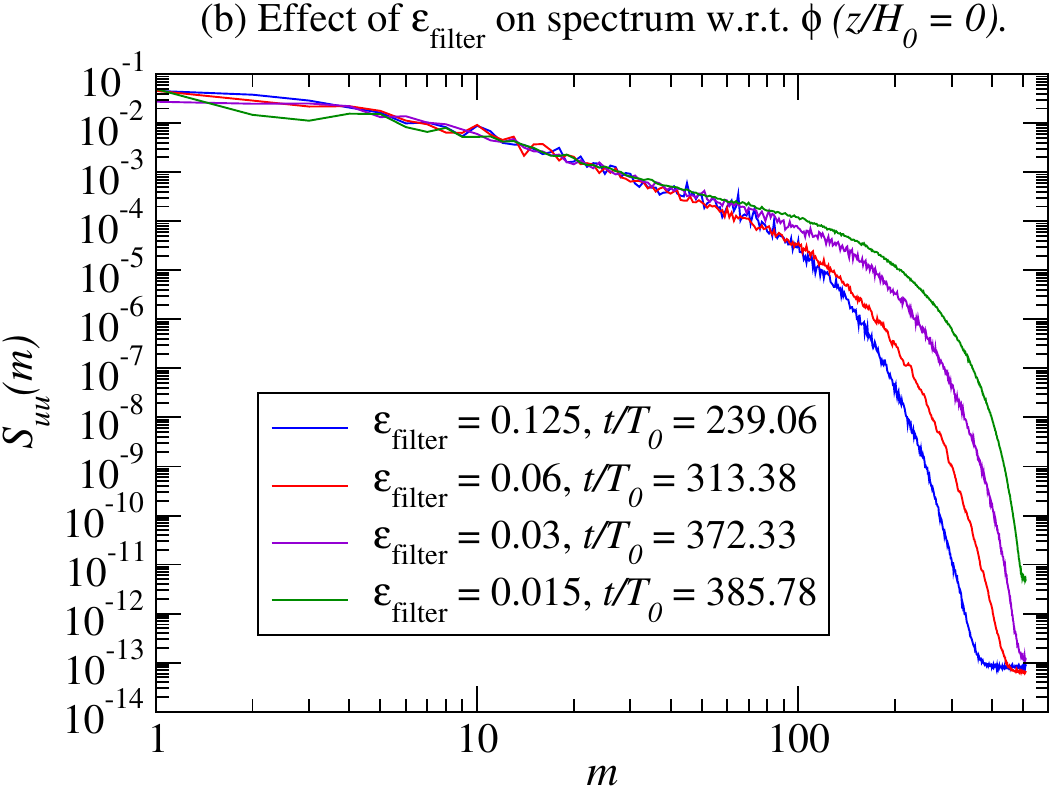}
\caption{Effect of the strength of the Pad\'e filter ($\epsfil$) on specific energy spectra in the midplane.  (a) With respect to the radial wavenumber, $k_r$.  (b) With respect to the azimuthal wavenumber, $m$.
The units of $S_{uu}$ are those of velocity squared, i.e., $(H_0/T_0)^2$.}
\label{fig:effect_of_filter_on_spec}
\end{figure}
Figure~\ref{fig:effect_of_filter_on_spec} shows the effect of the filter strength $\epsfil$ on spectra.  When $\epsfil$ is lowered, more of the power law region (inertial range) is resolved without a significant upturn near the Nyquist wavenumber, which indicates that $2\Delta$ oscillations are under control.  This was confirmed by zooming in on vorticity contour plots: there are some isolated $2\Delta$ oscillations, but they are not widespread.

Next, we investigate the anisotropy of the turbulence by plotting spectra of the different velocity components for the smallest filter strength ($\epsilon_\mathrm{filter} = 0.015$). To reduce statistical noise, spectra were computed for fifty fields in the period $t/T_0 \in [400.20, 452.55]$ and averaged.
\begin{figure}
 \centering
\vskip 0.4truecm
\includegraphics[width=3.3truein]{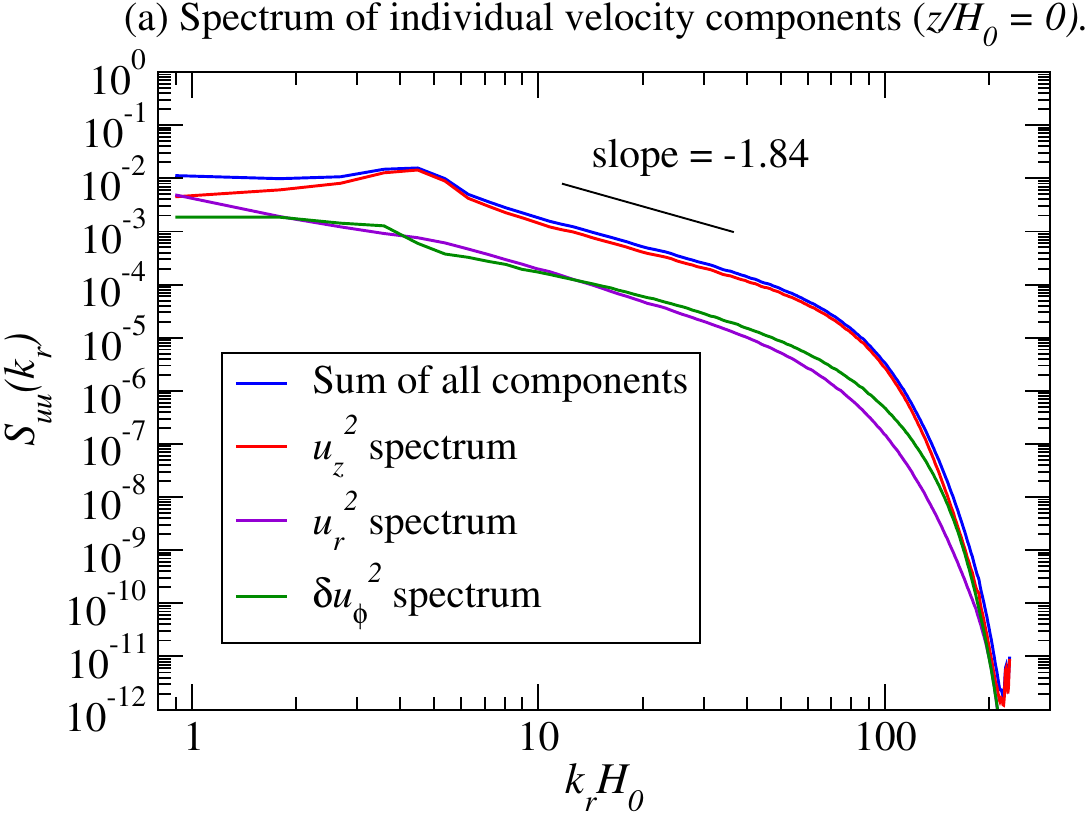}
\vskip 0.4truecm
\includegraphics[width=3.3truein]{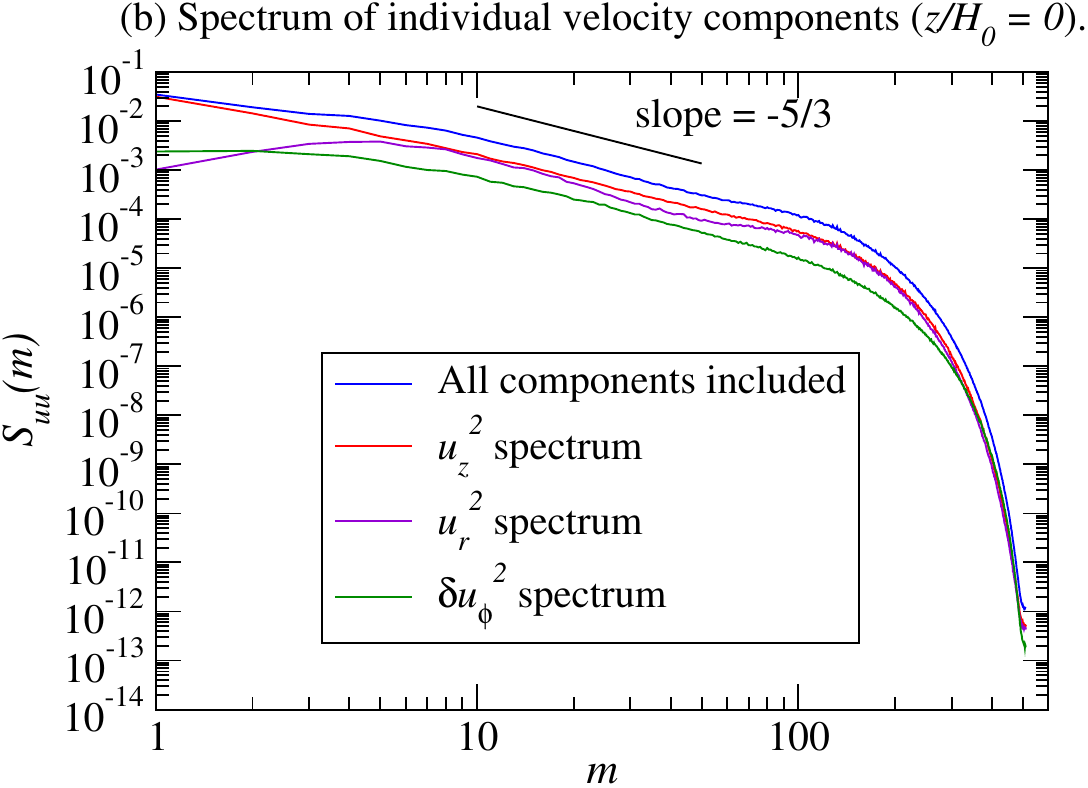}
\caption{Specific energy spectra of individual velocity components in the midplane.  Averaged over the time interval $t/T_0 \in [400.20, 452.55]$; $\epsfil = 0.015$.  (a) With respect to the radial wavenumber, $k_r$.  (b) With respect to the azimuthal wavenumber, $m$.}
\label{fig:spec}
\end{figure}
Figure~\ref{fig:spec}(a) displays the spectrum with respect to $k_r$ and shows that $u_z$ dominates the kinetic energy for most of the wavenumber range.  A bump corresponding to energy injection by VSI is clearly observed at $k_r H_0 \approx 4.5$ which corresponds to a wavelength of $\ell_r \approx 1.4 H_0$.  On the other hand, the most linearly unstable wavelength is $\lammax = 0.31$ at the midradius of the domain.
This bump is followed by a short (about 0.7 of a decade) power law range with a slope of $\approx -1.84$.  This is close to the $-2$ slope that a jump in velocity across a shear-layer would produce \citep{Townsend_and_Taylor_1951}.  Figure~\ref{fig:spec}(b) displays spectra with respect to the azimuthal wavenumber $m$.  The total energy (blue) has a slope close to the Kolmogorov value of $-5/3$.  However, there is a small dip (relative to a straight line) at $m \approx 40$ which corresponds to a wavelength of $\approx 9^\circ$; we see that the dip arises from the radial component (purple line).  Again, $u_z$ (red) is the dominant component though not as strongly as was the case for $k_r$ spectra. Radial velocity fluctuations (violet) are not much smaller than vertical velocity fluctuations (red) in the inertial (power law) range.  Azimuthal velocity fluctuations (green) have the least amplitude throughout the wavenumber range.

The azimuthal spectrum of \citet[][their Figure 13]{Manger_and_Klahr_2018} shows a broken power law consisting of a steeper than $-5/3$ slope followed by a slope of $-5$.  On the other hand, we observe a a single power law slope $\approx -5/3$ without a region of slope $= -5$.  Also, whereas our spectra have a dissipation range in which the spectra fall-off faster than any power law, the spectra of \cite{Manger_and_Klahr_2018} lack such a range.  The reasons for these discrepancies remain to be investigated.

\subsection{Specific kinetic energy spectra at different heights}\label{sec:spectra_vs_z}
\begin{figure}
 \centering
\includegraphics[width=3.3truein]{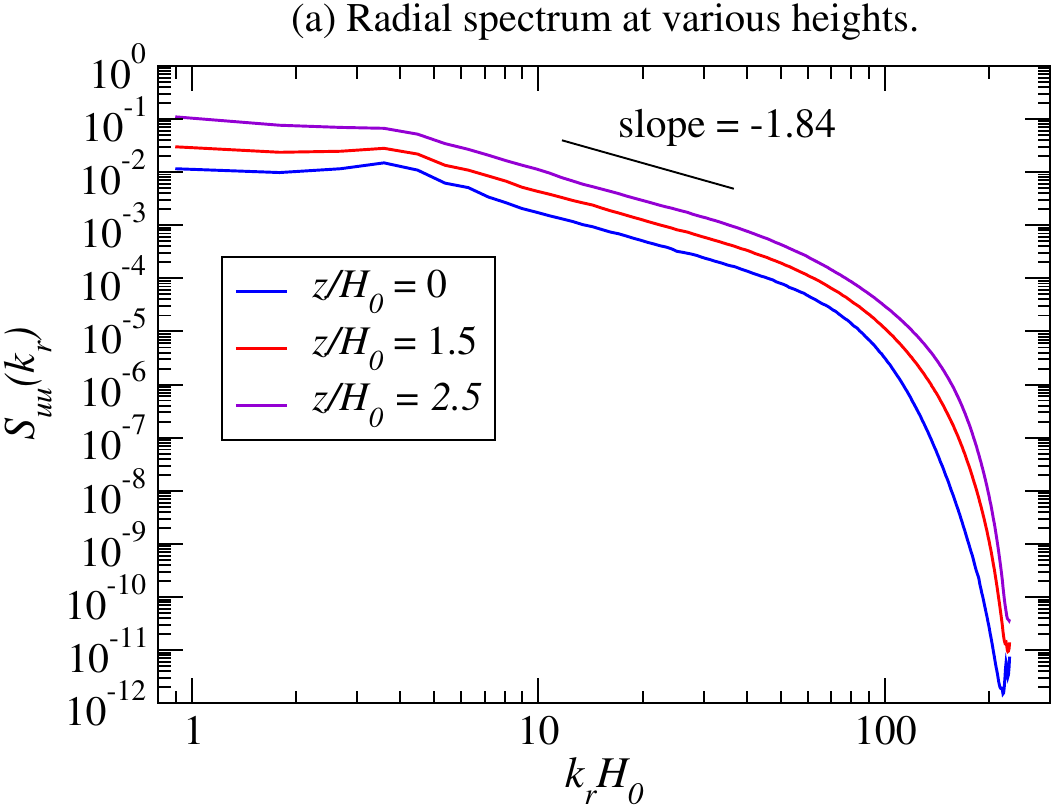}
\vskip 0.4truecm
\includegraphics[width=3.3truein]{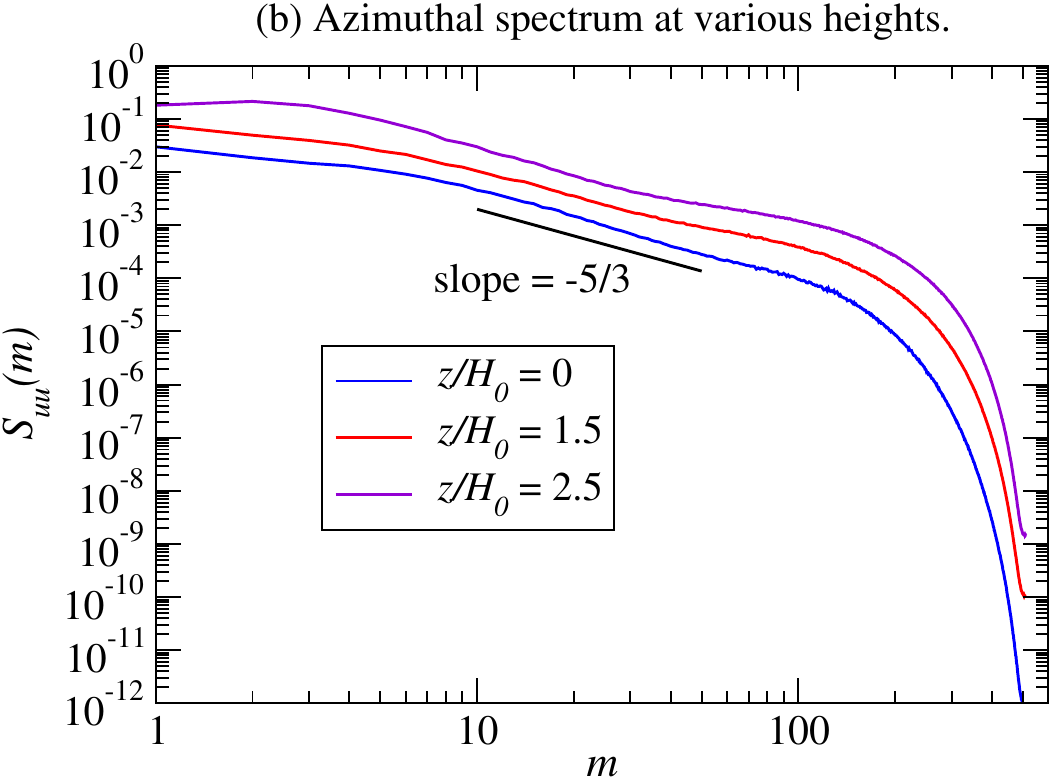}
\caption{Specific energy spectra (all velocity components included) at different heights averaged over samples in the time interval $t/T_0 \in [444.6, 479.3]$ with spacing $\delta t/T_0 = 0.5$.  Strength of Pad\'e filter $\epsfil = 0.015$.  (a) With respect to the radial wavenumber, $k_r$.  (b) With respect to the azimuthal wavenumber, $m$.}
\label{fig:spec_vs_z}
\end{figure}
Figure~\ref{fig:spec_vs_z} shows radial and azimuthal spectra (with all velocity components included) at three different heights.  The specific energy increases with height at each radial and azimuthal wavenumber.  The radial spectrum has the same slope in the power law region at all three heights.  On the other hand, at higher $|z|$, the azimuthal spectrum displays a greater dip at about $m = 30$.

\begin{figure*}
\centering
 \includegraphics[width=2.3truein,trim=3 3 3 3,clip]{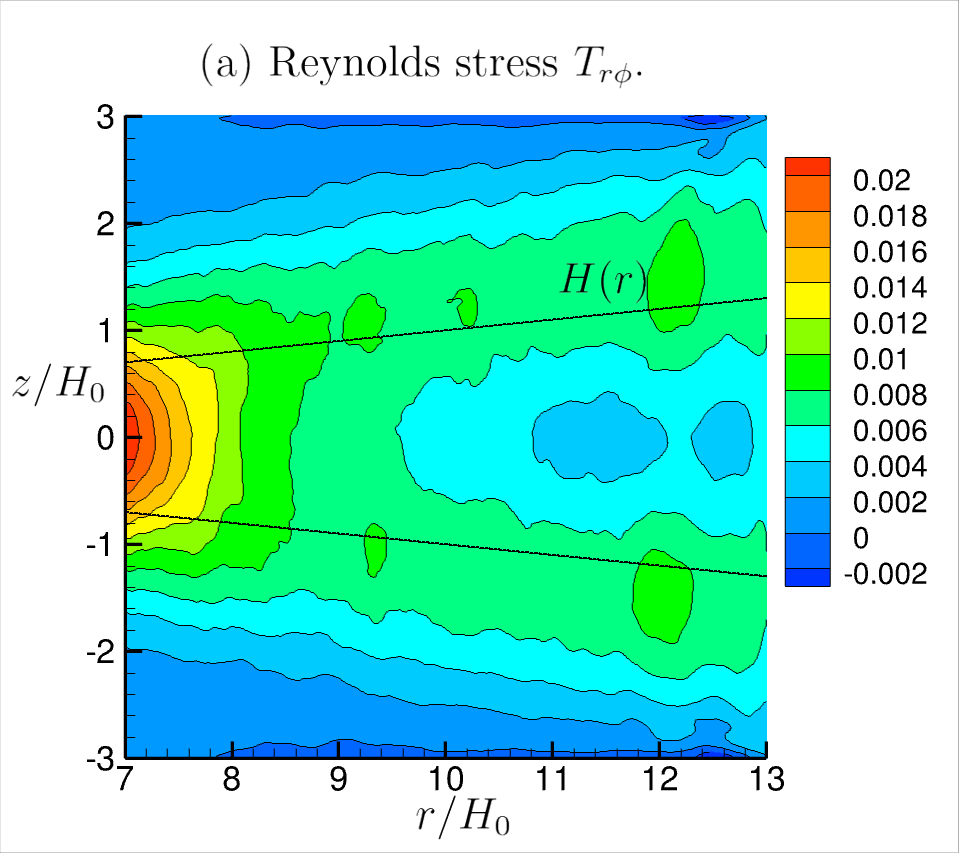}
 \hfill
\includegraphics[width=2.3truein,trim=3 3 3 3,clip]{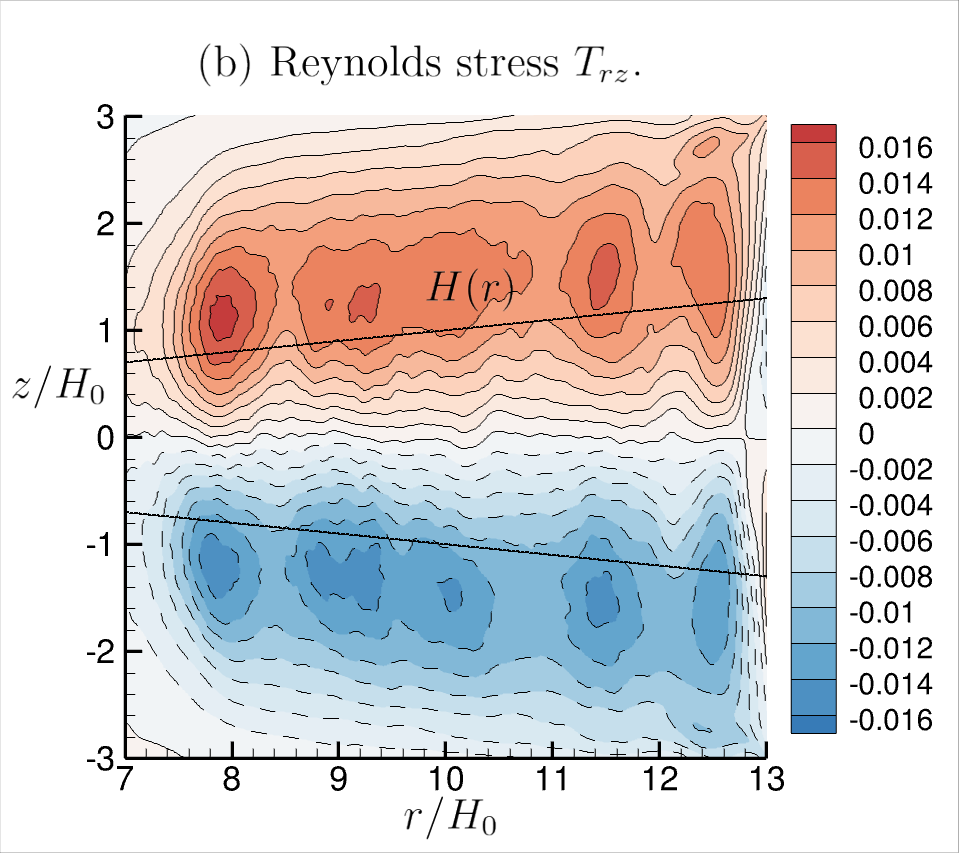} 
\hfill
\includegraphics[width=2.3truein,trim=3 3 3 3,clip]{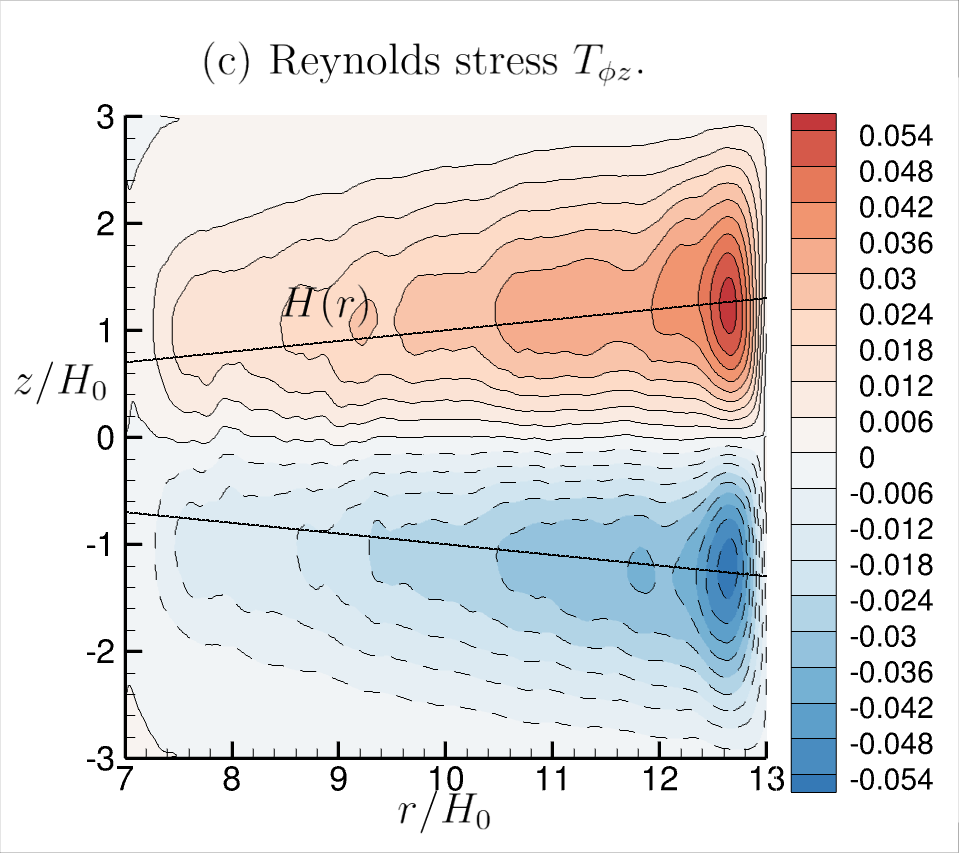}                 
\hfill
\includegraphics[width=2.3truein,trim=3 3 3 3,clip]{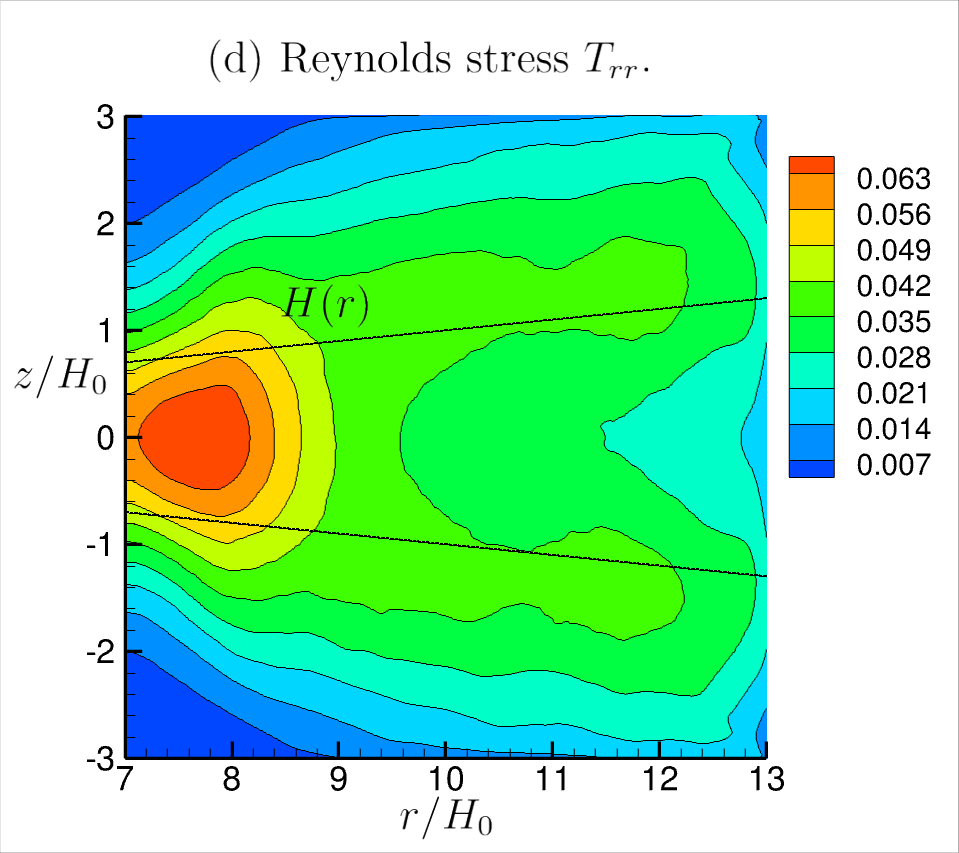} 
 \hfill
 \includegraphics[width=2.3truein,trim=3 3 3 3,clip]{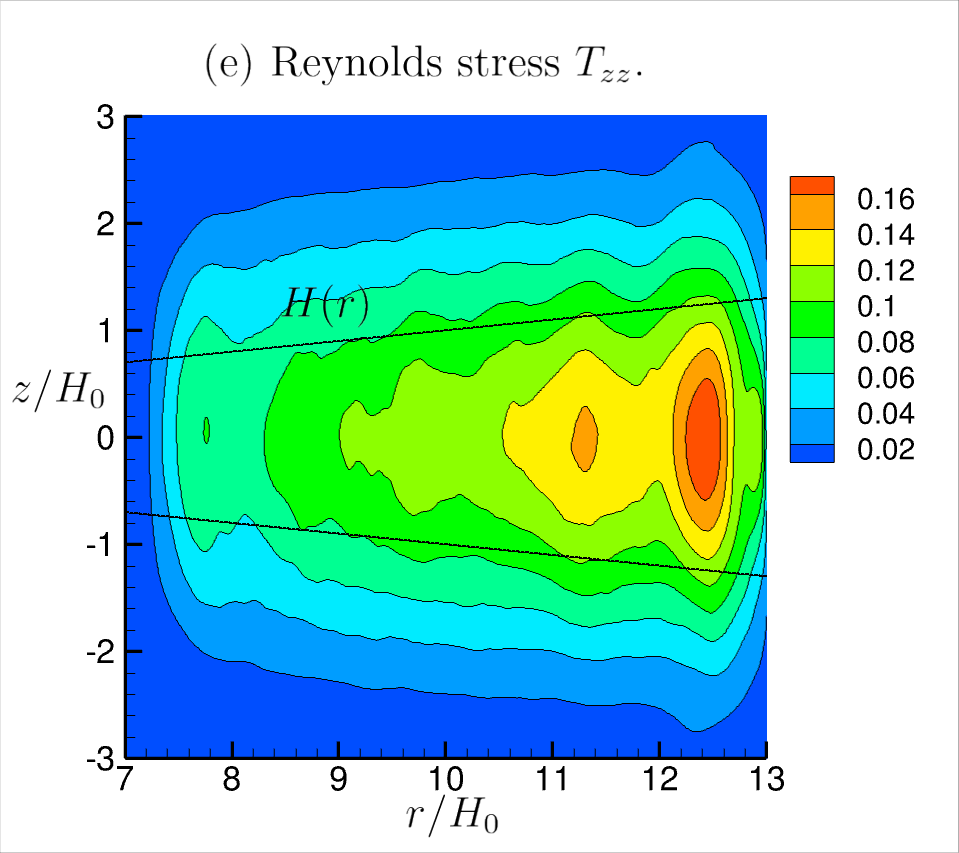}
 \hfill
\includegraphics[width=2.3truein,trim=3 3 3 3,clip]{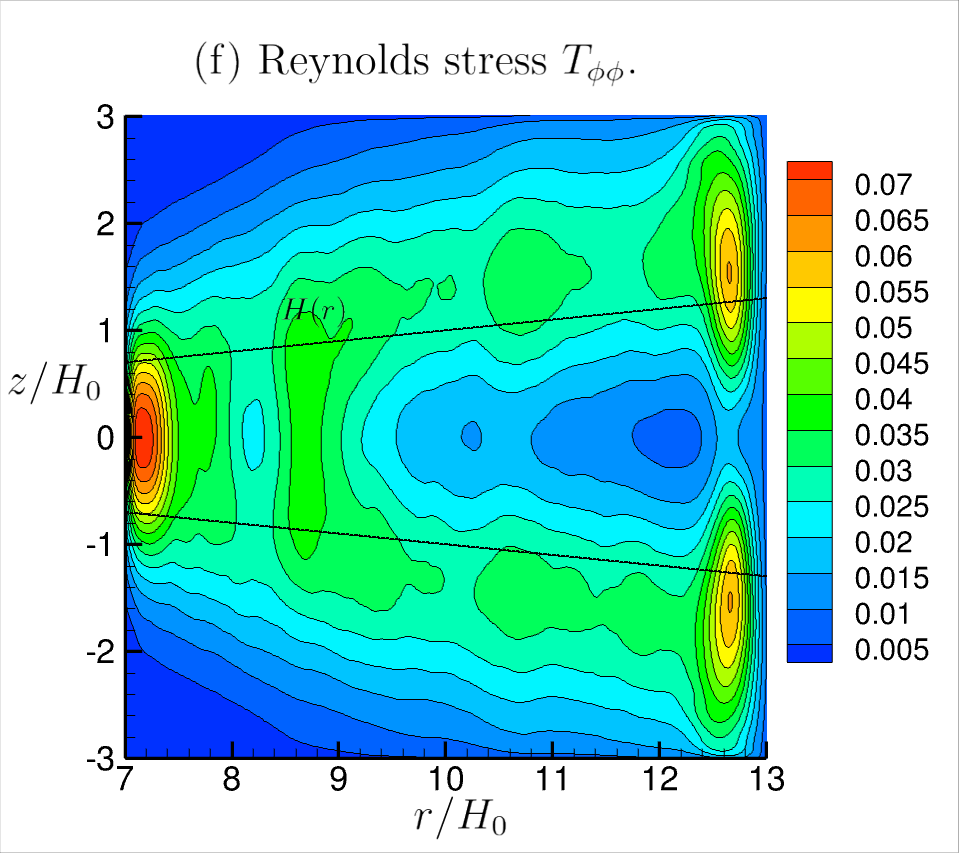} 
\hfill
\caption{Reynolds stress tensor.  Averaging was performed in the period $t/T_0 \in [54.53,300.19]$.  A region of $\Delta r/H_0 \approx 1$ at each radial boundary should be considered contaminated by boundary conditions.}
\label{fig:stresses}
\end{figure*}

\subsection{Reynolds stresses}\label{sec:stresses}
Here we present maps of the Reynolds stress tensor, $T_{ab}$, in the meridional plane.  
The role of $T_{ab}$ in the mean flow equations and the calculation of $T_{ab}$ from simulation data is described in Appendices \ref{sec:ra} and \ref{sec:cra}, respectively.  Time averaging was performed in the interval $t \in [54.53, 300.19]$ with samples spaced $\Delta t/T_0 \approx 0.5$ apart.  Off-diagonal components of the Reynolds stress tensor are shown in the top row of Figure~\ref{fig:stresses}.  Note that that the component $T_{r\phi}$ enters the mean angular momentum equation \eqp{ra:amb}.  Away from the radial ends of the computational domain, where the flow is influenced by artificial numerical boundary conditions, $T_{r\phi}$ peaks away from the midplane at slightly above $z/H(r) \approx 1$.  An off-midplane peak was also observed in \citet[][their Figure 5]{Stoll_and_Kley_2014}.  As will be seen in \S\ref{sec:mrmf}, this does not mean that the starward accretion mass flow peaks away from the midplane.  The largest off-diagonal component is $T_{\phi z}$.
To explain its sign pattern, positive and negative above and below the midplane, respectively, we invoke the mechanism given in \S\ref{sec:axi_vel} for the sign pattern of the product $\delta u_z \delta u_\phi$ in axisymmetric VSI, namely, the transport of a vertically varying mean specific angular momentum ($\p_z \overline{j}_\phi \neq 0$) by vertical jets.

Figure~\ref{fig:stresses} plots the normal components.  One observes that the peak value of $T_{zz}$ is about three times as large as that of $T_{rr}$ and $T_{\phi\phi}$, reflecting the dominance of vertical jet-like fluctuations.  In addition, $T_{zz}$  peaks at the midplane.  This is due to the presence of density ($\rho$) in the definition of the stresses, which rises faster toward the midplane than vertical velocity rms decreases.  On the other hand, $T_{rr}$ and $T_{\phi\phi}$ peak above the midplane at $z \approx 1.5 H_0$. This is consistent with the peak in $T_{r\phi}$ above the midplane.  In fact, the correlation coefficient
\be
   C_{r\phi} \equiv \frac{T_{r\phi}}{(T_{rr}T_{\phi\phi})^{1/2}},
\ee
was found to be almost constant at $\approx 0.2$-$0.3$ throughout the domain.  This suggests that the shapes of the structures responsible for $T_{r\phi}$ are similar at all locations of the domain; only their intensity and/or size varies.
The trace $T_{kk}$ (not shown), which is twice the kinetic energy, peaks at the midplane but maintains high values up to $z \approx 1.5 H_0$.  Strong peaks in all the stresses near the left and right radial boundaries are considered to be artifacts due to boundary conditions.

\subsection{The turbulence $\alpha$ parameter} \label{sec:alpha}

The $\alpha$ parameter introduced by \cite{Shakura_and_Sunyaev_1973}, is a non-dimensional turbulent viscosity in vertically integrated accretion disk theory.  Specifically,
\be
   \alpha(r) \equiv \frac{\nut(r)}{\cs(r) H(r)}, \eql{alpha}
\ee
where $\nut(r)$ is the turbulent viscosity.  In our calculation, we took $\cs(r)$ and $H(r)$ to be the isothermal sound speed and scale height of the basic state, respectively. In vertically integrated accretion disk analysis, which uses cylindrical coordinates, the vertically integrated turbulent torque appears as
\be
   \tau = \int_{-\infty}^{\infty} dz\, \frac{1}{r} \frac{\p}{\p r}\left(-r^2 T_{r\phi}\right). \eql{tau_def}
\ee 
The torque is then modelled using a turbulent viscosity as:
\be
\tau = \frac{1}{r} \frac{\p}{\p r} \left(r^2 \Sigma \nut r \frac{\p\overline{\Omega}}{\p r}\right), \eql{tau_model}
\ee
where $\overline{\Omega}$ is the Reynolds averaged vertically integrated angular rotation rate; the difference in $\alpha(r)$ was found to be very small when $\overline{\Omega}$ was replaced by the Keplerian value.  Comparing \eqp{tau_def} and \eqp{tau_model} one obtains
\be
   \nut = \frac{- \int_{-\infty}^{\infty} dz\, T_{r\phi}}{\Sigma r \left(\p\overline{\Omega}/ \p r \right)}, \eql{nut}
\ee
which when substituted into \eqp{alpha} allows us to calculate $\alpha(r)$.

\begin{figure}
\centering
\includegraphics[width=3.2truein]{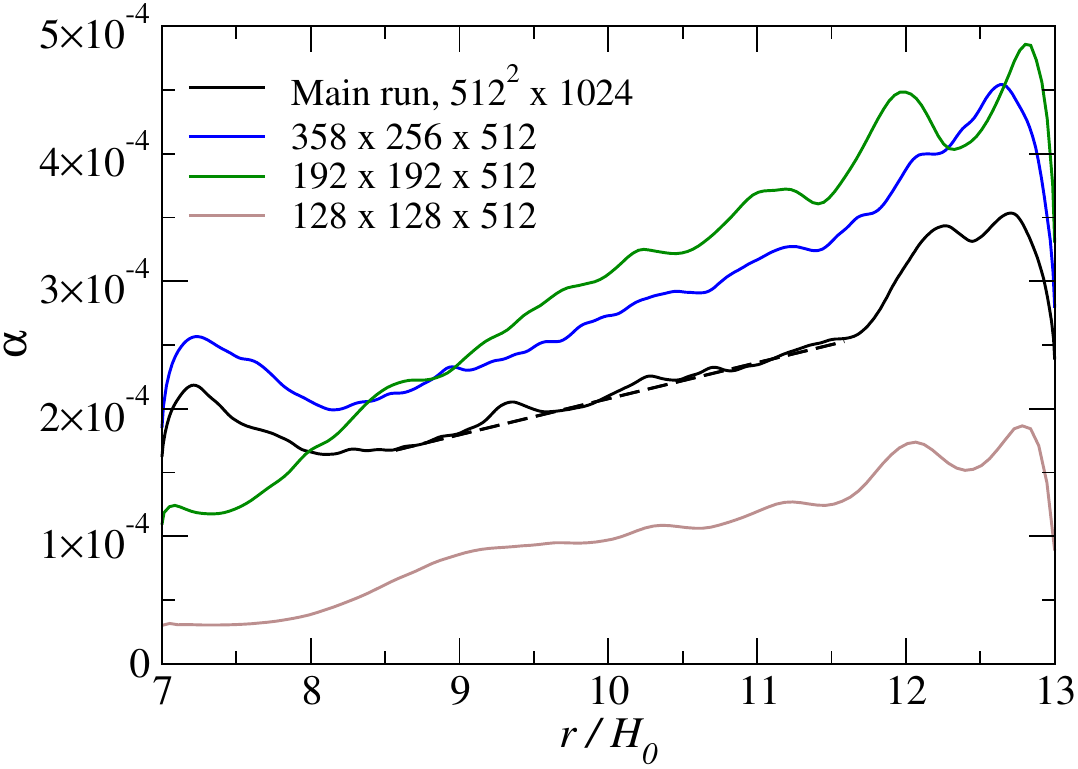}
\caption{Shakura-Sunyaev turbulent viscosity parameter $\alpha(r)$ for runs with different resolutions.  The straight dashed black line is $\alpha(r) = \left[2.0 + 0.28(r/H_0 - 10)\right] \times 10^{-4}$.}
\label{fig:alpha}
\end{figure}
The solid black curve in Figure~\ref{fig:alpha} shows $\alpha(r)$ for the most refined simulation.  Away from the region affected by radial boundaries, we obtain the linearly increasing behavior:
\be
   \alpha(r) = \left[2.0 + 0.28(r/H_0 - 10)\right] \times 10^{-4}.
\ee
An explanation one might initially suggest for the increase is the fact that the appropriate velocity scale for $\nut$ should be the driving vertical shear $\Delta u_\phi(r) = |u_\phi(r, z = H) - u_\phi(r, z = 0) |$ rather than the sound speed as assumed by \cite{Shakura_and_Sunyaev_1973}.  Unfortunately, however,
$\Delta u_\phi(r) \propto \ci(r)$ for the VSI basic state.  Another explanation one might suggest is that the radial wavelength $\ell_r(r)$ increases with $r$ faster than $H(r)$ does (\S\ref{sec:3D_wavelength}).

Figure~\ref{fig:alpha} shows that with increasing resolution (starting with the brown curve) $\alpha$ initially rapidly increases to the green curve but then decreases to the blue and black curves.  Inspection of the vorticity field for the brown case reveals minimal radial waviness of shear layers, which would contribute to a suppression of $T_{\phi r}$.  Why should the highest resolution case (black curve) have lower $\alpha(r)$ than the less resolved blue and green cases?  We conjecture that large scale waviness present at lower resolution breaks down into smaller vortices at higher resolution, leading to a smaller $T_{r\phi}$ correlation.

M20 use an alternate alpha parameter defined as
\be
   \alpha_1(r) \equiv \frac{\int_{-\infty}^{\infty} dz\, T_{R\phi}}{c_\rmi^2(r) \int_{-\infty}^{\infty} dz\, \overline{\rho}}, \eql{alpha1}
\ee
where $T_{R\phi}$ is a stress in \textit{spherical} coordinates 
\footnote{We had not noticed this until Dr. J. David Melon Fuksman pointed it out to us.}
and $\overline{\rho}$ is the mean density.  To calculate $\alpha_1(r)$ from our data in cylindrical coordinates we use the fact that
\be
   T_{R\phi} = T_{r\phi} \sin\theta + T_{z\phi}\cos\theta,
\ee
where $\theta$ is the polar angle in spherical coordinates.
\begin{figure}
\centering
\includegraphics[width=3.2truein]{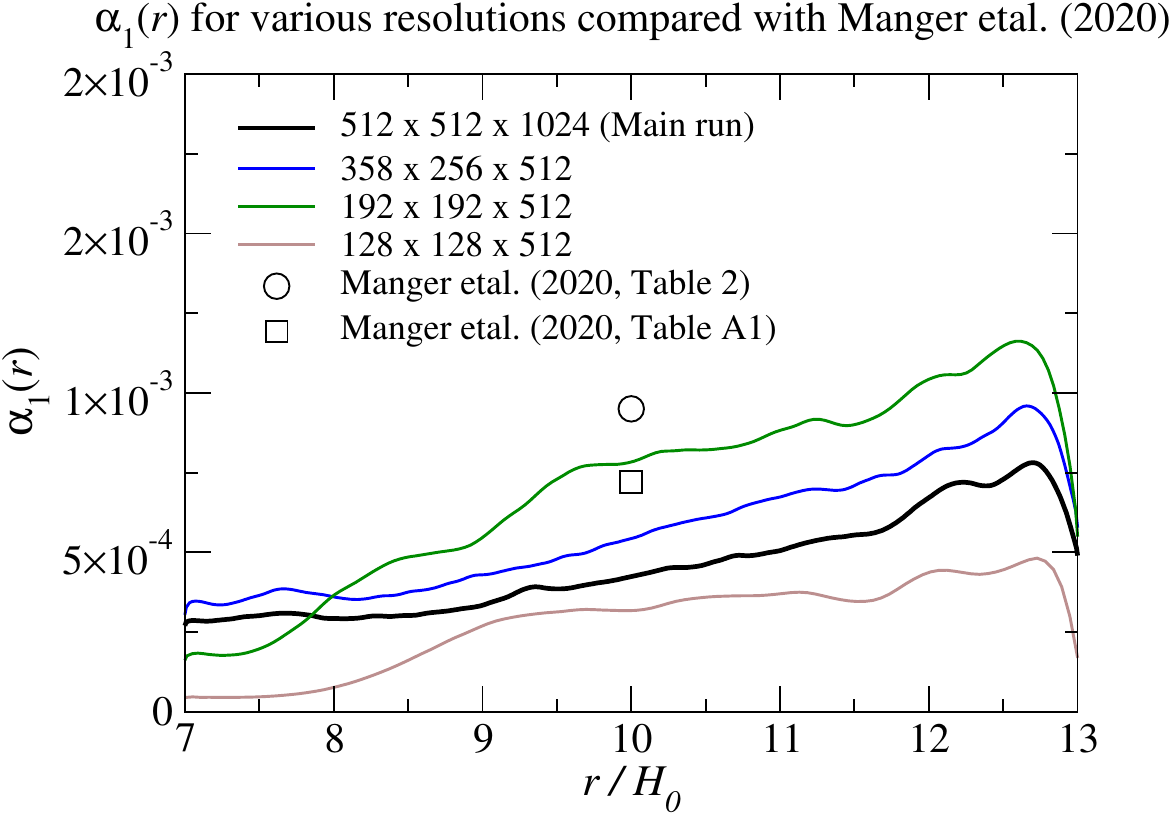}
\caption{Turbulent viscosity parameter $\alpha_1(r)$ (in spherical coordinates) as defined 
by M20; see Equation \eqp{alpha1}.  The lines are for the same cases as in Figure \ref{fig:alpha}.
$\Circle$: Value from Table 2 of M20. $\Box$: Value from Table A1 of M20.}
\label{fig:alpha1}
\end{figure}
Figure \ref{fig:alpha1} plots $\alpha_1(r)$ for the same cases as in Figure \ref{fig:alpha}.
Run p.1.5h0.1 of M20 has the same disk parameters as us.
In their Table 2 they report a radially averaged value of $\alpha_1 = (9.5 \pm 2.1) \times 10^{-4}$ where the $\pm 2.1 \times 10^{-4}$ represents the amplitude of temporal fluctuations.
This value corresponds to the period from 600 to 1000 orbits in their simulation.  The temporal average $\alpha_1 = 9.5 \times 10^{-4}$  is shown as the $\Circle$ symbol in Figure \ref{fig:alpha1}.
In their Table A1, they report a value of $\alpha_1 = (7.2 \pm 1.5) \times 10^{-4}$ which corresponds to the value in a putative stationary state at an earlier time of the simulation (H. Klahr, Private Communication); this value also corresponds to the values plotted in their Figure 1 (bottom).
Figure \ref{fig:alpha1} shows that both of these values are close to our value at midradius for a lower resolution of $192 \times 192 \times 512$ gridpoints (green curve).
Note that $\alpha_1(r) > \alpha(r)$ due to the additional contribution of the $T_{z\phi}$ stress component.

Finally, we mention in passing that \cite{Klahr_and_Bodenheimer_2003}, \citet{Stoll_and_Kley_2016}, and M20 calculate $T_{R\phi}$ using (their Equations 23, 6, and 4, respectively):
\be
T_{R\phi} = \overline{\rho u_R u_\phi} - \overline{\rho u_R}\,\overline{u_\phi}, \eql{TrpM}
\ee
which does not obey the symmetry property of the Reynolds stress.
On the other hand,  Equation \eqp{TabFinal} gives:
\be
 T_{r\phi} = \overline{\rho u_r u_\phi} - \frac{\overline{\rho u_r}\,\overline{\rho u_\phi}}{\overline{\rho}}, \eql{Trp}
\ee
which is symmetric.  When density fluctuations are small, which is true in the present case, the error in \eqp{TrpM} is small.

\subsection{Mean radial mass flux}\label{sec:mrmf}

\begin{figure}
\centering
\includegraphics[width=3.1truein,trim=2 2 2 2,clip]{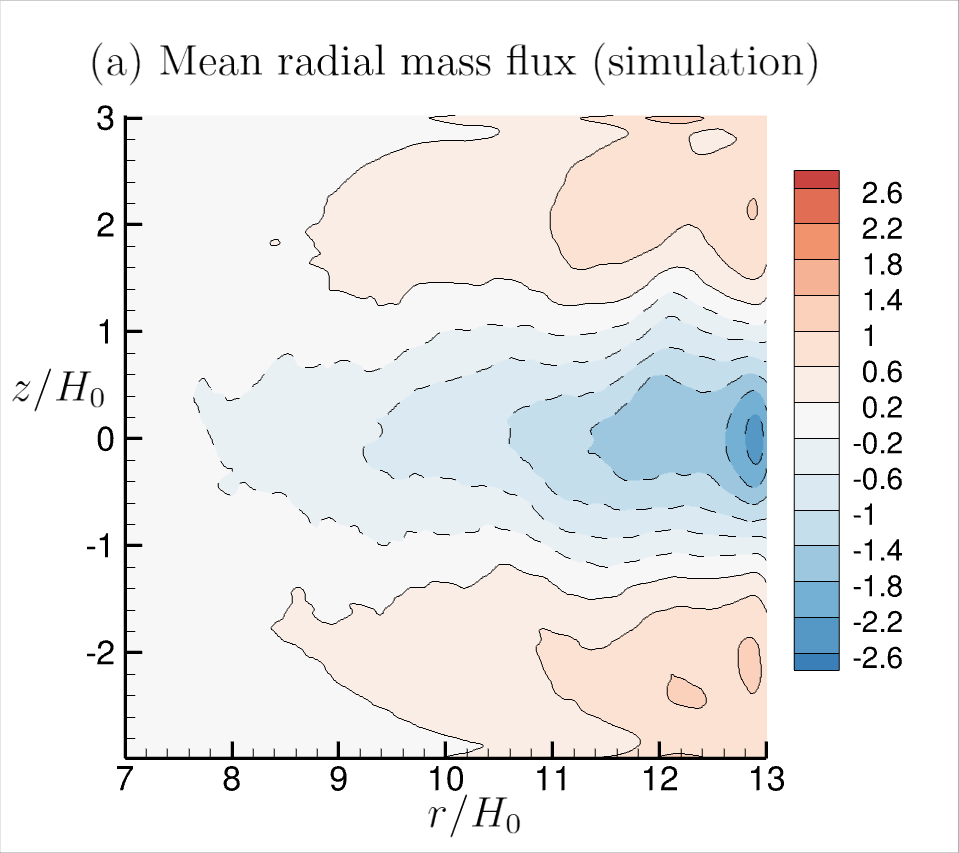}
\hfill
\includegraphics[width=3.1truein,trim=2 2 2 2,clip]{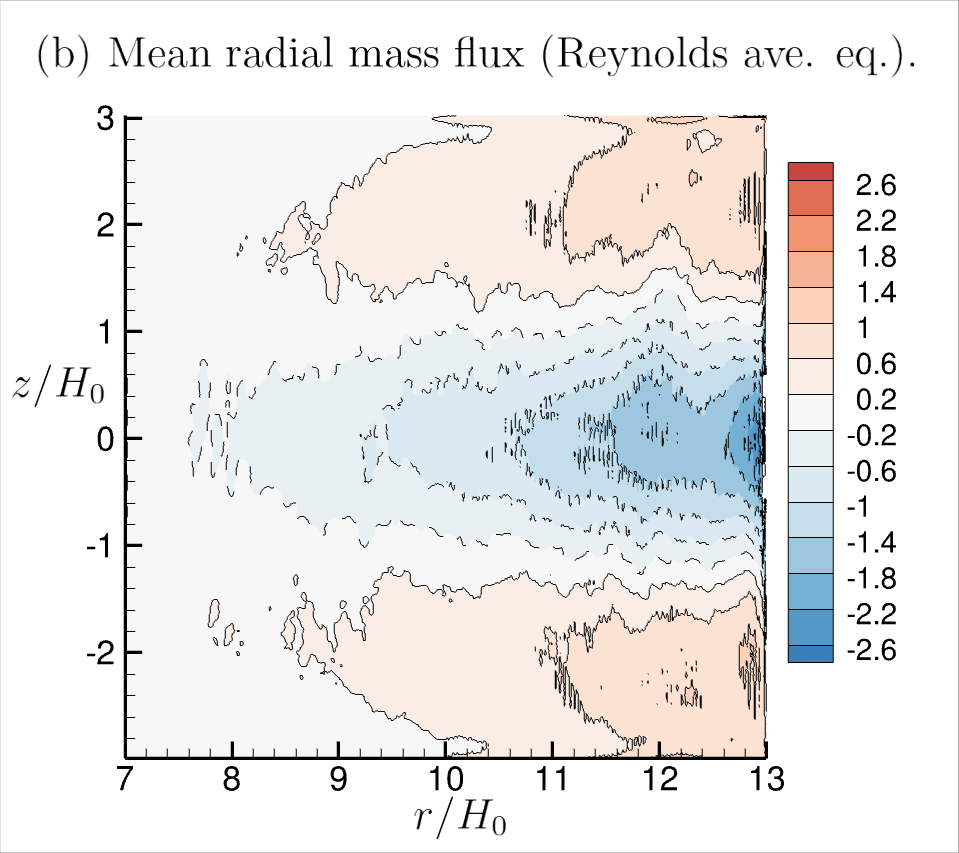} 
\hfill
\includegraphics[width=3.1truein,trim=2 2 2 2,clip]{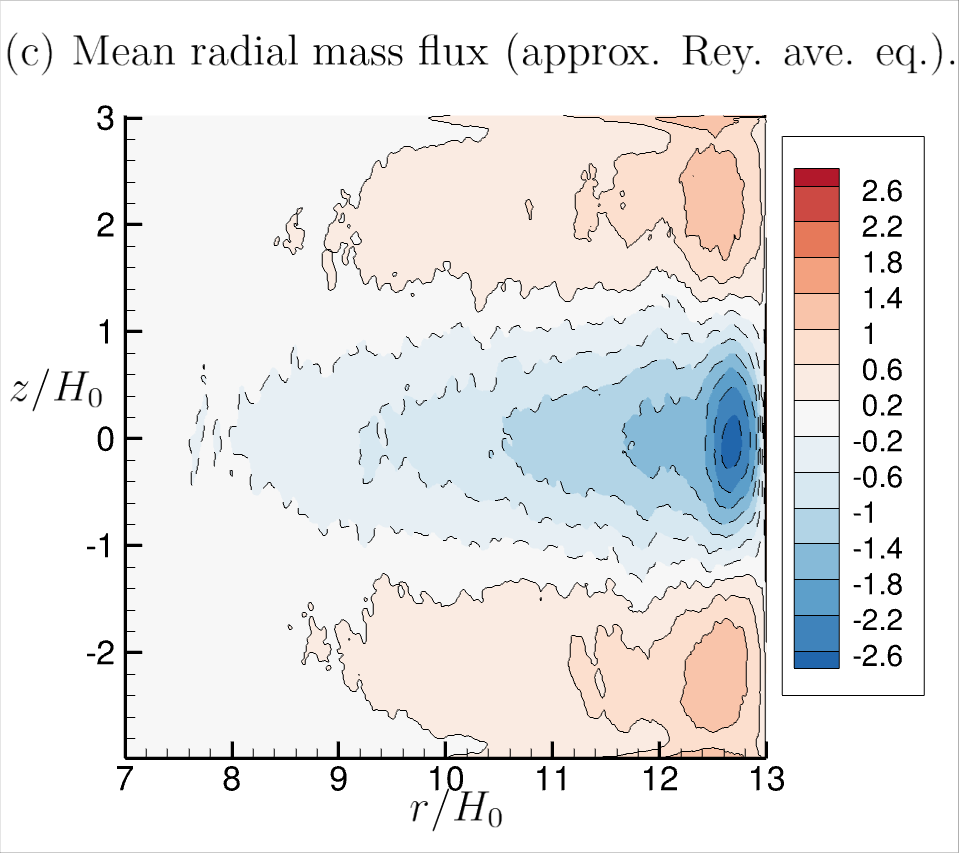}
\caption{Mean radial mass flux $\mdot = 2\pi r \overline{\rho u_r}$. (a) Simulation. (b) From \eqp{mdot}, a consequence of the Reynolds-averaged angular momentum equation.  (c) From \eqp{mdot} after assuming that the Favre averaged velocity $\upt$ is Keplerian.}
\label{fig:mass_flux}
\end{figure}
Figure~\ref{fig:mass_flux}(a) shows the mean radial mass flux (per unit height)
\be
   \dot{m} \equiv 2\pi r \overline{\rho u_r} \eql{mdot}
\ee
obtained from the simulation.  This flux is starward ($< 0$) for approximately $|z| < H_0$ and radially outward otherwise.  Such a mean radial flow was first identified by \citet[][their Figure 1]{Stoll_etal_2017}.

This flux arises from turbulent stress gradients and can be derived from the Reynolds-averaged angular momentum equation \eqp{ra:amb}.  For stationary turbulence in which statistics do not depend on time, this equation becomes
\be
   \frac{\p}{\p r}\left[
   r^2\left(
   \rhob\upt\urt + T_{\phi r}
   \right)\right]+
   r^2\frac{\p}{\p z}\left(
   \rhob\upt\uzt + T_{\phi z}
   \right) = 0.
\ee
Taking the derivative with respect to $r$, using the Reynolds-averaged mass conservation equation,
\be
   \frac{1}{r}\ppr\left(r\rhob\,\urt\right) + \ppz\left(\rhob\,\uzt\right) = 0,
\ee
solving for $\urt$, and using the definition \eqp{mdot}, one obtains
\be
   \mdot = \frac{2\pi}{\p_r(\upt r)}
   \left[
   -\ppr\left(r^2 \Tpr\right) -
   \ppz\left(r^2 \Tpz\right) -
   r^2 \rhob\,\uzt \frac{\p\upt}{\p z}
   \right]. \eql{mdot_ra}
\ee
Figure~\ref{fig:mass_flux}(b) shows that the mass flux given by \eqp{mdot_ra} agrees with the simulation (as it should) apart from statistical error.  This gives us confidence in the computation of the stress tensor.  If one introduces the assumption that $\upt$ varies weakly with $z$, then the last term in \eqp{mdot_ra} disappears.  If one further assumes that it is nearly Keplerian, i.e., $\upt \approx u_\rmK = (GM/r)^{1/2}$, then we get
\be
   \mdot \approx \frac{4\pi}{u_\rmK}
   \left[
   -\ppr\left(r^2 \Tpr\right) -
   \ppz\left(r^2 \Tpz\right)
   \right]. \eql{mdot_ra_approx}
\ee
Equation~\eqp{mdot_ra_approx} is analogous to Equation (7) in \cite{Stoll_etal_2017}.
Figure~\ref{fig:mass_flux}(c) shows that the result of \eqp{mdot_ra_approx} agrees well with the exact result except near the right end of the domain where it over predicts the mass flux.

Note that because $\Tpz(z)$ is antisymmetric about $z=0$, the second term in \eqp{mdot_ra_approx} is non-zero even at the midplane.  Therefore gradients of both $\Tpr$ and $\Tpz$ determine the accretion flow even at the midplane.  If one integrates \eqp{mdot_ra_approx} with respect to $z$ one gets the result that the net mass flow rate depends only on $\Tpr$, which is consistent with vertically integrated accretion disk theory.

The present result implies that turbulence can lead to secondary meridional flows that can be more complex than merely a starward accretion flow.  This fact is important for the transport of solids in disks; however, we emphasize, as have \cite{Stoll_etal_2017}, that solids can be transported by turbulent diffusion alone without the presence of a mean flow.

\section{Late time artifacts in the axisymmetric simulation}\label{sec:axi_artifacts}

Two artifacts arise at late times when the flow is constrained to be axisymmetric.  These are: (i) The specific angular momentum, $j_z = u_\phi r$, develops a high amplitude and vertically coherent staircase structure, first observed by \citet{Klahr_etal_2023} and MF24a,b.
This is present only very weakly in the 3D 
case (Figure~\ref{fig:jz_axi_early}(b), blue curve) in the upper layers. (ii) Vortical structures merge into larger structures, and the flow resembles the inverse cascade of 2D turbulence.  We now discuss these features.
\begin{figure}
\centerline{
\includegraphics[width=3.4truein]{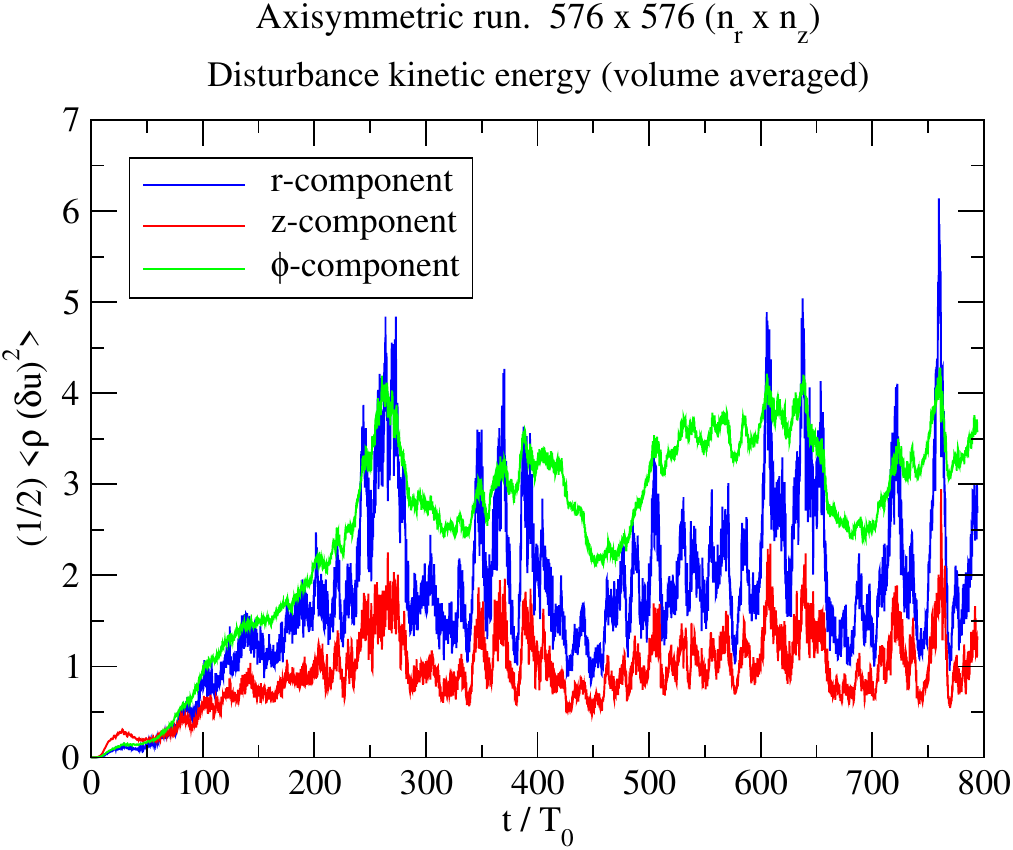}
                }
\caption{Time history of the volume averaged perturbation kinetic energy for the axisymmetric run.}
\label{fig:ke_axi}
\end{figure}

Figure \ref{fig:ke_axi} shows the domain averaged fluctuation kinetic energy components relative to the basic state.  Compared to the 3D run (Figure \ref{fig:ke}), a statistically stationary state is reached very much later (at about $t/T_0 \approx 300$).  This observation is not without precedent.  The high resolution axisymmetric VSI simulation ({\tt dg3c4\_1024}) of MF24a (their Figure 12, red curve)  does not reach a stationary state even after about 1300 orbits when the simulation stops.  It should also be noted that this case has a false and temporary stationary state up to about 600 orbits, after which the energy begins to rise again.  Their lower resolution case ({\tt dg3c4\_512}) exhibits a false period of stationarity up to 3000 orbits, after which the kinetic energy begins to rise again.  
The axisymmetric runs of \cite{Flores-Rivera_etal_2020} reach a stationary state after 100 orbits and were run to a final time of 200 orbits.  Given the potential for false and temporary periods of stationarity in axisymmetric runs, caution must be exercised before concluding that a true stationary state has been reached.

Given that flow length scales increase with time in our simulation (see below) as well as in the simulations of M24a, i.e., that there is an inverse cascade, the time to reach stationarity in any given simulation would depend on the radial domain size.

One also observes that compared to the 3D run (see Figure \ref{fig:ke}), the volume averaged perturbation energy values are much larger, and the individuals components are less isotropic. The vertical component, which is the strongest for $t/T_0 < 25$, now takes a back seat to the other two.

\begin{figure}
\centering
\includegraphics[width=3.4truein,trim=2 2 2 2,clip]{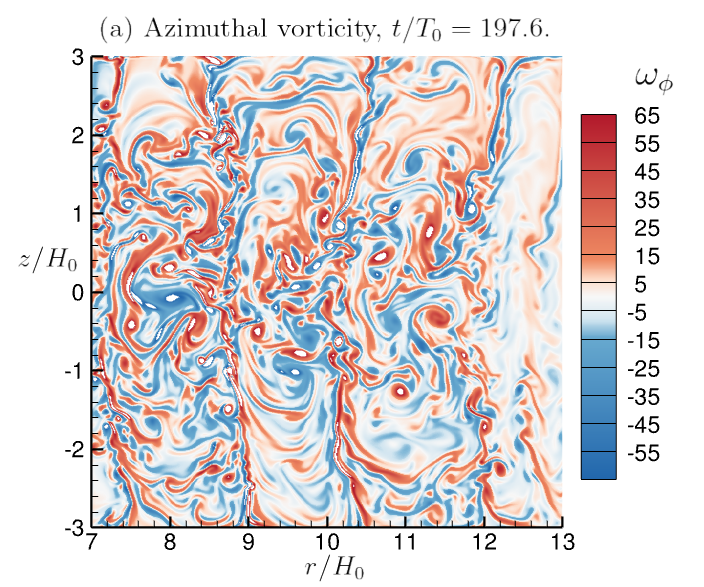} 
\includegraphics[width=3.4truein,trim=2 2 2 2,clip]{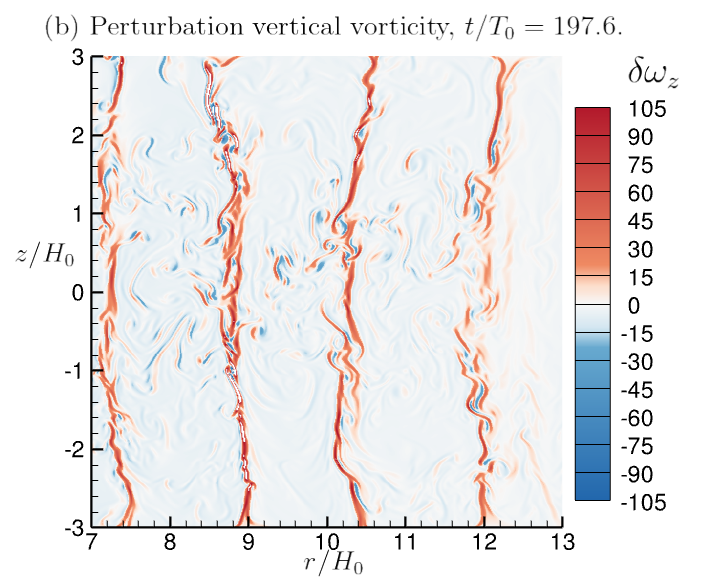} 
\includegraphics[width=3.4truein,trim=2 2 2 2,clip]{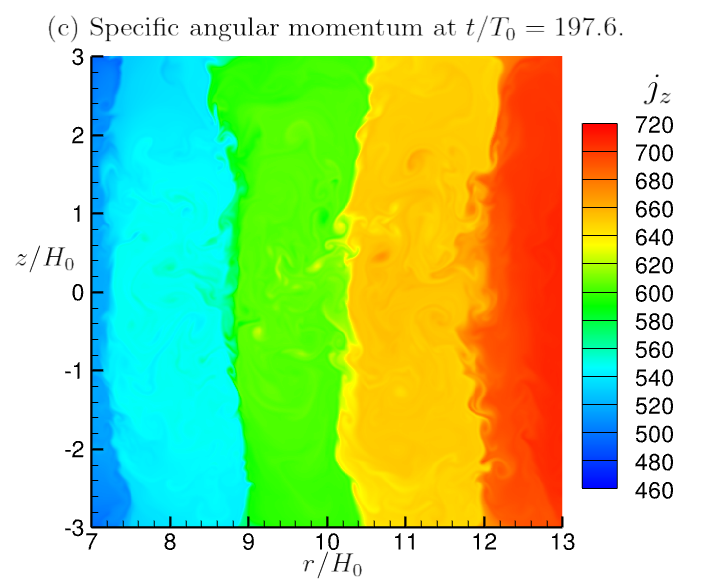}
\caption{Axisymmetric simulation at $t/T_0 = 197.6$.  This is a late time but is prior to the stationary state.  In panel (a), the completely white regions have vorticity values beyond the range shown in the legend.}
\label{fig:vor_axi2}
\end{figure}
Figure~\ref{fig:vor_axi2} shows the perturbation azimuthal and vertical vorticity components at $t/T_0 = 197.6$, which is at a late time but prior to the stationary state.  Their structure is very different from the 3D case and what is observed at early times in the axisymmetric case. The dominant component now is $\delta\omega_z$ rather than $\omega_\phi$, which translates to a larger magnitude of the azimuthal velocity perturbation compared to the vertical velocity.  The shear layers of $\omega_\phi$ that formed the edges of vertical jets at earlier times 
are still present as is the pattern $\omega_\phi \lessgtr 0$ for $z  \gtrless 0$ within these shear layers.  However, the flow between the shear layers consists of large eddies and is reminiscent of 2D turbulence.  Panel (b) shows that the $\approx 7$ layers of vertical vorticity seen at $t/T_0 = 36.1$ (Figure \ref{fig:vor_axi}) have now intensified, and only 4 remain.  At a later time ($t/T_0 = 403.7$, not shown) only 2 such layers remain.  In other words, flow length scales increase with time.  The layers of strong $\delta\omega_z$ are associated with jumps in the total specific angular momentum $j_z = u_\phi r$ (panel (c)), which has a strong vertically coherent staircase profile with respect to $r$.  
\begin{figure}
\centering
\includegraphics[width=3.3truein]{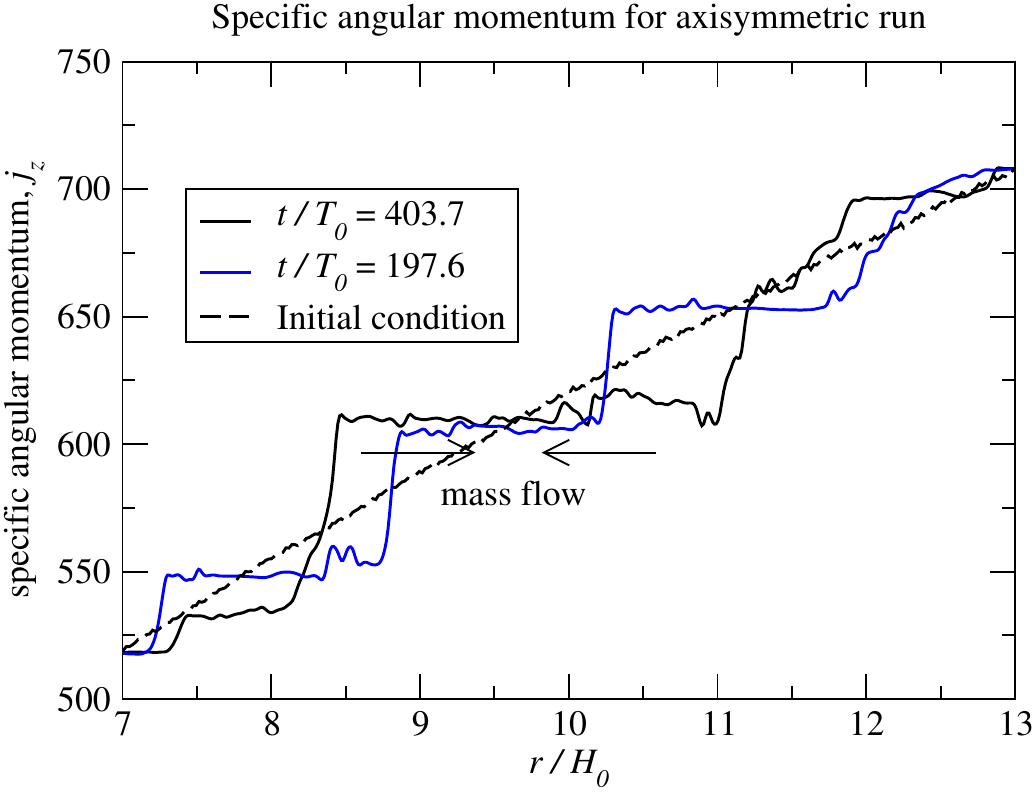}
\caption{Axisymmetric run at late times.  Specific angular momentum profile, $j_z = u_\phi r$, at the midplane.}
\label{fig:jz}
\end{figure}
This can be seen more clearly in Figure~\ref{fig:jz}, which shows the midplane profiles of $j_z(r)$ at $t/T_0 = 197.6$ and $t/T_0 = 403.7$.  The latter is in the stationary state and has only 2 jumps in $j_z$.
As shown earlier (Figure~\ref{fig:jz_axi_early}, blue line), in 3D there is no large scale rearrangement of angular momentum, rather only a weak one in the upper layers of the disk.
The circulation,  $\Gamma_z = 2\pi j_z$, of circular material lines is proportional to $j_z$.  Hence by Stokes' theorem, a region of nearly constant $\Gamma_z(r)$ must be irrotational.
For inviscid barotropic flow, Kelvin's theorem asserts that that $D\Gamma_z/Dt = 0$ for each circular material line.  
Hence there are three possible mechanisms for the formation of the staircase structure: vertical transport of $j_z(z)$ given its vertical gradient (this being the mechanism that operates at early times), radial mixing of $j_z(r)$ by eddies between the shear layers, or baroclinic torque.  For a circular material line we have that
\be
  \frac{D\Gamma_z}{Dt} = \int_{A} \frac{1}{\rho^2} \left(\nabla\rho \times \nabla p\right)\cdot \widehat{\mathbf{z}}dS,
\ee
where $A$ is the circular surface enclosed by the circle and $\widehat{\mathbf{z}}$ is the unit vector normal to the surface.  However, the $z$-component of the baroclinic term in the integrand is zero for axisymmetric flow.  The vertical gradient of $j_z$ is weaker than its radial gradient, and radial velocity fluctuations are larger than vertical ones at late times.  Hence, we conclude that radial mixing of $j_z$ by eddies in regions bounded by vertical shear layers is responsible for the staircase profile of $j_z$ at late times.

Accretion disk theory tells us that when angular momentum is redistributed, so is mass.  Specifically, a fluid element that has a super-Keplerian $j_z$ will move radially outward, and the opposite for sub-Keplerian $j_z$.  A $j_z(r)$ profile that has flattened from a radially increasing Keplerian profile will cause mass to accumulate.  
\begin{figure}
\centering
\includegraphics[width=3.4truein,trim=2 2 2 2,clip]{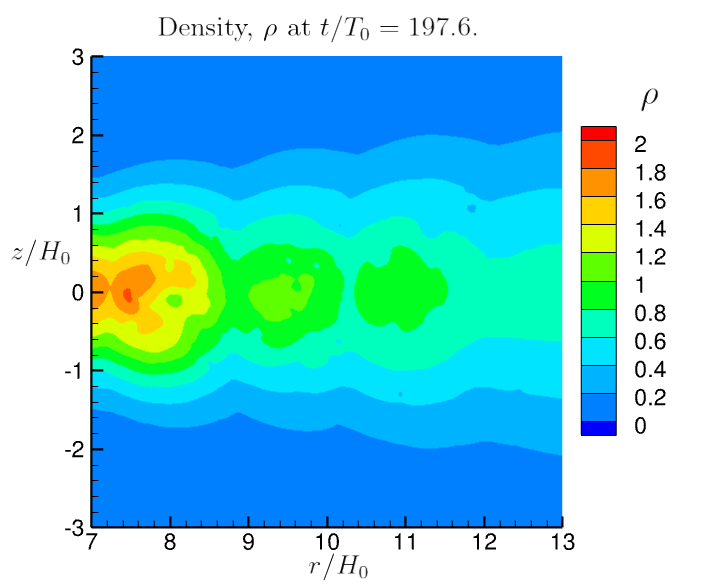} 
\caption{Density field for the axisymmetric simulation at $t/T_0 = 197.6$.}
\label{fig:rho_197p6}
\end{figure}
Figure~\ref{fig:rho_197p6} shows that mass becomes concentrated in the uniform-$j_z$ regions observed in Figure~\ref{fig:jz} 

In the axisymmetric simulations of MF24a there is also a clear difference between early and late times.  Let us focus on their {\tt dg3c4\_2048} simulation.  Their Figure 1 (left hand plots) and Figure 2 (top) shows very many small steps in the $j_z$ staircase at an earlier time.  On the other hand, their Figure 11 shows the same run at 1360 orbits.  Clearly, an increase of length scales has taken place leading to only 3 jumps in $j_z$.

\section{Present lack of large persistent vortices in the interior of the domain}
\label{sec:long_lived}
Previous VSI simulations \citep{Manger_and_Klahr_2018, Manger_etal_2020, Flock_etal_2020,Pfeil_and_Klahr_2021} have reported the existence of one or more large persistent vortices (LPVs) and attributed their formation to a secondary Rossby wave instability (RWI).  We do not observe such vortices in the region not influenced by radial boundary conditions.  The purpose of this section is to discuss this inconsistency in more detail.

%
%

\subsection{Caveats for applying RWI to VSI simulations}

RWI was first studied by \citet[][henceforth L99]{Lovelace_etal_1999} assuming a {\it razor thin} disk and conservation of the entropy 
\be
   S = P / \Sigma^\gamma
\ee
following fluid particles.  Here $P$ is the vertically integrated pressure, $\Sigma$ is the surface density, and $\gamma$ is the adiabatic exponent.  A necessary condition for RWI is that, in the basic state, the function
\be
   \calL(r) = \frac{\Sigma(r)}{2\omega_z(r)} \left[S(r)\right]^{2/\gamma}
\ee
has a local maximum.  The thin disk assumption is valid only for motions whose horizontal length scale $\ell \gg H(r)$.  On the other hand, the radial size of LPVs observed in simulations is about $H(r)$.  In addition, the locally isothermal equation of state, $p = \rho \ci^2(r)$, assumed in the present and other VSI simulations implies that radial motions \textit{fail to conserve} $S$ following fluid particles. Likewise, it remains to be shown whether \textit{radiative} hydrodynamic VSI simulations that invoke RWI \cite[e.g.,][]{Flock_etal_2020} possess an effective adiabatic exponent.
Despite the lack of adiabaticity, previous works plot a vertically averaged $\calL(r)$, presumably setting $\gamma = 1.4$.  In spite of these caveats, it appears that the presence of LPVs in VSI simulations is correlated with some (but not all) local maxima in $\calL(r)$ and we will discuss these simulations in the next subsection.

In previous VSI simulations, a diagnostic version of $\calL(r)$ is defined using averages.  For our case of a locally isothermal equation of state, $p = \rho c_\rmi^2(r)$, we define the diagnostic
\newcommand\Sigmaave{\left<\Sigma\right>_{\phi t}}
\newcommand\omegazave{\left<\omega_z\right>_{\phi z t}}
\be
   \calL_\mathrm{diag}(r) =  \frac{\Sigmaave}{2\omegazave}
   \left[\frac{\Sigmaave c_\rmi^2(r)}{\Sigmaave^\gamma}\right]^{2/\gamma},
\ee
with $\gamma = 1.4$ and where, for example, $\left<.\right>_{\phi z t}$ denotes an average with respect to $\phi$, $z$, and $t$.

\subsection{Previous VSI simulations that produced LPVs}

The simulations of \cite{Flock_etal_2020} employed radiative hydrodynamics and produced a peak in $\alpha(r)$ (their Figure 2) in a region adjacent to the inner radial boundary.  They explain the peak as being due to a radial decrease in thermal relaxation time (their Figure 13) to values that make VSI viable.  
The peak in $\alpha(r)$ led to a local dip in $\Sigma(r) $ (their Figure 1) and a pronounced maximum in the vertical vortensity $\omega_z(r, z = 0)/\Sigma$ (their Figure 5).  These features in turn led to pronounced extrema in the RWI indicator function $\calL(r)$ (their Figure B1), which is associated with one LPV.   No \textit{other} LPVs are generated in the rest of the domain.  In summary, the trigger for an LPV in \cite{Flock_etal_2020} is a local $\alpha(r)$ peak induced by a favorable gradient in radiative relaxation time rather than by VSI dynamics.

More relevant is the work of M20 who,
for the same disk parameters as us, observed the formation of multiple LPVs far from radial boundaries; see the lower right panels in their Figures 8 and 10.
Their $t$-$R$ vortex trajectory plot (their Figure 12, lower right panel) shows that the vortices have lifetimes of hundreds of orbits.
Their Figure B1 (right-hand column) show that the radial locations of these vortices are close to some but not all of the many maxima of their $\calL$ diagnostic.

\subsection{Present simulations}

\begin{figure}
\centering
\includegraphics[width=3.4truein]{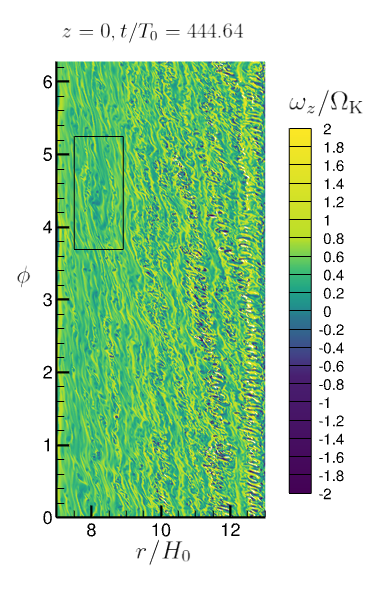}
\caption{$\omega_z/\Omega_\rmK(r)$ in the midplane for the highest resolution run.}
\label{fig:Manger_style}
\end{figure}
Figure~\ref{fig:Manger_style} plots $\omega_z/\Omega_\rmK(r)$ in the midplane for our main run; this is the same quantity plotted by M20.
Note that their color map is such that values of $\omega_z / \Omega_\rmK(r) > 0.5$ have the same color (yellow) as $\omega_z / \Omega_\rmK(r) = 0.5$, and values of $\omega_z / \Omega_\rmK(r) < 0$ have the same color (dark blue) as $\omega_z / \Omega_\rmK(r) = 0$.  As a result, their vortices appear darker than in our rendering.
The background Keplerian value is $\omega_z/\Omega_\rmK(r) = 0.5$ and is rendered light green in Figure~\ref{fig:Manger_style}.  The only large scale coherent structure present is indicated by the box; it consists of several little whirls and filaments.  It is centered at $r/H_0 \approx 8.2$ which is close to a local minimum in $\alpha(r)$; see black curve in Figure~\ref{fig:alpha}.  This minimum is a result of proximity to the left radial boundary.  
A similar structure is present at this radial location as early as $t/T_0 = 69.5$.

\begin{figure}
\centering
\includegraphics[width=3.2truein]{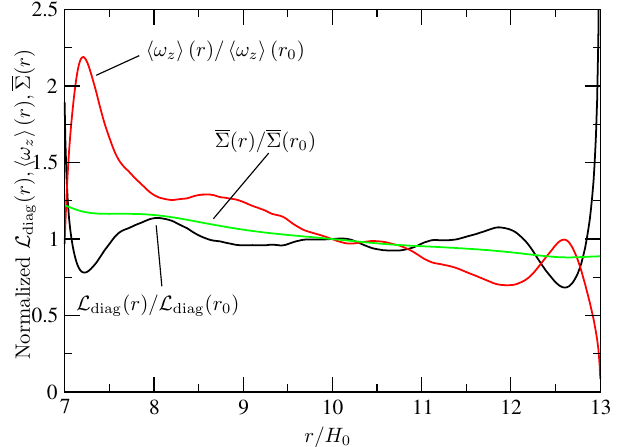}
\caption{Black line: The diagnostic $\mathcal{L}_\mathrm{diag}(r)$ for the Rossby wave instability (RWI).  Red line: Vertical vorticity $\omega_z$ averaged with respect to $t, \phi$ and $z$.  Green line: Surface density $\Sigma$ averaged with respect to $t$ and $\phi$. (Main run with a resolution of $512^2 \times 1024$.) }
\label{fig:L_Rossby}
\end{figure}
Figure \ref{fig:L_Rossby} plots the RWI diagnostic $\mathcal{L}_\mathrm{diag}(r)$ (black line) normalized by its value at $r_0$: it has local maximum at $r/R_0 = 8.03$, which is close to the radial location of the vortex structure.  The red line in Figure \ref{fig:L_Rossby} shows that the surface density is a local minimum at this location.  Another local maximum in  $\mathcal{L}_\mathrm{diag}(r)$ is present at $r/H_0 \approx 12$ but is not associated with a large vortex.

Besides RWI, another mechanism known to produce LPVs is the subcritical baroclinic instability (SBI).  The first requirement for instability is that the radial Brunt-V\"ais\"al\"a frequency squared
\be
   N_r^2 = -\frac{1}{\gamma\rho} \frac{\p p}{\p r} \frac{\p}{\p r} \ln\left(\frac{p}{\rho^\gamma}\right),
\ee
be negative.  For the present basic state (\S\ref{sec:basic_state}) we have at the midplane that
\be
   N_r^2 = \frac{c_0^2}{\gamma r_0^2} (p + q) \left[p(\gamma-1) - q\right] (r/r_0)^{q - 2}. \eql{Nr2}
\ee
For our case of $p = -3/2$ and $q = -1$, the coefficient in \eqp{Nr2} involving $p$ and $q$ equals $-1$; the first requirement is therefore satisfied. However, SBI also requires intermediate cooling times ($t_\mathrm{rad}$) \citep{Lesur_and_Papaloizou_2010,Barge_etal_2016}, whereas we have assumed that $t_\mathrm{rad} = 0$.

\subsection{Speculation}

Why do M20 observe LPVs in the interior of the domain for the same disk parameters but we do not?  The answer may lie in the link, discussed by MF24b, between maxima in $\calL(r)$ and the $j_z(r)$ staircase that is prominent in axisymmetric $j_z(r)$.  Specifically, a region of flattened $j_z(r) = u_\phi r$ corresponds to more rrotational flow ($u_\phi \approx \mathrm{const}/r$) and therefore suppressed $\omega_z$.  Since mass also accumulates in these regions, a local maximum is created in the first factor ($\Sigma/ 2 \omega_z$) of $\calL$.  Therefore, one possibility is that M20 has a stronger $j_z(r)$ staircase in 3D than the present simulation, which leads to several $\calL(r)$ maxima in the interior of the domain.  It is tempting to attribute this to M20's lower resolution; 
however, our low resolution simulations also failed to produce LPVs away from boundaries.  Further effort is required to ascertain the reasons for this difference between our and the M20 simulation.
 
\section{Closing remarks}

This work studied the vortical structure, Reynolds stresses, and midplane spectra of 3D turbulence driven by vertical shear instability (in the locally isothermal limit) for disk parameters $h = 0.1, p = -1.5$, and $q = -1$.  The salient points are:
\begin{enumerate}
\item \label{item:1} \textit{Radial wavelength.} In axisymmetric VSI, the radial wavelength at early times corresponds to the most amplified linear mode, but then inexorably increases in time.  In 3D, the ratio
\be
   \frac{\ell_r(r)}{\lammax(r)} = \left[\pi |q| (H_0/R_0)\right]^{-1} \frac{\ell_r(r)}{H(r)}
\ee
of the radial wavelength, $\ell_r(r)$, to the most linearly amplified wavelength ($\lammax$)
increases from 2.4 to 4.4 as $r$ increases from $8.6 H_0$ to $11.6 H_0$.
\item \textit{Layers of positive $\delta\omega_z$.} Axisymmetric VSI  has coherent vertical layers of $\delta\omega_z > 0$ that increase in strength with $|z|$ and are coincident with interfaces between the upward and downward jets (Figure \ref{fig:vor_axi}).  
Their formation and the preference for $\delta\omega_z > 0$ can be explained by the vertical transport of specific angular momentum $j_z \equiv u_\phi r$ by the vertical jets.  This phenomenon is equivalent to the formation of a staircase profile of $j_z(r)$ first discovered by \cite{Klahr_etal_2023} and MF24a,b in their axisymmetric simulations.
We find that these layers are also present in 3D but with much weaker jumps in specific angular momentum compared to the axisymmetric case (Figure~\ref{fig:jz_axi_early}(b)).  These layers become stronger and more azimuthally coherent with increasing $|z|$ and undergo KH-like instability (Figure \ref{fig:vort_2H}a), likely modified by background shear and rotation.  The resulting vortices are small and possess all three vorticity components.
\item \textit{Lack of large persistent $\omega_z$ vortices.} Absent are the large persistent RWI-driven vortices in the interior of the domain obtained by M20 for the same disk parameters.  Rather, we observe only one large coherent structure (containing finer scale features) near the left boundary of the domain where $\alpha(r)$ has an anomalous minimum due to boundary effects; the RWI diagnostic $\calL_\mathrm{diag}(r)$ has a local maximum at this location (Figure \ref{fig:L_Rossby}).  Flat regions in a $j_z(r)$ staircase, which is prominent in axisymmetric VSI, have a suppressed $\omega_z$ and accumulate mass which leads to peaks in the RWI indicator $\calL(r)$ \citep{Melon_Fuksman_etal_2024_I}.  One may speculate that, due to lower resolution, M20 obtain a more pronounced $j_z$ staircase in 3D than we do.  However, our lower resolution simulations also fail to display large persistent vortices.  Therefore, further effort is needed to sort out this issue.
\item \textit{Azimuthal vorticity.} In the axisymmetric case at early times, there is the following sign preference in the jet shear layers: $\omega_\phi \lessgtr 0$ for $z  \gtrless0$.  
By examining the two terms in the transport equation for $\omega_\phi$, we conclude that the perturbation in the vertical shear term is the dominant contributor to the production of $\omega_\phi$.  
The sign preference noted above is also visually noticeable in 3D.
\item \textit{Midplane vorticity.} In 3D, the midplane contains 
many sheared filamentary vortical layers as well as many compact (non-filamentary) vortices of roughly elliptical shape, having $\delta\omega_z < 0$ and comparable values of $\delta\omega_r$ and $\delta\omega_\phi$.
\item \label{item:alpha} \textit{Variation of $\alpha$ with $r$.} There is a \textit{linear increase of $\alpha(r)$} with radius with a slope (at the highest resolution) of about 14\% the midradius value per scale height.  This increase currently lacks an explanation; 
it may be related to the increase of radial spacing, $\ell_r(r)$,  with $r$ that is faster than $H(r)$ (see item \ref{item:1}, which itself requires an explanation).
\item \textit{Variation of $\alpha$ with  resolution.} As the resolution is increased from a low value, $\alpha$ at first increases but then decreases.  M20 (their Table A1)  observe an increase in $\alpha$ as resolution in increased for the same disk parameters as in the present work.  Our value of $\alpha$ for an intermediate resolution case is close to the value obtained by M20 for their highest resolution case.  It would be desirable in the future to obtain an $\alpha(r)$ that is converged as the grid is refined.
\item \textit{Meridional flow.} We confirm the finding of \cite{Stoll_etal_2017} that there is a mean meridional radial flow toward the star for $|z| < H_0$ and outward for $|z| > H_0$ and that this flow is created by gradients of Reynolds shear stresses.
\item \textit{Radial spectrum.} The spectrum of specific kinetic energy (velocity squared) at the midplane with respect to the radial wavenumber ($k_r$) was found to have a power law region with an exponent of $-1.82$, close to the value of $-2$ for shear layers \citep{Townsend_and_Taylor_1951}.  The spectrum is strongly dominated by the vertical velocity component.
\item \textit{Azimuthal spectrum.} The spectrum with respect to the azimuthal wavenumber has $-5/3$ power law range, but with a small depression due to radial velocity fluctuations.  Unlike M20 (their Figure 6), we do not find a $-5$ range. 
\item \textit{Late time artifacts in axisymmetric VSI.} As discussed above, some features of the 3D simulation can be understood by appealing to physics in the early stage of the axisymmetric simulation.  However, at late times, the axisymmetric simulations have artifacts that are not present in 3D.  These include increasing flow length scales and a strong staircase pattern in the specific angular momentum $j_\phi = u_\phi r$ with an accompanying redistribution of mass.
\end{enumerate}

We are grateful to Dr. J. David Melon Fuksman (Max Planck Institut f\"ur Astronomie, Heidelberg) for his detailed comments, to Drs. Paul Estrada and Ted Manning at NASA Ames for performing a careful internal review, and to the referee for many useful suggestions that led to an improved manuscript.
OMU acknowledges partial support from NASA's TCAN (Theoretical and Computational Astrophysical Networks) grant \#80NSSC21K0497.

%


%
\appendix

\section{Pad\'e filter}\label{PadeFilterDetails}

This briefly describes the fourth-order Pad\'e filter \citep{Lele_1992} which is applied after every time step as an implicit sub-grid treatment.  More details can be found in \cite{Shariff_2024}.  The attractiveness of Pad\'e filters is their sharp cut-off. For a given unfiltered quantity $u_i$, its filtered outcome $U_i$ on either a periodic domain, or within the interior section of a non-periodic domain, is given by the solution of the following system of equations for every line of data in the mesh:
\be
a U_{j-1} + U_j + aU_{j+1} = P(u_{j-2}+u_{j+2}) + Q(u_{j-1}+u_{j+1}) + Ru_j,
\eql{filter}
\ee
for $j = 1, \ldots n$, where $n$ is the number of points along the line.
For non-periodic directions, the equations to be used at two points adjacent to each boundary are given in \citet{Shariff_2024}.
Equations \eqp{filter} constitute a tridiagonal system that is solved along each data line in the mesh using the Thomas algorithm.

Regarding the coefficient $Q$ in \eqp{filter} as a free parameter, the condition that the leading order difference between $U_j$ and $u_j$ be $h^4$, where $h$ is the grid spacing, gives the rest of the coefficients as
\be
a = -\frac{1}{2} + 2Q, \hskip 0.4truecm
R = \frac{1}{2}\left(2 + 3a - 3Q\right), \hskip 0.4truecm
P = \frac{1}{4}\left(a - Q\right).  \eql{fc} 
\ee
There is no filtering when $Q = 1/2$.  To specify the strength of the filter, the code uses the parameter $\epsilon_\mathrm{filter}$ such that
\be
   Q = \frac{1}{2} - \frac{1}{4}\epsilon_\mathrm{filter}. \eql{q}
\ee
The transfer function $T(k)$ of the filter versus wavenumber can be obtained by substituting $u_j = e^{\rmi k x_j}$ and $U_j = T(k) e^{\rmi k x_j}$ (with $x_j = jh$) into \eqp{filter}.
\begin{figure}
\centering
\includegraphics[width=3.0truein]{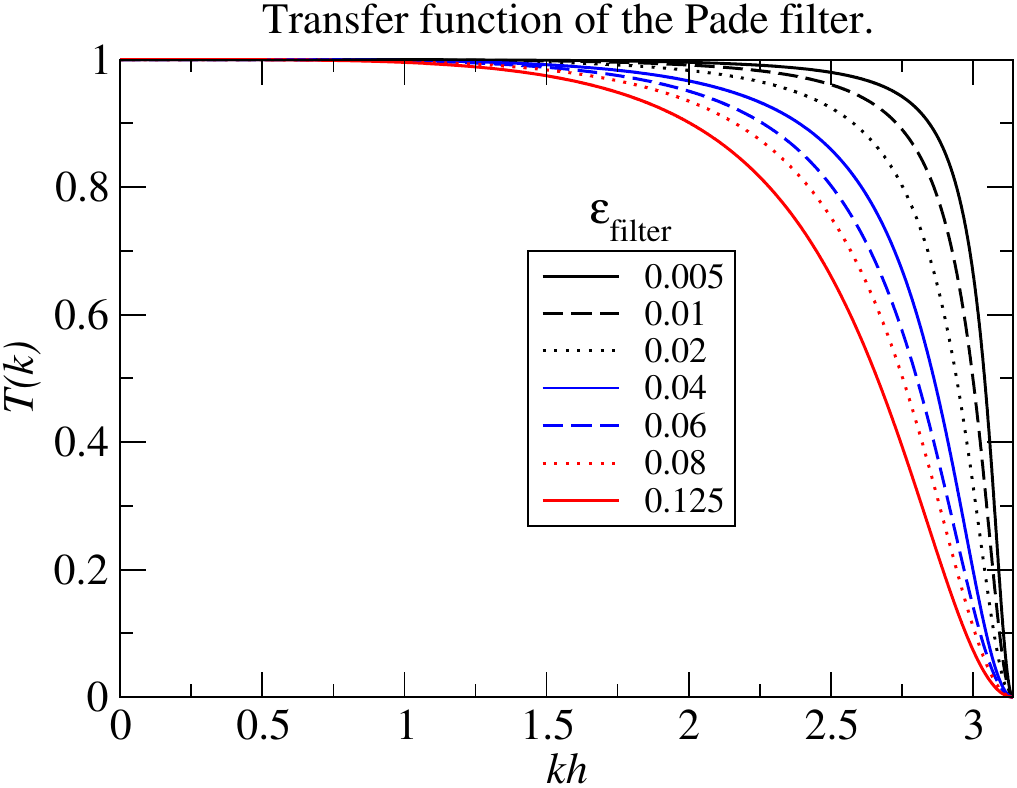}
\caption{Transfer function for the Pad\'e filter for various $\epsfil$ covering the range used in the present simulations.}
\label{fig:keff}
\end{figure}
Figure~\ref{fig:keff} shows $T(k)$ for values of $\epsfil$ that cover the range,
$\epsfil \in [0.015, 0.125]$, used in the present simulations.

\section{Vorticity equation for axisymmetric VSI}

For conservative body forces, the vorticity equation is given by
\be
\frac{\p\bfomega}{\p t} + \bfu{\cdot}\nabla\bfomega = -\bfomega\nabla{\cdot}\bfu + \bfomega{\cdot}\nabla\bfu + \frac{\nabla\rho \times \nabla p}{\rho^2}. \eql{vort_general}
\ee
The terms on the rhs of \eqp{vort_general} represent the change of vorticity by volumetric compression or expansion, tilting and stretching, and baroclinic torque.  The compression/expansion term
can be eliminated by introducing $\bfzeta \equiv \bfomega/\rho$.  \eqp{vort_general} then becomes
\be
\frac{\p\bfzeta}{\p t} + \bfu{\cdot}\nabla\bfzeta = \bfzeta{\cdot}\nabla\bfu + \frac{\nabla\rho \times \nabla p}{\rho^3}. \eql{vort_over_rho}
\ee
Using the expressions for $\bfn{\cdot}\nabla\bfF$ in cylindrical coordinates given in \citet[Appendix 2]{Batchelor_1967} we have
\be
\bfu\cdot\nabla\bfzeta = (\bfu\cdot\nabla\zeta_z)\,\bfe_z  + 
\left(\bfu\cdot\nabla\zeta_r + \frac{u_\phi\zeta_\phi}{r}\right)\bfe_r +
\left(\bfu\cdot\nabla\zeta_\phi + \frac{u_\phi\zeta_r}{r}\right)\bfe_\phi
\ee
and
\be
\bfzeta\cdot\nabla\bfu = (\bfzeta\cdot\nabla u_z)\,\bfe_z + 
\left(\bfzeta\cdot\nabla u_r - \frac{\zeta_\phi u_\phi}{r}\right)\bfe_r +
\left(\bfzeta\cdot\nabla u_\phi + \frac{\zeta_\phi u_r}{r}\right)\bfe_\phi
\ee
For $p = \rho \ci^2(r)$, the baroclinic term becomes
\be
\frac{\nabla\rho \times \nabla p}{\rho^3} = 
\frac{1}{\rho^2}\frac{\p\rho}{\p z}\frac{\p \ci^2}{\p r} \,\bfe_\phi.
\ee
Note that the baroclinic term acts only in the azimuthal direction.  Let us begin by working on the $\phi$ vorticity equation.  Using the fact that
\be
\zeta_r = - \frac{1}{\rho}\frac{\p u_\phi}{\p z} \hskip 0.25truecm \mathrm{and}
\hskip 0.25truecm \zeta_z = \frac{1}{\rho r}\frac{\p}{\p r}(r u_\phi),
\eql{zeta_defs}
\ee
we get that
\be
\bfzeta\cdot\nabla u_\phi = \zeta_z\frac{\p u_\phi}{\p z} + 
\zeta_r \frac{\p u_\phi}{\p r} = \frac{1}{2\rho r}\frac{\p u_\phi^2}{\p z}.
\ee
This term, which survives below, represents the tilting of $\zeta_r$ and $\zeta_z$ into $\zeta_\phi$.  Since $\zeta_z$ includes the strong Keplerian vorticity, this is an important effect.  So far, therefore, the $\phi$ vorticity equation becomes
\be
\frac{D\zeta_\phi}{Dt} + \frac{u_\phi\zeta_r}{r} = \frac{\zeta_\phi u_r}{r} + \frac{1}{2\rho r}\frac{\p u_\phi^2}{\p z} + 
\frac{1}{\rho^2}\frac{\p\rho}{\p z}\frac{\p \ci^2}{\p r}.
\ee
This can be simplified further by noting that
\be
\frac{D}{Dt}\left(\frac{\zeta_\phi}{r}\right) =
\frac{1}{r}\frac{D\zeta_\phi}{Dt} - \frac{\zeta_\phi u_r}{r^2}.
\ee
This gives
\be
r \frac{D}{Dt}\left(\frac{\zeta_\phi}{r}\right) + \frac{u_\phi \zeta_r}{r} = \frac{1}{2\rho r}\frac{\p u_\phi^2}{\p z} + 
\frac{1}{\rho^2}\frac{\p\rho}{\p z}\frac{\p \ci^2}{\p r}.
\ee
Finally, introducing the expression in \eqp{zeta_defs} for $\zeta_r$ into this gives the $\phi$ vorticity equation:
\be
\boxed{
\frac{D}{Dt}\left(\frac{\omega_\phi}{\rho r}\right) = \frac{1}{\rho r^2}\frac{\p u_\phi^2}{\p z} + \frac{1}{\rho^2 r}\frac{\p\ci^2}{\p r}\frac{\p\rho}{\p z}. \eql{vort_phi_eq}
}
\ee
Note that if the flow is nearly incompressible, i.e., $D\rho/Dt \approx 0$, then $\rho$ can be removed from under the $D/Dt$.
The hint for writing \eqp{vort_phi_eq} in this form is an exercise in \citet[][p.~517]{Batchelor_1967}.  The presence of $r$ in the quantity being transported simply reflects the fact that for axisymmetric flow, the stretching of $\omega_\phi$ is purely geometric, i.e., as a circular vortex line increases in radius by a certain ratio, its $\omega_\phi$ would increase by the same ratio, if the rhs of \eqp{vort_phi_eq} were zero.
Equation \eqp{vort_phi_eq} implies that the creation of azimuthal vorticity, which takes the form of vertical jets in VSI, is influenced by both vertical shear and baroclinic torque.  In the basic state, the two terms on the rhs of \eqp{vort_phi_eq} oppose each other to give a zero lhs, and the signs of the two terms above and below the midplane are
\be
   0 = \frac{---}{+++} + \frac{+++}{---},
\ee
where a horizontal line represents the miplane.
The $z$ and $r$ vorticity equations are
\begin{align}
    \frac{D\omega_z}{Dt} &= \omega_r\frac{\p u_z}{\p r} + \omega_z\frac{\p u_z}{\p z} \eql{vort_eq_z} - \omega_z\nabla\cdot\bfu\\
    \frac{D\omega_r}{Dt} &= \omega_z\frac{\p u_r}{\p z} + \omega_r\frac{\p u_r}{\p r} - \omega_r \nabla\cdot\bfu. \eql{vort_eq_r}
\end{align}
The three terms on each rhs represent tilting, stretching, and volumetric compression, respectively.  Note again that the baroclinic term enters only the $\omega_\phi$ equation.

\section{Reynolds-averaged conservation equations using the Favre decomposition}\label{sec:ra}

The purpose of Appendices \ref{sec:ra} and \ref{sec:cra} is to rigorously derive the residual stress that arises when the governing equations are Reynolds averaged.  The presentation is necessary given the different expressions used in the simulation literature.

In the absence of magnetic torques and when the gravitational potential ($\Phi$) is axisymmetric, the inviscid equations for conservation of mass and momentum equations read:
\begin{align}
&\frac{\p\rho}{\p t} + \nabla\cdot\left(\rho\bfu\right) = 0, \eql{mass}\\
   &\frac{\p}{\p t}\left(\rho u_\phi r\right) + 
   \nabla\cdot\left(\rho u_\phi r \bfu\right) = 
 - \frac{\p p}{\p \phi}, \eql{amom} \\
   &\ppt\left(\rho u_\rc\right) + \nabla\cdot\left(\rho u_\rc\bfu\right) -
\rho\frac{u_\phi^2}{\rc} = 
-\frac{\p p}{\p \rc} - \rho\frac{\p\Phi}{\p \rc}, \eql{rmom} \\
& \ppt\left(\rho u_z\right) + \nabla\cdot\left(\rho u_z \bfu\right) = -\frac{\p p}{\p z} - \rho \frac{\p\Phi}{\p z}, \eql{zmom}
\end{align}
where the divergence operator is
\be
   \nabla\cdot\bfF \equiv \frac{\p F_z}{\p z} + \frac{1}{r}\frac{\p}{\p r}\left(r F_r\right) + \frac{1}{r}\frac{\p F_\phi}{\p\phi} = 0. \eql{div}
\ee

We take the average of these equations with respect to azimuth and time over a period $[t_1, t_2]$ long enough to ensure converged statistics and during which the flow has attained statistical stationarity in time.  The averaging operator, denoted by a bar is, therefore:
\be
   \overline{f}(r, z)= \frac{1}{2\pi (t_2 - t_1)} \int_0^{2\pi} d\phi \int_{t_1}^{t_2} dt f(r, z, \phi, t).
\ee
This average satisfies all the axioms required for being a \textit{Reynolds} average.  Throughout, one can omit the time average if desired; the only change this results in is the appearance of $\p/\p t$ of mean quantities in \eqp{ra:massbar}--\eqp{ra:z}.

Consider the Reynolds average, $\overline{\rho u
v}$, of a generic triple product which appears in the momentum equations.  If we introduce the
Reynolds decomposition $u = \ubar + \up$ and $v = \vb + \vp$ we get
\be
   \overline{\rho u v} = \rhob\,\ubar\,\vb + \overline{\rho\up\vp} + 
                          \ubar\,\overline{\rho\vp} + \vb\,\overline{\rho\up}. \eql{unwieldy}
\ee
The last two terms make this unwieldy.  Note that \citet[][Equation 15]{Stoll_and_Kley_2014} and \citet[][p. 2623]{Nelson_etal_2013}  ignore the last two terms in their computation of the $r\phi$ Reynolds stress and perform the averaging over the inhomogeneous radial direction.

To allow averages of triple products involving the density $\rho$ to be written compactly, one introduces a density weighted average, denoted by a tilde, which is known as a \citet{Favre_1969} average:
\be
   \widetilde{f} \equiv \frac{\overline{\rho f}}{\rhobar}.
\ee
Fluctuations with respect to the Favre average are denoted by double primes:
\be
   f = \ftilde + \fpp. \eql{decomp}
\ee
An important relation which we shall use below is that
\ba
   \overline{\rho \fpp} = \overline{\rho \left(f - \ft\,\right)} = 
   \overline{\rho f} - \overline{\rho\ft} &=&
   \overline{\rho f} - \ft\,\rhob \\
   &=& \overline{\rho f} - \frac{\overline{\rho f}}{\rhob} \rhob = 0, \eql{relation}  
\ea
since $\ft$ is already an average.
Specifically, \eqp{relation} allows one to write triple products as:

\ba
   \overline{\rho u v} &=& \overline{\rho\,\ut\,\vt} + \overline{\rho\upp\vpp} + \overline{\rho\ut\vpp}
                       + \overline{\rho\upp\vt} \\
 &=& \rhob\,\ut\,\vt + \overline{\rho\upp\vpp} + \ut\,\overline{\rho\vpp} + \vt\,\overline{\rho\upp}\\
 &=& \rhob\,\ut\,\vt + \overline{\rho\upp\vpp},
\ea
which is a little more compact than \eqp{unwieldy}.

Decomposing flow variables as in \eqp{decomp}, applying the bar average to the conservation equations \eqp{mass}--\eqp{zmom}, using the above relations, assuming statistical stationarity and that the gravitational potential has no fluctuation, gives
\begin{align}
   &\nabla_{\rc z} \cdot (\rhob \bfut) = 0, \eql{ra:massbar}\\
   &\nabla_{\rc z}\cdot\left(\rhob\, \ut_\phi r \bfut + 
                         \overline{\rho \uppp r \bfupp}\right) = 0, \eql{ra:amb}\\
   &\nabla_{\rc z}\cdot
   \left(\rhob\,\ut_\rc\,\bfut+\overline{\rho\,\urpp\,\bfupp}\right) -
   \rhob\frac{\left(\ut_\phi\right)^2}{\rc} - \frac{\overline{\rho {\uppp}^2}}{\rc} = 
   -\frac{\p \pb}{\p \rc}
   - \rhob\frac{\p \Phi}{\p r}, \eql{rmb}\\
   &\nabla_{\rc z}\cdot\left(\rhob\,\ut_z\bfut +
                        \overline{\rho\,\uzpp\,\bfupp}\right) = -\frac{\p \pb}{\p z} - \rhob \frac{\p\Phi}{\p z}, \eql{ra:z}
\end{align}
where $\nabla_{rz}\cdot$ denotes the divergence operator \eqp{div} restricted to the $rz$ plane.  The above equations contain the Reynolds stresses
\be
   T_{ab} = \overline{\rho\,\uapp\,\ubpp}, \hskip 0.5truecm a,b \in \left\{z, r, \phi\right\}.
\ee

\section{Calculation of Reynolds stresses}\label{sec:cra}

This appendix explains how to efficiently calculate Reynolds stresses with fluctuations defined with respect to the time-$\phi$ averaged flow.  By efficiently we  mean without having to store entire flow fields in order to obtain the time mean.
The Reynolds stress is defined as
\be
   T_{ab} \equiv \overline{\rho\uapp\ubpp},
\ee
where the bar denotes an average with respect to $\phi$ and $t$, while $a$ and $b$ represent the  cylindrical coordinate indices $r$, $z$, or $\phi$.  A double prime denotes a Favre fluctuation, e.g.,
\be
   \uapp = u_a - \ut_a, \eql{uapp}
\ee
where the tilde signifies the Favre average:
\be
   \ut_a \equiv \overline{\rho u_a} / \overline{\rho}.
\ee

As a simulation proceeds, one does not have the time average and therefore cannot calculate the temporal fluctuations needed to accumulate $T_{ab}$.  However, the following simple algebra allows one to store only $\phi$ averages at sample times and calculate the time average as a post-processing step. 
Using \eqp{uapp} we have
\be
   T_{ab} \equiv \overline{\rho\uapp\ubpp} = \overline{\rho (u_a - \ut_a)(u_b - \ut_b)}. \eql{Tab}
\ee
Expanding \eqp{Tab} gives
\be
   T_{ab} = \overline{\rho u_a u_b} + \overline{\rho \ut_a \ut_b} - \overline{\rho \ut_a u_b} - \overline{\rho u_a \ut_b}.
\ee
Now, every Favre average such as $\ut_a$ is independent of $\phi$ and $t$ and can be pulled out from under the bar operation, giving
\be
   T_{ab} = \overline{\rho u_a u_b} + \rhob\ut_a\ut_b - \ut_a \overline{\rho u_b} - \ut_b \overline{\rho u_a}.
\ee
Next, using the definition of the Favre average gives
\be
   T_{ab} = \overline{\rho u_a u_b} + \rhob \frac{\overline{\rho u_a}}{\rhob} \frac{\overline{\rho u_b}}{\rhob} - \frac{\overline{\rho u_a}}{\rhob}\overline{\rho u_b} - \frac{\overline{\rho u_b}}{\rhob}\overline{\rho u_a}.
\ee
Finally, the second and third terms cancel giving the final expression:
\be
   T_{ab} = \overline{\rho u_a u_b} - \frac{1}{\rhob} \overline{\rho u_a}\,\,\overline{\rho u_b}. \eql{TabFinal}
\ee
To compute \eqp{TabFinal} we store $\phi$ averages of $\rho u_a u_b$, $\rho u_a$, and $\rho$
at a sample of times during the simulation, during the period when statistical stationarity has been achieved.
Then, as post-processing step, a further time average is computed to complete the bar operation.  Finally, the Reynolds stresses are obtained using \eqp{TabFinal}.  Equation \eqp{TabFinal} can also be used when the time average is skipped and only a $\phi$ average is included.

\bibliography{disks.bib}
\bibliographystyle{aasjournal}



\end{document}